\documentclass[apj,iop,twocolappendix,numberedappendix]{emulateapj}
\usepackage{times,mathptmx}
\usepackage{amsmath,amssymb}
\usepackage{hyperref}
\def\substitute@command#1#2{% 
 \ClassWarning{aastex}{% 
  Command \protect#1\space is deprecated in aastex. 
  Using \protect#2\space instead (please fix your document). 
 }% 
 #2% 
}%  

\newcommand{\footlabel}[2]{%
    \addtocounter{footnote}{1}%
    \footnotetext[\thefootnote]{%
        \addtocounter{footnote}{-1}%
        \refstepcounter{footnote}\label{#1}%
        #2%
    }%
    $^\mathrm{\scriptsize\ref{#1}}$%
}

\newcommand{\footref}[1]{%
    $^\mathrm{\scriptsize\ref{#1}}$%
}

\def \mthi {\tau_\mathrm{eff,HI}}
\def \thi {$\mthi$}
\def \mtheii {\tau_\mathrm{eff,HeII}}
\def \theii {$\mtheii$}

\def \lya  {Ly$\alpha$}

\slugcomment{ApJ, in press; received 2015 December 18; accepted 2016 May 28}

\begin{document}

\shorttitle{Extended Helium Reionization}
\shortauthors{Worseck et al.}

\title{Early and Extended Helium Reionization over more than 600 Million Years of Cosmic Time\footnotemark[0]}

\footnotetext[0]{
Based on observations made with the NASA/ESA \textit{Hubble Space Telescope}, obtained at
the Space Telescope Science Institute, which is operated by the Association of Universities
for Research in Astronomy, Inc., under NASA contract NAS5-26555. These observations are
associated with Program \#11742. Archival \textit{Hubble Space Telescope} data 
(\#7575, 9350, 11528, 12178, 12249) were obtained from the Mikulski Archive for
Space Telescopes (MAST). Several \textit{Hubble Space Telescope} programs
provided ancillary calibration data (\#11860, 11895, 12414, 12423, 12716, 12775, 12870, 13108).
Some of the data presented herein were obtained at the W.M. Keck Observatory, which is operated
as a scientific partnership among the California Institute of Technology, the University of California and NASA;
it was made possible by the generous financial support of the W.M. Keck Foundation.
Based on observations made with ESO Telescopes at the La Silla Paranal Observatory
under program IDs 166.A-0106, 071.A-0066 and 083.A-0421.
}

\author{
G\'abor Worseck\altaffilmark{1,2}, J.~Xavier Prochaska\altaffilmark{1},
Joseph~F. Hennawi\altaffilmark{2}, Matthew McQuinn\altaffilmark{3}
}

\altaffiltext{1}{Department of Astronomy and Astrophysics, UCO/Lick Observatory,
University of California, 1156 High Street, Santa Cruz, CA 95064, USA}
\altaffiltext{2}{Max-Planck-Institut f\"{u}r Astronomie, K\"{o}nigstuhl 17, 69117 Heidelberg, Germany}
\altaffiltext{3}{Department of Astronomy, University of Washington, 3910 15th Ave NE, Seattle, WA 98195, USA}
\email{gabor@mpia-hd.mpg.de}

\begin{abstract}
We measure the effective optical depth of \ion{He}{2} \lya\ absorption \theii\
at $2.3<z<3.5$ in 17 UV-transmitting quasars observed with UV spectrographs
on the \textit{Hubble Space Telescope} (\textit{HST}).
The median \theii\ values increase gradually from $1.95$ at $z=2.7$ to $5.17$
at $z=3.4$, but with a strong sightline-to-sightline variance.
Many $\simeq 35$ comoving Mpc regions of the $z>3$ intergalactic medium (IGM)
remain transmissive ($\mtheii<4$), and the gradual trend with redshift appears
consistent with density evolution of a fully reionized IGM.
These modest optical depths imply average \ion{He}{2} fractions of
$x_\mathrm{HeII}<0.01$ and \ion{He}{2} ionizing photon mean free paths of
$\simeq 50$ comoving Mpc at $z\simeq 3.4$, thus requiring that a substantial
volume of the helium in the Universe was already doubly ionized at early times;
this stands in conflict with current models of \ion{He}{2} reionization driven by
luminous quasars.  
Along 10 sightlines we measure the coeval \ion{H}{1} Ly$\alpha$ effective
optical depths, allowing us to study the density dependence of \theii\ at
$z\sim 3$. We establish that the dependence of \theii\ on increasing
\thi\ is significantly shallower than expected from simple models of
an IGM reionized in \ion{He}{2}. This requires higher \ion{He}{2} photoionization
rates in overdense regions or underdense regions being not in photoionization equilibrium.
Moreover, there are very large fluctuations in \theii\ at all \thi\, which greatly
exceed the expectations from these simple models. These data present a
distinct challenge to scenarios of \ion{He}{2} reionization -- an IGM
where \ion{He}{2} appears to be predominantly ionized at $z\simeq 3.4$, 
and with a radiation field strength that may be correlated with the density field,
but exhibits large fluctuations at all densities.
\end{abstract}

\keywords{
dark ages, reionization, first stars -- diffuse radiation -- intergalactic medium
-- quasars: absorption lines
}

\section{Introduction}

Despite the significant attention devoted to resolving the nature of
hydrogen reionization, the final major phase transition on cosmological scales
ended more than one Gyr later (at $z\sim 3$) when helium was stripped of its
second electron. This process, termed \ion{He}{2} reionization,  
required a radiation field with $h\nu > 54.4$\,eV photons which 
was probably driven by high-$z$ quasars
\citep[e.g.][]{miralda-escude00,mcquinn09a,haardt12,compostella13,compostella14}.
This follows empirically from the observation that the emissivity of
$z\sim 3$ quasars is sufficient to reionize \ion{He}{2}
\citep[e.g.][]{furlanetto08,faucher09,haardt12}.

The reionization of \ion{He}{2} has several important cosmological consequences.
The photoionization of \ion{He}{2} deposits thermal energy into the
intergalactic medium (IGM), heating the gas that gives rise to the
\ion{H}{1} Ly$\alpha$ forest
\citep[e.g.][]{hui97,furlanetto08b,bolton09,mcquinn09a,becker11,compostella13,compostella14,puchwein15}.
The temperature changes in the IGM -- with time and overdensity -- depend
on the duration of \ion{He}{2} reionization, the distribution of
sources, and their spectral energy distributions
\citep{tittley07,bolton09,mcquinn09a,compostella13,compostella14,puchwein15}.
The gradual rise in the IGM temperature from $z=5$ to $z\simeq 2.8$ suggests that
\ion{He}{2} reionization may have been an extended process \citep{becker11,boera14}
that could have started around the first luminous quasars at $z\simeq 6$
\citep{bolton12,madau15}. 

Another consequence is the hardening of the extragalactic UV
background radiation field. This impacts the ionization states of
metals in the IGM \citep[e.g.][]{madau09,bolton11} and possibly gas
within galaxies \citep{vladilo03}. Photoionization modeling of metal
line systems broadly constrains the spectral shape of the UV
background \citep{agafonova05,agafonova07,fechner11}, but there are
remaining degeneracies with the absorber metallicity and relative
abundances \citep{bolton11,fechner11}. In turn, the derived physical
properties of metal line systems rely on the adopted UV background
spectrum \citep[e.g.][]{faucher09,haardt12}, and a fluctuating UV
background during \ion{He}{2} reionization
\citep[e.g.][]{furlanetto09} will induce systematic uncertainties in
the derived absorber properties.
As UV background models rely on the adopted source emissivities and
the Lyman continuum absorption in the IGM, the relative contributions of quasars
and star-forming galaxies to the UV background, as well as the absorber properties
derived from photoionization models remain significantly uncertain even in the
post-reionization IGM \citep{kollmeier14,shull14,shull15,khaire15,madau15}.

Several groups have explored the physics that governs \ion{He}{2} reionization
-- photoionization/recombination, cosmological expansion, gas heating/cooling --
with analytic and numerical techniques
\citep{fardal98,miralda-escude00,gleser05,tittley07,furlanetto08,faucher09,
furlanetto09,furlanetto10,mcquinn09a,haardt12,tittley12,compostella13,compostella14}.
These studies generally reproduce an IGM in which \ion{He}{2} was reionized at
$z\sim 3$, driven by large ($\sim 10$\,Mpc) \ion{He}{3} bubbles around luminous
quasars. These bubbles percolate during a time-interval of $\sim 1$\,Gyr
($3\la z\la 5$), typically requiring a couple of phases of quasar activity
in a given region \citep{compostella13,compostella14}.

There remains significant uncertainty in the precise timing and morphology of
\ion{He}{2} reionization, however, because several key inputs are poorly
constrained or have not been adequately modeled. These include
(1) the duty cycle, spectral energy distribution, and opening angles of quasars;
(2) the number density of faint quasars at $z_\mathrm{em}>3$
(e.g.\ \citealt{glikman11} and \citealt{giallongo15} vs.\ \citealt{masters12});
(3) the incidence and distribution of absorbers that self-shield to 54\,eV photons; and
(4) the possible contribution from more exotic sources such as
thermal emission from shocked gas in galaxy halos \citep{miniati04},
X-ray emission from stellar binaries or massive black holes \citep[e.g.][]{venkatesan01,ricotti04,power09,mcquinn12},
or \ion{He}{2}-ionizing emission from $z_\mathrm{em}\sim 3$ galaxies \citep{furlanetto08}.
Uncertainties in these areas of the $z\ga 3$ Universe lead to significant
differences in the \ion{He}{2} reionization and its impact on the $z\la 3$ IGM.
In contrast to \ion{H}{1} in the $z\ga 6$ Universe, however, it is possible to
independently constrain several of these unknowns observationally.
In particular, one can (1) characterize the properties of
$z_\mathrm{em}\sim 3$ quasars, and (2) probe the density field of the $z\sim 3$
IGM with spectroscopy of the optically thin \ion{H}{1} Ly$\alpha$ forest.

The only direct means of studying \ion{He}{2} reionization, however, is through
absorption spectroscopy of the \ion{He}{2} Ly$\alpha$ transition at
rest wavelength $\lambda_\mathrm{rest} = 303.78$\,\AA, accessible to
far-UV (FUV) sensitive space telescopes at $z>2$. Analogous to
\ion{H}{1} Ly$\alpha$ studies of the IGM near the putative epoch of
\ion{H}{1} reionization \citep[e.g.][]{becker01,white03,fan06}, models predict
a transition from a forest of \ion{He}{2} absorption lines at $z\sim 2$ to
troughs of complete absorption as the number of IGM patches with significant
\ion{He}{2} fractions rises with redshift, signaling the epoch of helium reionization.
However, for the first decade of \textit{HST} operations only seven \ion{He}{2}
sightlines probed this transition, because for most $z_\mathrm{em}>2.7$ quasars
the spectral range covering \ion{He}{2} absorption is extinguished by intervening
optically thick \ion{H}{1} absorbers \citep{picard93,worseck11}.
\textit{HST} low-resolution ($R=\lambda/\Delta\lambda<2,000$) spectra of the four sightlines covering $z>3$
(Q~0302$-$003, PKS~1935$-$692, SDSS~J2346$-$0016, SDSS~J1711$+$6052)
generally reveal large \ion{He}{2}
effective optical depths $\tau_\mathrm{eff,HeII}>3$, indicating an
incomplete \ion{He}{2} reionization
\citep{jakobsen94,hogan97,heap00,anderson99,zheng04b,zheng08}.
At $2.7\la z\la 2.9$, the patchy \ion{He}{2} absorption recorded in three
sightlines (HE~2347$-$4342, Q~0302$-$003, HS~1157$+$3143) hint at the `overlap'
of \ion{He}{3} zones \citep{reimers97,reimers05,heap00,smette02}, akin to the case of
hydrogen at $z\ga 6$ \citep{gnedin00}. At $z<2.7$, the low \ion{He}{2}
absorption in the sightlines to HE~2347$-$4342 and HS~1700$+$6416
has been resolved into an emerging \ion{He}{2}
forest with the \textit{Far Ultraviolet Spectroscopic Explorer}
(\textit{FUSE}, $R\approx 20,000$),
indicating that \ion{He}{2} reionization ended at $z\sim 2.7$
\citep{kriss01,zheng04,shull04,fechner06}.
Higher-quality observations of HE~2347$-$4342 and HS~1700$+$6416
with the Cosmic Origins Spectrograph \citep[COS;][]{green12} have confirmed these results \citep{shull10,syphers13}.
The $R\approx 18,000$ COS spectrum of Q~0302$-$003 for the first time
resolved the onset of the \ion{He}{2} \lya\ forest and the still patchy
$z>2.9$ \ion{He}{2} absorption in this sightline \citep{syphers14}.

For the five \ion{He}{2} sightlines recorded at sufficiently high
spectral resolution ($R\ga 800$), coeval spectra
of the optically thin \ion{H}{1} Ly$\alpha$ forest provide estimates
of the number density ratio $n_\mathrm{HeII}/n_\mathrm{HI}$. In a
fully reionized IGM $n_\mathrm{HeII}/n_\mathrm{HI}$ probes the
spectral shape of the UV background and its source population
\citep[e.g.][]{miralda-escude90,madau94,fardal98,haardt12}, so
observational constraints are of great astrophysical interest.  
The patches of strong quasi-continuous \ion{He}{2} absorption at
$z>2.7$ typically require $n_\mathrm{HeII}/n_\mathrm{HI}\ga 300$
\citep{reimers97,reimers05,heap00,smette02,shull10,syphers14}, whereas
the $z<2.7$ \ion{He}{2} Ly$\alpha$ forest revealed order-of-magnitude
fluctuations around $n_\mathrm{HeII}/n_\mathrm{HI}\sim 100$ on scales
down to $\sim 1$\,Mpc
\citep{kriss01,zheng04,shull04,fechner06,fechner07}. However, these
studies were affected by various systematics, most importantly by
uncertainty in the zero level of the \textit{FUSE} data
\citep{fechner07} and \ion{H}{1} continuum uncertainty
\citep[e.g.][]{heap00}. We recently showed that by accounting for
\ion{H}{1} continuum systematics with realistic mock spectra from
numerical simulations, the higher-quality COS spectra from the two
sightlines sampling $z<2.7$ are consistent with
$n_\mathrm{HeII}/n_\mathrm{HI}\simeq 100$ without the need for
fluctuations exceeding a factor of two \citep{mcquinn14}. This implies
that in the post-reionization IGM, radiative transfer effects
\citep{maselli05,tittley07,tittley12} or Poisson fluctuations in the
number density of quasars \citep{bolton06} generate only modest UV
background fluctuations. 
However, all existing inferences on the epoch of \ion{He}{2}
reionization have been tempered by the very small sample of sightlines
available to study \ion{He}{2} \lya\ absorption. Indeed, expanding the
dataset has been the most pressing need to advance our understanding
of \ion{He}{2} reionization. 

Over the past five years, the confluence of three astronomical
advances have led to an almost tenfold increase in the number of
quasars sightlines available for \ion{He}{2} Ly$\alpha$ absorption
studies. These advances were (1) the discovery of $\sim 40,000$
new quasars at $z_\mathrm{em}>2.7$ by the Sloan Digital Sky Survey (SDSS)
and the Baryon Oscillation Spectroscopic Survey (BOSS; \citealt{paris14});
(2) far-UV and near-UV imaging of almost the entire extragalactic sky by
the \textit{Galaxy Evolution Explorer} (\textit{GALEX}) satellite;
(3) the installation of COS during \textit{HST} Servicing Mission 4,
which enabled high-quality FUV spectroscopy of sources ten times fainter
than any previous instrument. Efficient pre-selection techniques based
on \textit{GALEX} imaging resulted in many tens of $z_\mathrm{em}\sim
3$ quasars suitable for \ion{He}{2} absorption studies
\citep{syphers09a,syphers09b,worseck11}, the FUV-brightest of which
have been followed up with \textit{HST}/COS in recent cycles
\citep{worseck11b,syphers11,syphers12,zheng15}. These new statistical
samples have revealed a large sightline-to-sightline variance in the
\ion{He}{2} effective optical depths at $2.7<z<3$ that is primarily
due to variations in the \ion{He}{2} fraction and the \ion{He}{2}
photoionization rate, implying that \ion{He}{2} reionization completed
at $z\simeq 2.7$ \citep{worseck11b}. The large variance in \ion{He}{2}
absorption might persist at $z>3$ \citep{syphers11,syphers12,zheng15},
but precise measurements of the increasing \ion{He}{2} effective optical
depths require high-S/N spectra of the UV-brightest
\ion{He}{2}-transparent $z_\mathrm{em}>3$ quasars.

Recognizing the technological advances that have enabled these discoveries,
our group has mounted a dedicated campaign to study \ion{He}{2}
reionization, the Helium Reionization
Survey\footlabel{note:hers}{We will publish the reduced \ion{He}{2} and \ion{H}{1} spectra as a MAST high-level science product upon acceptance of the manuscript.}
%\footnote{\label{note:hers}We will publish the reduced \ion{He}{2} and \ion{H}{1} spectra as a MAST high-level science product upon acceptance of the manuscript.}
(HERS). Our survey touches every aspect of the problem:
(1) Discovery of new quasars transmissive at \ion{He}{2} Ly$\alpha$
(Worseck et al.\ in prep.);
(2) Statistical surveys of \ion{He}{2} Ly$\alpha$ absorption with
\textit{HST} spectroscopy (\citealt{worseck11b}; this manuscript);
(3) High-quality echelle spectroscopy of the coeval \ion{H}{1} Ly$\alpha$ forest
that probes the underlying density field along the sightline
(\citealt{worseck11b}; this manuscript);
(4) Uniform, customized reduction and analysis of the absorption line spectra as
required for \ion{He}{2} reionization studies (\citealt{mcquinn14}; this manuscript);
(5) Statistical analysis of the highly ionized proximity zones of
the background quasars \citep{khrykin15};
(6) A survey for faint foreground quasars that could ionize \ion{He}{2} along
the sightlines, advancing on previous results from three well-studied sightlines
(\citealt{jakobsen03,worseck06,worseck07,syphers13}; Schmidt et al.\ in prep.);
(7) Development of advanced statistical methods to compare our data to
predictions from a suite of cosmological simulations that explore the physical
parameters governing the timing and morphology of \ion{He}{2} reionization.
The main data products will be released to the public via our online
repository\footref{note:hers}.
%\footnotemark[\ref{note:hers}].

The focal point of this effort are the \ion{He}{2} Ly$\alpha$ absorption spectra
obtained with \textit{HST}/COS or \textit{HST}/STIS at scientifically useful
spectral resolution ($R\ga 800$) and signal-to-noise ratio
(S/N$>2$) in various programs before \textit{HST} Cycle~20.
\citep{heap00,reimers05,shull10,worseck11b,syphers11,syphers12,syphers13,syphers14,zheng15}.
In this manuscript, we provide the first public data release of HERS,
with emphasis on \ion{He}{2} Ly$\alpha$ and \ion{H}{1} Ly$\alpha$ spectroscopy
(Sections~\ref{sect:obsdatared} and \ref{sect:he2overview}).
Scientifically, we examine the redshift evolution in \ion{He}{2} Ly$\alpha$
effective optical depth at $2.3<z<3.5$ to infer characteristics of \ion{He}{2}
reionization over a cosmic time of $\sim 600$\,Myr (Section~\ref{sect:taueff}).
In a subset of sightlines with coeval \ion{H}{1} Ly$\alpha$ spectroscopy,
we test for a density dependence in the progression of \ion{He}{2} reionization
(Section~\ref{sect:he2h1}), before concluding in Section~\ref{sect:concl}.

We adopt a flat cosmology with $H_0=70$\,km\,s$^{-1}$\,Mpc$^{-1}$ and
$\left(\Omega_\mathrm{m},\Omega_\Lambda\right)=(0.27,0.73)$ \citep{komatsu11}.
Unless otherwise noted, distances are quoted as proper. In our assumed cosmology
$\Delta z=0.04$ -- an interval that we will frequently use for our
measurements -- corresponds to a proper distance of $\approx 10$\,Mpc at $z=3$.
The object designations of quasars discovered by SDSS will be abbreviated to
SDSS~J\texttt{HHMM$\pm$DDMM}.
%\vspace*{6ex}

\section{Observations and data reduction}
\label{sect:obsdatared}
\subsection{\textit{HST} Far-Ultraviolet Spectra}
\subsubsection{Our \textit{HST} Cycle~17 Survey for \ion{He}{2}-transparent Quasars}
In \textit{HST} Cycle~17 we employed COS \citep{green12} in a spectroscopic survey
for intergalactic \ion{He}{2} absorption in the sightlines to eight UV-bright
$z_\mathrm{em}\sim 3$ quasars (Program 11742). Our targets were identified
by cross-matching the \textit{GALEX} GR3 source catalog covering
$\sim 19,000$\,deg$^2$ \citep{morrissey07} to published quasar catalogs
\citep{worseck11}.
This cross-matching yielded eight $z_\mathrm{em}>2.73$ quasars securely detected
in the \textit{GALEX} FUV band (S/N$>3$ in the GR3 catalog) and bright enough
($m_\mathrm{FUV}<21.5$) to simultaneously verify quasar flux at \ion{He}{2}
Ly$\alpha$ in the quasar rest frame and to obtain science-grade (S/N$\sim 4$)
spectra of the \ion{He}{2} absorption along their sightlines in a modest amount
of observing time. Early results on the first two targets have been presented in
\citet{worseck11b}. Here we present the complete dataset homogeneously reduced
and analyzed. All spectra are accessible at our project's data
archive\footref{note:hers}.
%\footnotemark[\ref{note:hers}].

Our survey was performed with the COS grating G140L in the 1105\,\AA\ setup
($\lambda\lambda$1110--2150\,\AA, $R\sim 2000$ per
resolution element at 1150\,\AA) at two focal plane offset positions to
reduce fixed-pattern noise and to correct for the COS grid wire shadows.
The eight targets were observed in single visits of 2--3 orbits between
January 2010 and January 2011 (Table~\ref{tab:he2qsolist}).
Focal plane offset positions were varied between successive orbits of a
visit to maximize individual exposure times using the entire visibility period.
Wavelength calibration spectra were recorded in parallel with the science
spectra in time-tag mode.

\subsubsection{Archival \ion{He}{2} Spectra}
We supplemented our survey dataset by retrieving all 11 science-grade
(S/N$\ge 2$) \ion{He}{2} absorption spectra available in the \textit{HST} archive
as of December 2012. Among the 13 \ion{He}{2}-transparent sightlines discovered
by \citet{syphers12} (Program 12178, PI Anderson), three quasars allow for a
quantitative analysis: HS~1024$+$1849, Q~1602$+$576 and HS~0911$+$4809
(Table~\ref{tab:he2qsolist}).
In Program 12249 (PI Zheng) follow-up \textit{HST}/COS spectroscopy was
obtained for four $z_\mathrm{em}>3.4$ quasars previously verified to show
flux at \ion{He}{2} Ly$\alpha$ \citep{syphers11,zheng15}. All quasars from
Programs 12178 and 12249 were observed with the COS G140L grating in the
1105\,\AA\ setup, while three (HS~0911$+$4809, SDSS~J2346$-$0016 and SDSS~J1253$+$6817)
have additional coverage at $\lambda<1100$\,\AA, obtained in the 1280\,\AA\ setup.
Here we focus on the \ion{He}{2} Ly$\alpha$ absorption fully covered by COS
detector Segment A in both setups. For SDSS~J2346$-$0016 we have not included
the last 2 orbits of G140L 1280\,\AA\ exposure taken in December 2011, as these
provided marginal coverage of the \ion{He}{2} Ly$\alpha$ absorption region.

HE~2347$-$4342 \citep{shull10} and HS~1700$+$6416 \citep{syphers13} had been
observed as part of the COS GTO program (Program 11528, PI Green).
For HE~2347$-$4342 we retrieved both the low-resolution G140L data and the
high-resolution ($R\sim 16,000$) G130M data.
Finally, we also included the archival \textit{HST}/STIS G140L
$R\sim 1000$ spectra of Q~0302$-$003
\citep[Program 7575,][]{heap00} and HS~1157$+$3143
\citep[Program 9350,][]{reimers05}. 

\tabletypesize{\footnotesize}
\begin{deluxetable*}{lcclrlrrrrrrr}
\tablewidth{0pt}
\tablecaption{\label{tab:he2qsolist}Analyzed UV-bright Quasars}
\tablehead{
\colhead{Object}&\colhead{RA (J2000)}&\colhead{DEC (J2000)}&\colhead{$z_\mathrm{em}$}&\colhead{Ref.\tablenotemark{a}}&\colhead{Instrument}&\colhead{$R$\tablenotemark{b}}&\colhead{$t_\mathrm{exp}$ [s]}&\colhead{S/N\tablenotemark{c}}
&\colhead{$f_{1500\text{\AA}}$\tablenotemark{d}}&\colhead{$\alpha$\tablenotemark{e}}&\colhead{$z_\mathrm{abs}$}&\colhead{$\log N_\mathrm{HI}$\tablenotemark{f}}
}
\startdata
\object[CTS 0216]{CTS~0216}                          &$02^\mathrm{h}16^\mathrm{m}23\fs05$ &$-39\degr07\arcmin55\farcs3$ &$2.740$ &11 &COS G140L &2000 &5005 &1 &$6.786$ &$-0.749$ &$0.2925$ &$17.76$\\
\object[HS 1700+6416]{HS~1700$+$6416}                &$17^\mathrm{h}01^\mathrm{m}00\fs61$ &$+64\degr12\arcmin09\farcs1$ &$2.751$ &2 &COS G140L &2000 &15705 &15 &$21.590$ &$-1.756$ &$0.8648$ &$16.05$\\
& & & & & & & & & & &$0.7222$ &$16.17$\\
& & & & & & & & & & &$0.5528$ &$15.87$\\
\object[CSO 0806]{CSO~0806}                          &$13^\mathrm{h}04^\mathrm{m}11\fs99$ &$+29\degr53\arcmin48\farcs8$ &$2.850$ &11 &COS G140L &2000 &4739 &0 &$4.078$ &$-3.664$ &$0.4119$ &$\sim 18.5$\\
\object[HS 1024+1849]{HS~1024$+$1849}                &$10^\mathrm{h}27^\mathrm{m}34\fs13$ &$+18\degr34\arcmin27\farcs5$ &$2.860$ &10 &COS G140L &2000 &2038 &4 &$4.975$ &$-0.560$ &\nodata &\nodata\\
\object[Q 1602+576]{Q~1602$+$576}                    &$16^\mathrm{h}03^\mathrm{m}55\fs92$ &$+57\degr30\arcmin54\farcs4$ &$2.862$ &10 &COS G140L &2000 &2408 &5 &$5.428$ &$-2.463$ &\nodata &\nodata\\
\object[HE 2347-4342]{HE~2347$-$4342}                &$23^\mathrm{h}50^\mathrm{m}34\fs21$ &$-43\degr25\arcmin59\farcs6$ &$2.887$ &3 &COS G140L &2000 &11557 &14 &$20.325$ &$-2.690$ &$0.5766$ &$<15.8$\\
& & & & &COS G130M &16000 &28458 &19 & & &$0.4215$ &$<15.8$\\
\object[PC 0058+0215]{PC~0058$+$0215}                &$01^\mathrm{h}00^\mathrm{m}58\fs39$ &$+02\degr31\arcmin31\farcs4$ &$2.89$  &11 &COS G140L &2000 &6212 &4 &$1.415$ &$-1.289$ &\nodata &\nodata\\
\object[FIRST J093643.5+292713]{SDSS~J0936$+$2927}   &$09^\mathrm{h}36^\mathrm{m}43\fs50$ &$+29\degr27\arcmin13\farcs6$ &$2.930$ &11 &COS G140L &2000 &4739 &4 &$1.086$ &$-2.314$ &$0.2121$ &blend\\
\object[SDSS J081850.01+490817.0]{SDSS~J0818$+$4908} &$08^\mathrm{h}18^\mathrm{m}50\fs01$ &$+49\degr08\arcmin17\farcs0$ &$2.957$ &11 &COS G140L &2000 &7598 &4 &$1.246$ &$-2.295$ &$0.2015$ &$\la 17.0$\\
\object[HS 1157+3143]{HS~1157$+$3143}                &$12^\mathrm{h}00^\mathrm{m}06\fs24$ &$+31\degr26\arcmin30\farcs8$ &$2.989$ &5&STIS G140L&1000 &26820 &11 &$0.541$ &$-7.346$ &\nodata &\nodata\\
\object[SDSS J092447.35+485242.8]{SDSS~J0924$+$4852} &$09^\mathrm{h}24^\mathrm{m}47\fs35$ &$+48\degr52\arcmin42\farcs8$ &$3.027$ &9 &COS G140L &2000 &7598 &8 &$2.432$ &$-2.085$ &$0.4570$ &$<16.0$\\
& & &  & & & & &     &           &         &$0.2280$ &blend\\
\object[SDSS J110155.74+105302.3]{SDSS~J1101$+$1053} &$11^\mathrm{h}01^\mathrm{m}55\fs74$ &$+10\degr53\arcmin02\farcs3$ &$3.029$ &9 &COS G140L &2000 &7157 &4 &$1.028$ &$-2.953$ &$0.3177$ &$\sim 16.5$\\
& & & & & & & & & & &$0.1358$ &$21.13$\\
\object[FIRST J123749.0+012607]{SDSS~J1237$+$0126}   &$12^\mathrm{h}37^\mathrm{m}48\fs99$ &$+01\degr26\arcmin07\farcs0$ &$3.154$ &11 &COS G140L &2000 &6212 &4 &$1.401$ &$-2.290$ &\nodata &\nodata\\
\object[Q 0302-003]{Q~0302$-$003}                    &$03^\mathrm{h}04^\mathrm{m}49\fs85$ &$-00\degr08\arcmin13\farcs5$ &$3.286$ &1 &STIS G140L&1000 &23281 &12 &$3.129$ &$-3.534$ &\nodata &\nodata\\
\object[HS 0911+4809]{HS~0911$+$4809}                &$09^\mathrm{h}15^\mathrm{m}10\fs01$ &$+47\degr56\arcmin58\farcs8$ &$3.350$ &10 &COS G140L &2000 &5520 &6 &$3.890$ &$-0.475$ &$0.3028$ &$<16.8$\\
& & & & & & & & & & &$0.1827$ &$\sim 18.5$\\
\object[SDSS J125353.71+681714.2]{SDSS~J1253$+$6817} &$12^\mathrm{h}53^\mathrm{m}53\fs71$ &$+68\degr17\arcmin14\farcs2$ &$3.481$ &8 &COS G140L &2000 &14095 &7 &$1.854$ &$-2.933$ &$0.6930$ &$16.17$\\
\object[SDSS J234625.66-001600.4]{SDSS~J2346$-$0016} &$23^\mathrm{h}46^\mathrm{m}25\fs66$ &$-00\degr16\arcmin00\farcs4$ &$3.512$ &4 &COS G140L &2000 &20737 &8 &$2.054$ &$-1.703$ &\nodata &\nodata\\
\object[SDSS J171134.41+605240.3]{SDSS~J1711$+$6052} &$17^\mathrm{h}11^\mathrm{m}34\fs41$ &$+60\degr52\arcmin40\farcs3$ &$3.834$ &6 &COS G140L &2000 &23951 &4 &$1.604$ &$-5.734$ &$0.7750$ &$16.66$\\
& & & & & & & & & & &$0.4370$ &$\la 18.0$\\
\object[SDSS J131914.20+520200.1]{SDSS~J1319$+$5202} &$13^\mathrm{h}19^\mathrm{m}14\fs20$ &$+52\degr02\arcmin00\farcs1$ &$3.930$ &7 &COS G140L &2000 &26643 &2 &$1.002$ &$-5.971$ &$0.7026$ &$17.33$
\enddata
%\tablecomments{}
\tablenotetext{a}{Discovery reference. 1: \citet{jakobsen94}, 2: \citet{davidsen96}, 3: \citet{reimers97}, 4: \citet{zheng04b}, 5: \citet{reimers05}, 6: \citet{zheng05}, 7: \citet{syphers09a}, 8: \citet{syphers09b}, 9: \citet{worseck11b}, 10: \citet{syphers12}, 11: this paper.}
\tablenotetext{b}{Spectral resolution $R\equiv\lambda/\mathrm{FWHM}$ at $\lambda=1150$\,\AA.}
\tablenotetext{c}{Signal-to-noise ratio per pixel (COS G140L: $\simeq 0.24$\,\AA\,pixel$^{-1}$, COS G130M: $\simeq 0.03$\,\AA\,pixel$^{-1}$, STIS G140L: $0.6$\,\AA\,pixel$^{-1}$) near \ion{He}{2} Ly$\alpha$ in the quasar rest frame. CTS~0216 and CSO~0806 are not considered further due to strong intervening Lyman limit systems.}
\tablenotetext{d}{Flux density at 1500\,\AA\ in $10^{-16}$\,erg\,cm$^{-2}$\,s$^{-1}$\,\AA$^{-1}$ corrected for Galactic extinction but not for identified \ion{H}{1} Lyman continuum absorption.}
\tablenotetext{e}{Power-law spectral index $\alpha$ for $f_\lambda=f_{1500\text{\AA}}\left(\lambda/1500\mathrm{\AA}\right)^{\alpha}$ including a correction for identified \ion{H}{1} Lyman continuum absorption in the FUV spectrum.}
\tablenotetext{f}{Logarithmic column density of identified intervening \ion{H}{1} absorber in cm$^{-2}$.}
\end{deluxetable*}

\subsubsection{Custom Data Reduction}
\label{sect:he2datared}
The \textit{HST}/COS spectra were homogeneously reduced using
CALCOS~v2.21\footnote{
The most recent CALCOS~v3.1 yields identical results, as it still employs boxcar extraction for
our dataset that was recorded at COS lifetime position 1. Currently, the spatially varying trace
of the new \texttt{TWOZONE} extraction algorithm in CALCOS~v3.1 has not been calibrated for all
COS lifetime positions.}
and custom software.
The raw data were retrieved from the \textit{HST} archive together with the
associated calibration files as of December 2014 to ensure calibration with
inflight data corrected for the degrading instrument sensitivity.

Several customizations of CALCOS were necessary to properly extract
and calibrate the spectra. By default, CALCOS employs pulse height
amplitude (PHA) screening to exclude part of the detector dark current at the
extremes of the pulse height distribution. For the COS G140L data we refrained
from pulse height screening because it may non-trivially modify the distribution
function of the detector counts, which can be modeled as a superposition of
several Poisson distributions (source signal and background contributions).
This considerably simplifies the construction of likelihood functions
for model fitting to the COS spectra, while almost preserving their quality.
In addition, pulse height screening is non-trivial in the presence of detector
gain sag due to continuous exposure on the same detector region
(Appendix~\ref{sect:darkappendix}). The four detector offset positions lead to
a spread of gain sag around geocoronal Ly$\alpha$ emission in our COS G140L
spectra, such that gain sag is particularly strong in the detector region of
interest covering \ion{He}{2} absorption at $2.8\la z\la 3.1$.
For the COS G130M data of HE~2347$-$4342 pulse height screening was necessary
($2\le\mathrm{PHA}\le 18$) to exclude hotspots on COS detector Segment B.
As HE~2347$-$4342 had been observed with an almost pristine detector in November
2009 these cuts included all source signal.

We also adjusted the source extraction windows to preserve
spectrophotometric accuracy while minimizing the background
contribution. COS G140L spectra of well-centered point sources in the
COS Primary Science Aperture have almost all flux enclosed within 25
pixels at the wavelengths of interest (1100\,\AA$<\lambda<$1800\,\AA),
which we chose as the width of our rectangular source extraction
box. For the G130M exposures of HE~2347$-$4342 we chose a width of 31 pixels
to account for their larger full width at half maximum (FWHM).

Background subtraction was performed with custom software, treating open-shutter
background (zodiacal light, earthshine, Galactic \& extragalactic UV
emission, scattered light, geocoronal emission lines) and COS detector dark
current separately (see Appendix~\ref{sect:backgroundappendix} for a discussion).
The standard practice to estimate the COS dark current in unilluminated detector
regions leads to systematic errors due to gain sag
(see \citealt{syphers12} and Appendix~\ref{sect:darkappendix}).
As detailed in Appendix~\ref{sect:darkappendix}, for each science exposure we
estimated the dark current in the COS aperture with appropriately smoothed and
scaled dark monitoring data obtained within $\pm 1.5$ months around the date of
observation, and in approximately the same environmental (space weather) conditions
as estimated from the pulse height distribution outside the COS aperture.
The narrow time window makes differential gain sag negligible, while the overall
rescaling and space-weather restrictions account for the variations of the COS
dark current with time and across the detector.
Extensive validation tests in which subsets of dark exposures were treated as
data show that on the scales of interest ($\Delta z=0.04$ corresponding to 150
native G140L pixels) our custom routine estimates the dark current with negligible
systematic error and a statistical error of a few percent.
Such accuracy is crucial for measurements of strong \ion{He}{2} absorption
($\tau_\mathrm{eff,HeII}>3$), and in general for the analysis of fluxes that are
comparable to the COS dark current
($f_\lambda\la 10^{-17}$\,erg\,cm$^{-2}$\,s$^{-1}$\,\AA$^{-1}$).

Due to the small circular aperture of COS (an out-of-focus field stop of
$\simeq 2.5\arcsec$ diameter; \citealt{green12}) quasi-diffuse open-shutter background
and geocoronal emission lines cannot be subtracted easily.
The intensity of the geocoronal emission lines varies due to solar activity and
\textit{HST}'s orbit parameters in a particular observation.
While \ion{H}{1} Ly$\alpha$ emission is always present, \ion{O}{1} and \ion{N}{1}
emission is typically negligible in orbital night, i.e.\ when \textit{HST} is in
the Earth's shadow. We carefully examined the time-tagged count lists as a
function of the two relevant orbit parameters
(solar altitude and \textit{HST}'s angle to the Earth's limb), and used only
time periods without visible extended emission from geocoronal lines in the
affected spectral regions (if available).
For most targets \ion{N}{1}\,$\lambda$1200\,\AA\ and \ion{O}{1}\,$\lambda$1304\,\AA\
vanished in orbital night, but some targets observed in 2011 required stricter
cuts to lower solar altitudes and/or higher target limb angles, probably due to higher
solar activity. Weaker geocoronal lines appearing during periods of high solar activity
(\ion{N}{1}\,$\lambda$1134\,\AA, \ion{N}{1}\,$\lambda$1243\,\AA, \ion{O}{1}]\,$\lambda$1356\,\AA)
were excluded with similar cuts. Regions with residual geocoronal emission
(usually very few counts above the background) were excluded from scientific
analysis. For HE~2347$-$4342 and HS~1700$+$6416 we also excluded geocoronal
\ion{H}{1} Ly$\beta$ emission contaminating the \ion{He}{2} Ly$\alpha$
forest at $z\simeq 2.37$.

The multi-component quasi-diffuse open-shutter background was estimated and subtracted
in post-processing. While zodiacal light is negligible at $\lambda<1500$\,\AA\
even at low helioecliptic latitudes
($f_\lambda<5\times 10^{-22}$\,erg\,cm$^{-2}$\,s$^{-1}$\,\AA$^{-1}$ over the COS aperture, e.g. \citealt{debes15}),
earthshine is non-negligible at low limb angles during orbital day.
We verified that for HS~1700$+$6416, the only target observed in these
conditions for a substantial amount of time, the count rate in the \ion{He}{2}
absorption region roughly doubled at limb angles $<21\degr$ compared to the
rest of the orbit. Consequently, only nighttime data was used in the \ion{He}{2}
absorption region of HS~1700$+$6416.

Dust-scattered Galactic UV starlight, \ion{H}{2} two-photon emission,
and the $z\simeq 0$ extragalactic UV background give rise to a non-negligible
diffuse UV emission \citep{seon11,murthy14}. 
In addition, there may be distinct emission lines from warm-hot Galactic halo gas
\citep[e.g.][]{martin90a,korpela06,welsh07} and H$_2$ Lyman-Werner fluorescence
\citep[e.g.][]{sternberg89,martin90b,korpela06} depending on the line of sight.
The H$_2$ Lyman-Werner fluorescence will appear as quasi-continuous unresolved
emission in COS G140L spectra ($R\simeq 160$).
As detailed in Appendix~\ref{sect:galuvbappendix}, we subtracted this `sky
background' adopting the exposure-time-weighted mean \textit{GALEX} FUV flux near
our targets from \citet{murthy14}, assuming $f_\lambda=\mathrm{const}$.
This approximately accounts for the diffuse emission and the H$_2$ fluorescence,
as the \textit{GALEX} FUV band covers the H$_2$ Lyman band. Metal emission lines
do not contaminate the $2.4<z<3.5$ \ion{He}{2} transmission, while potential
\ion{O}{6} emission was masked together with geocoronal Ly$\beta$.
Accounting for small FUV sky background fluxes
(4--11$\times 10^{-19}$\,erg\,cm$^{-2}$\,s$^{-1}$\,\AA$^{-1}$ over the COS aperture)
was required to prevent low-level flux leaks in the $z>3$ \ion{He}{2} Gunn-Peterson
troughs. We note that even the minimum diffuse FUV sky emission measured at high
Galactic latitude \citep{seon11,murthy14} is a factor $\sim 3000$ higher than the
contribution from zodiacal light and earthshine for typical \textit{HST}
observations\footnote{
\textit{HST} instrument handbooks considering only zodiacal light and earthshine
\citep[e.g.][]{debes15} need significant revision.}.

Close comparison of data taken during orbital day and night revealed two
COS background components that had been neglected in previous studies.
First, the out-of-focus COS aperture gives rise to extended wings of geocoronal
Ly$\alpha$ emission that are not accurately characterized at present, but
likely negligible $\ga 15$\,\AA\ away from Ly$\alpha$ in G140L spectra.
Second, analysis of ancillary archival data allowed for the first on-orbit
determination of scattered geocoronal Ly$\alpha$ emission in G140L spectra.
We present an empirical model for the sum of these two background components
in Appendix~\ref{sect:scatterappendix}. In total, the G140L grating scatters
$\simeq 0.055$\% of the geocoronal Ly$\alpha$ flux along the dispersion axis.
The modeled scattered light was subtracted from the science data, treating its
statistical uncertainty as systematic error to our measurements.
We encourage further improvements to the background calibration of the COS
instrument in a dedicated \textit{HST} calibration program.

Subexposures were coadded by summing the integer gross counts and the
post-processed time-variable background (sum of dark current, quasi-diffuse sky
emission and scattered light) per pixel on the CALCOS FUV wavelength grid,
accounting for varying pixel exposure times due to offsets in dispersion
direction, detector grid wires and geocoronal emission
(see Appendix~\ref{sect:bkgsumappendix} for examples).
Our coadding routine preserves integer counts obtained in the Poisson regime.
Near 1250\,\AA\ the coadded COS G140L background is dominated by dark current
(51--85\%), while the contribution from scattered light varies between
8\% and 45\%, depending on the fraction of exposure time spent in orbital day
and solar activity. The sky background is low, but non-negligible
(4--10\% of the total background).
Flux conversion was achieved via the pixel exposure time and the time-varying
flux calibration curve determined by CALCOS. Spectra taken at different central
wavelengths were coadded by summing nearest-neighbor counts.
The spectra were rebinned by a factor of three to yield approximate Nyquist
sampling of two pixels per resolution element
(G140L: $\simeq 0.24$\,\AA\,pixel$^{-1}$, G130M: $\simeq 0.03$\,\AA\,pixel$^{-1}$).
The S/N was calculated in the Poisson regime of the data and accounting for the
background \citep{feldman98}. For plotting purposes we computed an approximate
$1\sigma$ error array by adding in quadrature the larger of the asymmetric
Poisson $1\sigma$ error and the background error. Most COS G140L spectra reach
S/N$\simeq 4$ per binned pixel near \ion{He}{2} Ly$\alpha$ (Table~\ref{tab:he2qsolist}).
The Bayesian method of \citet{kraft91} yields similar results.

The extreme UV ($\lambda<1150$\,\AA) spectra of HS~1700$+$6416 and HE~2347$-$4342
recorded on COS detector Segment B in the G140L 1230\,\AA\ and 1280\,\AA\ setups
required wavelength recalibration \citep[e.g.][]{shull10,syphers13}.
Due to the combination of flux calibration uncertainties, low signal, and
\ion{He}{2} Ly$\beta$ absorption we restricted our analysis to $\lambda>1000$\,\AA.
We adopted the COS G140L dispersion given by \citet{shull10} and aligned the COS
spectra with their archival \textit{FUSE} spectra after convolving them with the
COS G140L line spread function. Considering the background subtraction problems
of \textit{FUSE} \citep{zheng04,fechner06} and possible source variability,
the spectra were aligned by eye using several sharp features in the emerging
\ion{He}{2} Ly$\alpha$ forest. For HE~2347$-$4342 we also used the higher quality
COS G130M spectrum in the overlapping spectral range. With respect to the CALCOS
reduction we adopted shifts in the COS Segment B wavelength zero point of $-0.6$\,\AA\footnote{
The wavelength shifts are smaller than in previous analyses \citep{shull10,syphers13}
due to updates of the CALCOS pipeline and/or the calibration files.}.
On Segment A interstellar absorption lines indicate smaller or negligible
wavelength shifts (maximum shift $-0.4$\,\AA\ for HE~2347$-$4342).

The archival STIS spectra of Q~0302$-$003 and HS~1157$+$3143 were reduced in the
same fashion as the COS data by customizing CALSTIS~v2.30 to our needs, i.e.\
adjusting the extraction and background subtraction windows \citep{heap00}.
The Poisson counts of individual exposures were coadded, yielding a continuum S/N
of 11 per 0.6\,\AA\ pixel near \ion{He}{2} Ly$\alpha$ in the quasar rest frame.

\subsubsection{Continuum Definition}
The strong unresolved intergalactic \ion{He}{2} absorption in the G140L spectra
precludes a local definition of the quasar continuum, which instead has to be
extrapolated from the spectral region redward of \ion{He}{2} Ly$\alpha$ in the
quasar rest frame. For simplicity, we modeled the quasar continuum as a power-law
$f_\lambda\propto\lambda^\alpha$, accounting for Galactic extinction,
identified low-redshift IGM/ISM absorption, weak extreme UV quasar emission lines
redward of \ion{He}{2} Ly$\alpha$ and
residual geocoronal contamination.
We emphasize that the fitted power-laws do not represent the intrinsic spectral energy
distributions due to partial Lyman limit system breaks in the NUV, which has not
been spectroscopically covered for any recently discovered \ion{He}{2} sightline
(see \citealt{syphers13};\,\citeyear{syphers14} for constraints on the
spectral index of two \ion{He}{2} quasars with complete spectroscopic coverage).
Furthermore, the poorly characterized extreme UV quasar continuum at
$\lambda_\mathrm{rest}<304$\,\AA\ may show weak emission lines similar
to the ones seen at $\lambda_\mathrm{rest}>400$\,\AA\
\citep{stevans14,tilton16}, but these are unlikely to significantly affect our
measurements of strong \ion{He}{2} absorption.

All UV spectra were corrected for Galactic
extinction using their line-of-sight selective extinction $E(B-V)$ derived by
\citet{schlegel98} and the \citet{cardelli89} extinction curve assuming the
Galactic average for the ratio between total $V$ band extinction and selective
extinction $R_V=3.1$. Each spectrum was searched for low-redshift \ion{H}{1}
Lyman limit systems whose Lyman continuum absorption modifies the power-law
continuum. Lyman limit systems (including partials) were identified by their
Lyman series transitions, with redshifts confirmed by at least two observed
transitions.

We then interactively selected regions redward of \ion{He}{2} Ly$\alpha$ deemed
free of obvious emission and absorption lines and fitted the power-law continuum
with the column densities of the identified Lyman limit systems as additional
free parameters if the spectral range redward of the break was required for a
satisfactory fit of the continuum. In most cases we did not consider
$\lambda>1800$\,\AA\ due to the dropping G140L efficiency.
The fit was performed via a maximum-likelihood
routine on the Poisson gross counts (see below) and continuum errors were
estimated by a Monte Carlo routine, refitting Poisson deviates of the inferred
continuum counts 10,000 times. The inferred $1\sigma$ statistical continuum error
in the \ion{He}{2} absorption region naturally varies with the considered
spectral range and the S/N, increasing from a few percent at the bright high-S/N
end to $\sim 10$\% in the lowest-S/N spectra. More importantly, the fitted
continuum depends on our ability to identify (partial) Lyman limit systems,
especially if their Lyman limit break occurs blueward of \ion{He}{2} Ly$\alpha$
of the background quasar. For these systems we estimated column densities from
the covered Lyman series lines, accounting for the instrument line-spread function.
The adopted continuum fit parameters and identified Lyman limit systems are
listed in Table~\ref{tab:he2qsolist} and details on individual objects are given
in Appendix~\ref{sect:noteappendix}.

\begin{figure*}[t]\centering
\includegraphics[width=\textwidth]{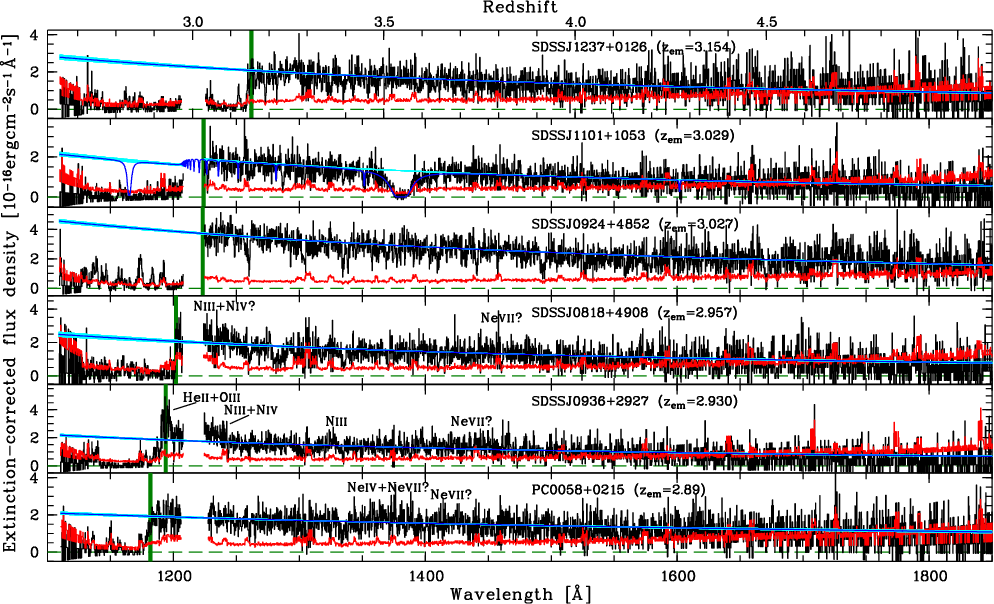}
\caption[]{\label{fig:he2spc_cycle17}
Extinction-corrected Nyquist-sampled \textit{HST}/COS G140L spectra
(black; $R\sim 2000$, S/N$\simeq 4$ per 0.24\,\AA\ pixel near \ion{He}{2} Ly$\alpha$)
and their corresponding $1\sigma$ error arrays (red) of the 6
\ion{He}{2}-transparent quasar sightlines from our Cycle~17 survey.
The redshift axis (top) is for \ion{He}{2} Ly$\alpha$. The spectral region at
$\lambda\sim 1215$\,\AA\ contaminated by residuals of geocoronal Ly$\alpha$ has
been omitted. The green dashed lines mark the zero level. Blueward of \ion{He}{2}
Ly$\alpha$ in the quasar rest frame (green vertical bars) there is strong,
but varying intergalactic \ion{He}{2} absorption. The blue lines show power-law
continuum fits to emission- and absorption-free regions redward of \ion{He}{2}
Ly$\alpha$ and their $1\sigma$ uncertainties (cyan shaded) estimated from
Monte Carlo simulations. For SDSS~J1101$+$1053 we include \ion{H}{1} Ly$\alpha$
and Ly$\beta$ absorption of a foreground damped Ly$\alpha$ absorber
\citep{worseck11b} and a partial Lyman limit system
($N_\mathrm{HI}\sim 10^{16.5}$\,cm$^{-2}$, $z_\mathrm{abs}=0.3177$).
Three spectra show at least tentative evidence for quasar emission lines.
}
\end{figure*}

\subsection{Optical High-Resolution Spectra}
Our Cycle~17 \textit{HST}/COS survey was complemented by an extensive ground-based
campaign to obtain optical high-resolution spectra covering the \ion{H}{1}
Ly$\alpha$ forests of our eight Cycle~17 targets. The four southern/equatorial
targets (CTS~0216, PC~0058$+$0215, SDSS~J1101$+$1053, SDSS~J1237$+$0126)
were observed with the Very Large Telescope (VLT) UV-Visual Echelle Spectrograph
\citep[UVES;][]{dekker00} in service mode between April 2009 and April 2010
(Program 083.A-0421). We used the 1\arcsec\ slit ($R\sim 45,000$) and the UVES
blue arm central wavelength setting 437\,nm to obtain continuous coverage of the
\ion{H}{1} Ly$\alpha$ forest of our targets ($\lambda\lambda$3758--4987\,\AA\
corresponding to $2.09<z<3.10$). For CTS~0216 and PC~0058$+$0215 we used UVES
dichroic \#2 at central wavelength 760\,nm to simultaneously probe metal
absorption redward of Ly$\alpha$. Total exposure times were chosen to yield a
homogeneous continuum S/N$\sim 20$ per 1.85\,km\,s$^{-1}$ pixel in the
Ly$\alpha$ forest (Table~\ref{tab:groundlog}). The data were reduced using the ESO UVES
pipeline\footnote{\anchor{http://www.eso.org/sci/software/pipelines/}{http://www.eso.org/sci/software/pipelines/}}
v.4.4.8 and normalized with an automatic cubic spline fitting routine
\citep{dallaglio08}.

The four northern targets of our Cycle~17 \textit{HST}/COS survey were observed
with the Keck~I High-Resolution Echelle Spectrometer \citep[HIRES;][]{vogt94} on
5 February 2010 (Table~\ref{tab:groundlog}). We used the C1 decker
(0\farcs86 slit, $R\sim 45,000$) and the blue cross-disperser at two echelle
angles to cover Ly$\alpha$ and Ly$\beta$ emission of the targets on the middle
CCD of the array, yielding almost continuous spectral coverage from the
atmospheric cutoff to $\simeq 5850$\,\AA. We supplemented our sample by
obtaining HIRES spectroscopy of the two high-redshift \ion{He}{2}-transparent
quasars SDSS~J1711$+$6052 and SDSS~J2346$-$0016 on UT dates 3--4 August 2011.
We used the C1 decker with the red cross-disperser at three echelle angles to
cover the wavelength range $\lambda\lambda$4060--7330\,\AA. All HIRES spectra
were reduced with the HIRedux
pipeline\footnote{\anchor{http://www.ucolick.org/~xavier/HIRedux/}{http://www.ucolick.org/\textasciitilde xavier/HIRedux/}},
yielding a characteristic continuum S/N$\sim 20$ per $2.6$\,km\,s$^{-1}$ pixel in
the Ly$\alpha$ forest. Individual echelle orders were normalized interactively by
low-order polynomials and weighted by inverse variance in their overlapping regions.

Archival high-quality (S/N$\sim 100$) spectra of the \ion{He}{2}-transparent
quasars HE~2347$-$4342 \citep[VLT/UVES, $R\sim 45,000$;][]{dallaglio08} and
HS~1700$+$6416 \citep[Keck/HIRES, $R\sim 38,500$;][]{fechner06} complemented
our data set of \ion{H}{1} forest spectra.

\tabletypesize{\footnotesize}
\begin{deluxetable}{lllrr}
\tablewidth{0pt}
\tablecaption{\label{tab:groundlog}\ion{H}{1} Ly$\alpha$ Forest Spectra}
\tablehead{
\colhead{Object}&\colhead{$z_\mathrm{em}$}&\colhead{Instrument}&\colhead{$t_\mathrm{exp}$ [h]}&\colhead{S/N\tablenotemark{a}}
}
\startdata
CTS~0216\tablenotemark{b}       &$2.740$ &UVES  &$1.7$  &20\\
HS~1700$+$6416\tablenotemark{c} &$2.751$ &HIRES &$23.4$ &100\\
HE~2347$-$4342\tablenotemark{c} &$2.887$ &UVES  &$20.0$ &100\\
CSO~0806\tablenotemark{b}       &$2.850$ &HIRES &$2.0$  &15\\
PC~0058$+$0215                  &$2.89$  &UVES  &$7.5$  &22\\
SDSS~J0936$+$2927               &$2.930$ &HIRES &$2.0$  &18\\
SDSS~J0818$+$4908               &$2.957$ &HIRES &$2.0$  &8\\
SDSS~J0924$+$4852               &$3.027$ &HIRES &$3.0$  &20\\
SDSS~J1101$+$1053               &$3.029$ &UVES  &$11.1$ &30\\
SDSS~J1237$+$0126               &$3.154$ &UVES  &$13.0$ &34\\
SDSS~J2346$-$0016               &$3.512$ &HIRES &$4.5$  &40\\
SDSS~J1711$+$6052               &$3.834$ &HIRES &$9.0$  &25
\enddata
%\tablecomments{}
\tablenotetext{a}{Characteristic continuum S/N per pixel in the Ly$\alpha$ forest.}
\tablenotetext{b}{Quasar not considered further due to lacking flux at 304\,\AA.}
\tablenotetext{c}{Archival spectra \citep{fechner06,dallaglio08}.}
\end{deluxetable}

%\vspace*{2ex}

\section{Ubiquitous \ion{He}{2} Ly$\alpha$ Absorption in 17 Quasar Sightlines}
\label{sect:he2overview}

\begin{figure*}[t]\centering
\includegraphics[width=\textwidth]{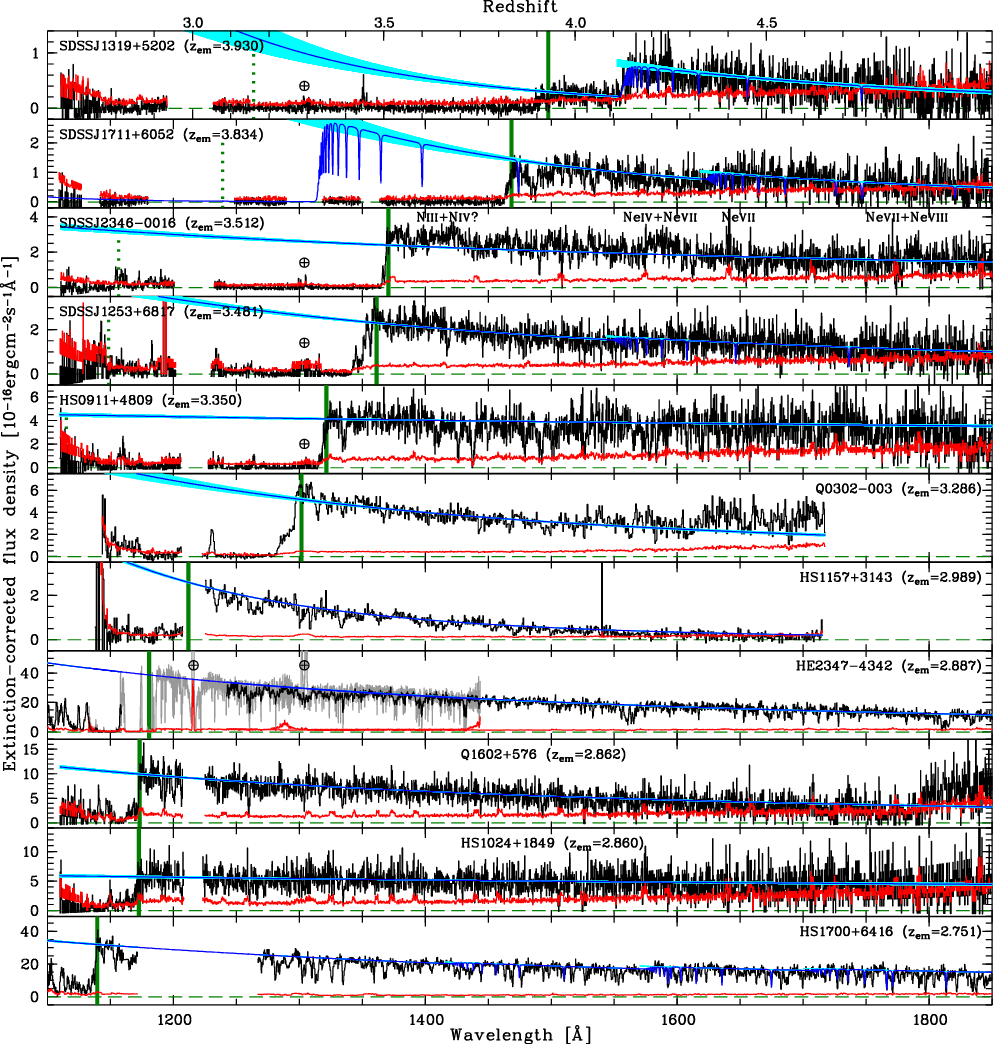}
\caption[]{\label{fig:he2spc_archive}Extinction-corrected \textit{HST} FUV spectra
(black) and corresponding $1\sigma$ error arrays (red) of 11 archival
\ion{He}{2}-transparent quasar sightlines, homogeneously reduced and analyzed.
The redshift axis (top) is for \ion{He}{2} Ly$\alpha$.
Q~0302$-$003 and HS~1157$+$3143 have been observed with STIS
(G140L, $R\sim 1000$, 0.6\,\AA\,pixel$^{-1}$),
whereas the remaining spectra have been taken with COS
(G140L, $0.24$\,\AA\,pixel$^{-1}$ in a Nyquist-sampled spectrum).
For HE~2347$-$4342 we also show its Nyquist-sampled G130M spectrum
(gray; $R\sim 16,000$, $0.03$\,\AA\,pixel$^{-1}$).
Spectral regions with remaining strong geocoronal emission during orbital night
have been omitted, and regions with residual emission have been marked
(Earth symbols). The green dashed lines mark the zero level.
\ion{He}{2} absorption occurs blueward of \ion{He}{2}
Ly$\alpha$ in the quasar rest frame (green vertical bars).
Most background quasars have a \ion{He}{3} proximity zone; i.e.\
reduced \ion{He}{2} opacity at small velocity separations from the quasar.
The dotted lines mark the onset of \ion{He}{2} Ly$\beta$. The blue lines show
power-law continuum fits to absorption-free regions redward of \ion{He}{2}
Ly$\alpha$ and its $1\sigma$ error (cyan shaded).
For SDSS~J1319$+$5202, SDSS~J1711$+$6052, SDSS~J1253$+$6817 and HS~1700$+$6416
the power-law continua include identified Lyman series and continuum absorption
from intervening Lyman limit systems (Table~\ref{tab:he2qsolist}), convolved
to COS resolution. SDSS~J2346$-$0016 show hints of the presence of extreme
UV emission lines (labeled).}
\end{figure*}

The quality of our Cycle~17 \textit{HST}/COS survey spectra
(continuum S/N$\simeq 4$ near \ion{He}{2} Ly$\alpha$) allow for detailed analysis
of intervening \ion{He}{2} Ly$\alpha$ absorption.
Figure~\ref{fig:he2spc_cycle17} presents the \textit{HST}/COS spectra of the 6
quasars from our survey that show significant flux at \ion{He}{2} Ly$\alpha$ in the
quasar rest frame (see Appendix~\ref{sect:noteappendix} for the two quasars with
strong intervening Lyman limit systems precluding \ion{He}{2} analysis).
Initial results on the two \ion{He}{2}-transparent quasars SDSS~J0924$+$4852 and
SDSS~J1101$+$1053 were presented in \citet{worseck11b}. Blueward of \ion{He}{2}
Ly$\alpha$ in the quasar rest frame we detect patchy intergalactic \ion{He}{2}
Ly$\alpha$ absorption on a range of spatial scales, varying from large-scale
strong absorption (e.g.\ $\sim 30$\,Mpc at $z\simeq 2.8$ toward SDSS~J0936$+$2927)
to alternating transmission and absorption features on scales of only a few Mpc.
Most sightlines exhibit low \ion{He}{2} absorption at the lowest covered
redshifts $z<2.7$, although the modest sensitivity of COS at
$\lambda\sim 1100$\,\AA\ results in just a few detected counts per pixel.

The archival \ion{He}{2} sightlines shown in Fig.~\ref{fig:he2spc_archive}
provide additional coverage of \ion{He}{2} Ly$\alpha$ absorption at the redshifts
$2.66<z\la 3$ probed by our Cycle~17 sample, as well as extending it to $z>3.2$.
At $z\la 2.7$ several sightlines show similar alternating small-scale \ion{He}{2}
transmission and absorption features
(e.g.\ HS~1700$+$6416, HE~2347$-$4342 and Q~1602$+$576).
In \textit{FUSE} spectra of HS~1700$+$6416 and HE~2347$-$4342 many of the lower
redshift spikes have been resolved into an emerging \ion{He}{2} Ly$\alpha$ forest
\citep{kriss01,zheng04,fechner06}. The lack of strong saturation in the COS
spectra indicates the onset of a \ion{He}{2} Ly$\alpha$ forest at $z\la 2.7$,
which is unresolved in the COS G140L spectra.

At $2.7\la z\la 3$ some sightlines show saturated \ion{He}{2} absorption on
$\ga 10$ Mpc scales (e.g.\ HE~2347$-$4342, Q~0302$-$003, SDSS~J0936$+$2927).
These long troughs are unlikely to exist in a fully reionized IGM except in
high-density regions \citep{furlanetto09,shull10}. Such overdensities may be
revealed through analysis of the coeval \ion{H}{1} \lya\ forest
(see Section~\ref{sect:he2h1}). Other sightlines show substantial \ion{He}{2}
transmission at the same redshifts (e.g.\ SDSS~J1253$+$6817 and SDSS~J2346$-$0016),
which likely indicate the final phase of patchy \ion{He}{2} reionization
\citep[e.g.][]{reimers97}.
Some of these transmission regions have been associated to foreground quasars near
the sightline \citep{worseck06,worseck07,syphers14}.

In general, the 7 sightlines probing $z>3$ show very strong \ion{He}{2}
absorption on large scales. However, significant transmission spikes are visible
in several spectra. The transmission spike in the Q~0302$-$003 sightline likely
corresponds to the \ion{He}{3} proximity zone of a foreground quasar at $z=3.05$
\citep{heap00,jakobsen03,syphers14}. Smaller but still significant spikes exist
in the sightlines toward HS~0911$+$4809 ($z\simeq 3.16$, $6.2\sigma$ significance),
SDSS~J1319$+$5202 ($z\simeq 3.45$, $8.8\sigma$ significance), and
SDSS~J1253$+$6817 ($z\simeq 3.15$, $6.2\sigma$ significance).
The length of the absorption troughs between these spikes is hard to determine
due to geocoronal emission and data quality. The flux spike in SDSS~J1253$+$6817 
is part of a longer shallow transmission region at $3.08<z<3.18$.
Another flux spike is revealed by the night portion of the data at $z\simeq 3.06$
that is clearly separated from the residual geocoronal Ly$\alpha$ emission.
However, in the vicinity of geocoronal line residuals only prominent flux spikes
can be unambiguously identified, such that these regions must be excluded
from a statistical analysis.

\section{The Redshift Evolution of the \ion{He}{2} Effective Optical Depth}
\label{sect:taueff}

\subsection{Measurement Technique}
\label{sect:tauml}

\begin{figure*}[t]
\includegraphics[width=\textwidth]{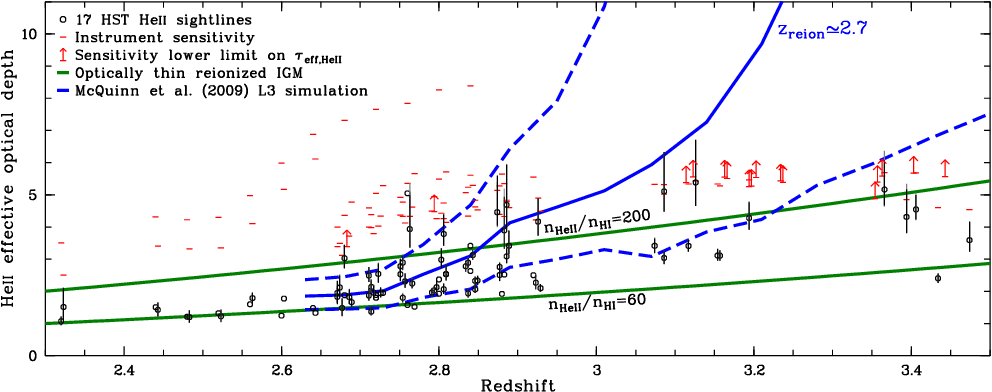}
\caption[]{\label{fig:he2tau}
\ion{He}{2} effective optical depth $\tau_\mathrm{eff,HeII}$ vs.\ redshift for
17 \ion{He}{2} sightlines in identical redshift bins of $\Delta z=0.04$
($\approx 10$ proper Mpc at $z\sim 3$), discovered in our Cycle~17 survey
(Fig.~\ref{fig:he2spc_cycle17}) or reanalyzed from the \textit{HST} archive
(Fig.~\ref{fig:he2spc_archive}). The measured $\tau_\mathrm{eff,HeII}$ values
are plotted as black circles with error bars distinguishing statistical errors
due to Poisson count statistics
(black, double-sided $1\sigma$ errors corresponding to a confidence level of $68.26$\%)
and additional systematic errors from background uncertainties (gray).
For clarity, the data are plotted slightly offset with respect to the identical
bin centers and total error bars smaller than the symbol size have been omitted.
For every measurement we also plot the $1\sigma$ instrumental sensitivity limit
(red horizontal dashes), which we adopt as measured values (arrow symbols)
if the upper confidence level includes infinite $\tau_\mathrm{eff,HeII}$
or if the signal is formally negative ($P>0.1587$).
Overplotted are predictions from a semianalytic model of a reionized IGM
matching low-redshift observations with two representative $n_\mathrm{HeII}/n_\mathrm{HI}$
ratios of 60 and 200 (green lines), and results evaluated in $\Delta z=0.04$
bins from a numerical simulation by \citet{mcquinn09a} in which \ion{He}{2}
reionization finishes at $z_\mathrm{reion}\simeq 2.7$
(blue; solid: median $\tau_\mathrm{eff,HeII}$, dashed: $1\sigma$ deviation).}
\end{figure*}

To quantitatively assess the wealth of structure and variance among the
17 \ion{He}{2} sightlines out to high redshifts, we quantified the \ion{He}{2}
absorption by computing the effective optical depth $\tau_{\mathrm{eff,HeII}}$.
Specifically, $\tau_\mathrm{eff,HeII}\equiv-\ln{\left<f_\lambda/E_\lambda\right>}$
represents the average ratio of observed quasar flux density $f_\lambda$
and the extrapolated continuum $E_\lambda$, which we compute
in fixed redshift bins of common size $\Delta z=0.04$
($\approx 10$ proper Mpc at $z\sim 3$).

The scatter in the \ion{He}{2} effective optical depth on a given scale length
depends on the \ion{He}{2} reionization history and on density fluctuations
in the sightlines. Our choice of $\Delta z=0.04$ is a compromise between
resolving the small-scale variance in the \ion{He}{2} absorption, preserving
the sensitivity to high $\tau_\mathrm{eff,HeII}$ at $z>3$ and maximizing
the individual sightline coverage in fixed redshift bins (e.g.\ regarding different
$z_\mathrm{em}$, proximity zone size, geocoronal emission).
Identical regular redshift bins enable an objective comparison of the sightlines
without prior emphasis on individual sightline peculiarities.  

Given the typical continuum S/N$\simeq 4$ of the survey spectra, any residual
flux $f_\lambda$ in the \ion{He}{2} absorption region will correspond to very
few detected counts in the Poisson regime of the photon-counting multichannel
plate detectors. Non-Poisson fixed-pattern noise due to COS detector effects
becomes significant only at continuum S/N$\gg 5$ \citep{syphers12,keeney12}.
Also the dark current is well described by a Poisson distribution
(Appendix~\ref{sect:darkappendix}).
All fitting was performed by maximizing the Poisson likelihood function 
\begin{equation}\label{eq:likelihood}
L = \prod_{j=1}^{n}\frac{\left(S_j+B_j\right)^{N_j}e^{-\left(S_j+B_j\right)}}{N_j!}
\end{equation}
of $n$ pixels with an integer number of registered counts $N_j$, the non-integer
multi-component background $B_j=B_{\mathrm{dark},j}+B_{\mathrm{sky},j}+B_{\mathrm{Ly}\alpha,j}$
(Appendix~\ref{sect:backgroundappendix}), and the unknown signal $S_j$.
To compute $\tau_\mathrm{eff,HeII}$ the signal was modeled as a constant
in \ion{He}{2} transmission over a segment of $n$ contiguous pixels,
converted to non-integer source counts via the pixel exposure time $t_j$,
the extinction-corrected flux calibration curve $C_j$, and the extrapolated
continuum $E_j$ as
\begin{equation}
S_j=t_jC_jE_je^{-\tau_\mathrm{eff,HeII}}.
\end{equation}
For our Nyquist-sampled COS G140L spectra, $\Delta z=0.04$ corresponds to $n=51$
pixels. Confidence intervals ($1\sigma$, $68.26$\% confidence) were computed via
ordering the Poisson likelihood ratio \citep{feldman98}, first applied to COS
data by \citet{syphers11}. For each redshift bin we also computed a $1\sigma$
lower limit on $\tau_\mathrm{eff,HeII}$ by refitting $\tau_\mathrm{eff,HeII}$ on
mock data generated from 100,000 Poisson deviates of the background assuming
zero source flux. \citet{feldman98} call this the sensitivity, which in our case
quantifies how large an effective optical depth could have been reliably measured
(i.e.\ with a finite upper confidence limit), given the expected number of
continuum counts and the background during the observations. If the \ion{He}{2}
transmission was formally negative ($\tau_\mathrm{eff,HeII}\rightarrow\infty$)
or if the upper confidence limit included infinite $\tau_\mathrm{eff,HeII}$ we
chose to quote the sensitivity limit as our measurement.
A related quantity is the probability of having measured more than
$N=\sum_j N_j$ counts in a given redshift bin with the total background
$B=\sum_j B_j$,
\begin{equation}\label{eq:prob}
P\left(>N|B\right) = 1-\sum_{k=0}^{N}\frac{B^k e^{-B}}{k!},
\end{equation}
which we use to estimate the significance of a measured Poisson signal given
the background. Our quoted $1\sigma$ sensitivity limit on
$\tau_\mathrm{eff,HeII}$ corresponds to a probability
$P=0.1587$\footnote{We ensured that our Monte-Carlo simulations to estimate the
$\tau_\mathrm{eff,HeII}$ sensitivity limit give the same result as the numerical
inversion of Equation~\ref{eq:prob}.}.
A statistically significant signal has $N\gg B$ such that $P$ goes to zero,
i.e.\ it is very unlikely to result from a Poisson fluctuation of the background.
On the other hand, $\tau_\mathrm{eff,HeII}$ values higher than the sensitivity
limit have $P>0.1587$, while downward Poisson background fluctuations
($N\rightarrow 0$ resulting in formally negative transmission) have $P\rightarrow 1$.

The above modeling assumes that the continuum and the background are perfectly
known and that contamination from other absorption lines (Galactic and extragalactic) is
negligible. The latter is a good approximation, since the \ion{He}{2} absorption
is generally strong and measured over a large spectral segment.
The only significant line contamination is due to the low-$z$ damped Ly$\alpha$
absorber toward SDSS~J1101$+$1053 (Fig.~\ref{fig:he2spc_cycle17}), which was masked out.
As for the continuum, a bootstrap analysis including the statistical continuum error
(Figs.~\ref{fig:he2spc_cycle17} \& \ref{fig:he2spc_archive}) increases the
Poisson statistical errors by a negligible amount ($\la 10$\%). Systematic
continuum error was minimized by screening for partial Lyman limit systems.

The main limitation of the \citet{feldman98} method is the assumption of a fixed
background, which obviously does not hold for our post-processed modeled
background. We estimated the error budget of all our background components in
the data reduction (Appendix~\ref{sect:backgroundappendix}), and incorporated it
into our measurements as a systematic error estimated by Monte Carlo simulations.
Specifically, for every $\Delta z=0.04$ bin we determined the mean relative
background error (2--8\%, see Appendix~\ref{sect:bkgsumappendix}), and drew
100,000 Gaussian deviates of it to determine a range of background scaling
factors. We then inferred $\tau_\mathrm{eff,HeII}$ for the modified background,
and generated a mock data sample by drawing from the Poisson distribution of
background and inferred signal. Measuring $\tau_\mathrm{eff,HeII}$ on these mock
samples yields an estimate of the total error from statistical Poisson shot noise
and systematic background error. We distinguished between statistical error
computed at fixed background \citep{feldman98}, and systematic error arising
from background variations. As expected, the inclusion of background
uncertainties results in more realistic error estimates.
In particular $\tau_\mathrm{eff,HeII}$ values higher than the formal sensitivity
limit for our fiducial mean background but with finite statistical errors
(i.e.\ those with $0.1587<P\la 0.25$) have infinite upper confidence limits
after accounting for background error. Therefore, in our approach,
sensitivity lower limits on $\tau_\mathrm{eff,HeII}$ can arise from statistical
and systematic errors.

Last, but not least, we masked the proximity zones of the background quasars
(usually estimated from the onset of saturated \ion{He}{2} absorption),
redshift bins partially covered by a given spectrum, and regions affected by
geocoronal emission residuals.

\subsection{Observational Results}

\begin{figure*}[t]\centering
\includegraphics[width=0.85\textwidth]{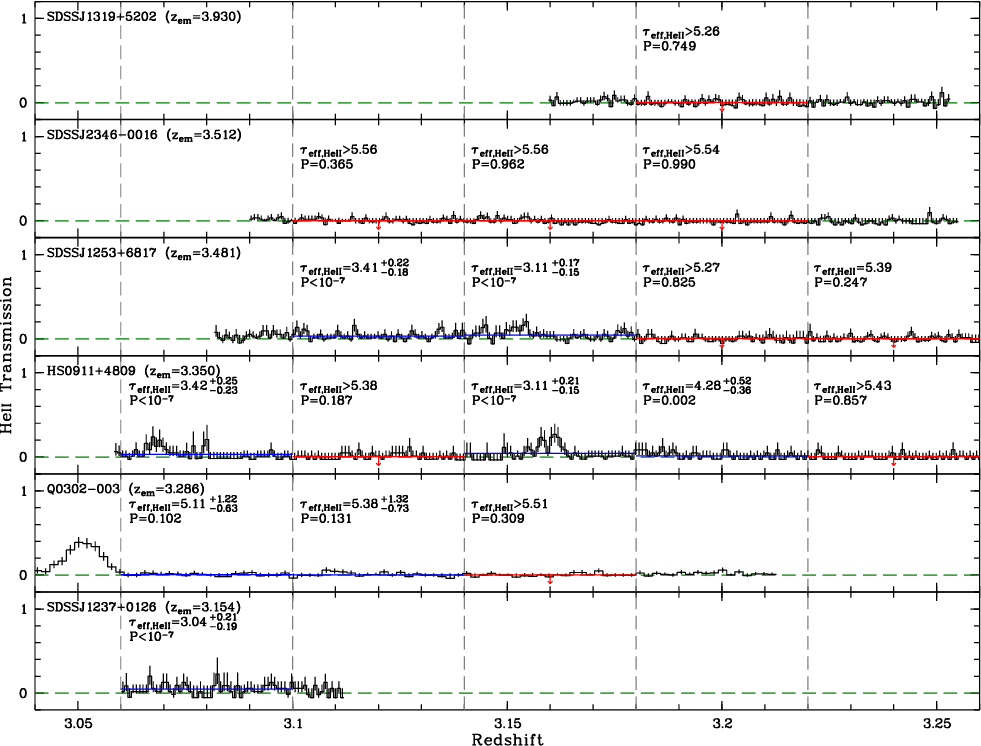}
\caption[]{\label{fig:highzspc2}
Our 17 measurements of the \ion{He}{2} absorption at $3.06<z<3.26$ covered by 6
sightlines of our sample. The normalized \ion{He}{2} spectra (black) have been
binned to two pixels per resolution element ($\simeq 0.24$\,\AA\,pixel$^{-1}$)
and individual Poisson errors are overplotted. Redshift ranges where we cannot
perform an unbiased measurement due to residual geocoronal emission,
the line-of-sight proximity zone, or \ion{He}{2} Ly$\beta$ absorption, are not shown.
Horizontal dashed lines mark the zero level while vertical dashed lines indicate
our regular $\Delta z=0.04$ bins. The $\tau_\mathrm{eff,HeII}$ measurements
(labeled with total error, Table~\ref{tab:he2tau}) have been converted to
\ion{He}{2} transmission (solid lines), with blue and red lines indicating
robustly measured values and sensitivity limits, respectively.
Incompletely covered redshift bins were not considered. The total error
(statistical $1\sigma$ error and estimated systematic error due to background uncertainty)
on the measurements are comparable to the line thickness, whereas arrows
indicate upper limits on the \ion{He}{2} transmission. We also indicate the
probability $P$ that the measured counts arise from a Poisson background
fluctuation (Equation~\ref{eq:prob}).
}
\end{figure*}
\begin{figure*}[t]\centering
\includegraphics[width=0.85\textwidth]{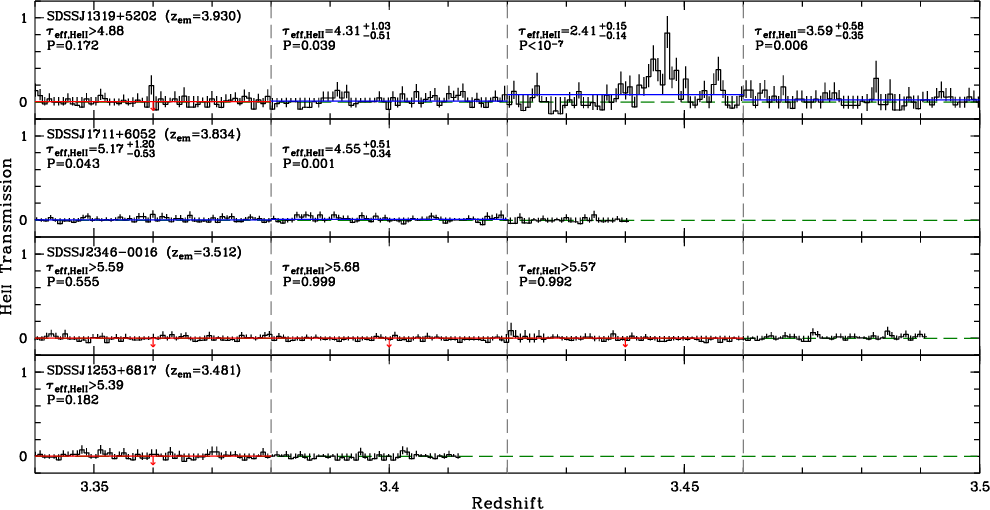}
\caption[]{\label{fig:highzspc}
Similar to Fig.~\ref{fig:highzspc2} but for the 10 redshift bins at $3.34<z<3.50$.
}
\end{figure*}

Figure~\ref{fig:he2tau} and Table~\ref{tab:he2tau} show the \ion{He}{2}
effective optical depths of the 17 \ion{He}{2}-transparent quasar sightlines
in the chosen regular $\Delta z=0.04$ bins at $z<3.5$.
We also plot the $1\sigma$ sensitivity limit for each measurement,
i.e.\ the highest $\tau_\mathrm{eff,HeII}$ value that could have been reliably
measured ($P=0.1587$). The effective optical depth increases with redshift,
but with significant scatter that also appears to increase with redshift.
The statistical and systematic errors become significant only near the
sensitivity limit (typically at $\tau_\mathrm{eff,HeII}\ga 4$).
Out of the 103 measurements, 16 are sensitivity limits, plotted as arrows in
Fig.~\ref{fig:he2tau}. Twelve of these arise due to statistical Poisson errors
(i.e.\ limited depth of the observations), while four are due to background
uncertainties (i.e.\ statistically marginally detected flux becoming consistent with zero).

At $z<2.66$ only two sightlines, HS~1700$+$6416 and HE~2347$-$4342,
sample the \ion{He}{2} Ly$\alpha$ absorption, yielding similar values of
$\tau_\mathrm{eff,HeII} \approx 1.5$. At $z\simeq 2.32$ the HS~1700$+$6416
sightline shows somewhat stronger \ion{H}{1} and \ion{He}{2} absorption
where it roughly intersects an overdensity of galaxies \citep{steidel05,simcoe06}.
Redshifts $z>2.66$ are very well sampled by our COS spectra. At $z=2.68$ 7/9
sightlines have $\tau_\mathrm{eff,HeII}\simeq 1.8$, while SDSS~J0924$+$4852 and
SDSS~J1101$+$1053 have $\tau_\mathrm{eff,HeII}\ga 3$. Uncertain \ion{H}{1}
Lyman continuum absorption at $z\simeq 0.3$ in these two sightlines
(Table~\ref{tab:he2qsolist}) cannot fully account for these large
$\tau_\mathrm{eff,HeII}$ values because we observe significant \ion{He}{2}
transmission at higher redshifts (Fig.~\ref{fig:he2spc_cycle17}). 
At $z=2.76$ we see a large scatter in $\tau_\mathrm{eff,HeII}$ from $1.52$
(Q~1602$+$576) to $5.05$ (HE~2347$-$4342). Six of the 11 sightlines sampling
the $z=2.80$ \ion{He}{2} absorption still have $\tau_\mathrm{eff,HeII}<2.5$,
while the fully saturated sightline to SDSS~J0936$+$2927 has a sensitivity limit
$\tau_\mathrm{eff,HeII}>4.48$ (Fig.~\ref{fig:he2spc_cycle17}).
At $z=2.88$ we observe a large dispersion around $\tau_\mathrm{eff,HeII}\simeq 3$
among 9 sightlines, ranging from $\tau_\mathrm{eff,HeII}=1.92$ to
$\tau_\mathrm{eff,HeII}=4.69$, close to the characteristic sensitivity limit for
the data at these redshifts. As most of the $\tau_\mathrm{eff,HeII}$ values at
$2.7<z<2.9$ are accurately measured (well below the sensitivity limit),
we conclude that there is a gradual increase in $\tau_\mathrm{eff,HeII}$ from
$\simeq 1.8$ at $z=2.68$ to $\simeq 3$ at $z=2.88$, but with considerable
sightline-to-sightline variance.

The sightline-to-sightline variance in \ion{He}{2} absorption persists out to
$z>3$ as probed by 7 science-grade quasar spectra. Figure~\ref{fig:highzspc2}
shows the data and our $\tau_\mathrm{eff,HeII}$ measurements at $3.06<z<3.26$.
We find a large spread between robustly measured values
$\tau_\mathrm{eff,HeII}\simeq 3.3$ and lower limits $\tau_\mathrm{eff,HeII}\ga 5.5$.
The five $\tau_\mathrm{eff,HeII}<4$ values at $3.06<z<3.26$ occur in three of
the six sightlines (SDSS~J1237$+$0126, SDSS~J1253$+$6817 and HS~0911$+$4809).
For each of these measurements the probability of finding a background
fluctuation consistent with the measured flux is $P<10^{-7}$ (Equation~\ref{eq:prob}),
with this value being robust to possible background systematics
(even a 20\% higher background would still result in significant positive flux).
The \ion{He}{2} transmission still appears patchy, occurring on smaller scales
than our chosen $\Delta z=0.04$, but limited S/N and spectral resolution prevent
a detailed characterization of these length scales. Toward SDSS~J1253$+$6817 we
measure low effective optical depths in two contiguous redshift bins ($3.10<z<3.18$),
possibly continuing to lower redshifts that were excised due to geocoronal residuals.
Other sightlines (e.g.\ SDSS~J2346$-$0016) show complete Gunn-Peterson troughs
at the same redshifts. This indicates that part of the spread in the data is due
to large-scale variance between the sightlines.

Figure~\ref{fig:highzspc} shows the 4 sightlines covering $3.34<z<3.5$. Half of
the redshift bins have $\tau_\mathrm{eff,HeII}\simeq 4$, although all sightlines
are sensitive to $\tau_\mathrm{eff,HeII}\simeq 5$. Again we see a strong
sightline-to-sightline variance, with the highest effective optical depths
measured toward SDSS~J2346$-$0016, whereas the absorption in two sightlines
remains low (SDSS~J1319$+$5202 and SDSS~J1711$+$6052). Our Ly$\alpha$ effective
optical depths are in good agreement with inferences from \ion{He}{2} Ly$\beta$
absorption at these redshifts \citep{syphers11}. The lowest \ion{He}{2} effective
optical depth at $z>3.3$ is robustly measured in a flux spike in the
SDSS~J1319$+$5202 sightline at $z=3.44$. Again we see that the \ion{He}{2}
transmission occurs on smaller scales than our $\Delta z=0.04$ redshift windows
($\delta z\la 0.02$ corresponding to $\la 4$ proper Mpc at $z=3.44$).
The other four detections occurring in the sightlines to SDSS~J1319$+$5202 and
SDSS~J1711$+$6052 are closer to the sensitivity limit, meaning that some of them
may be Poisson background fluctuations ($0.001\le P\le 0.043$).
Large-scale underestimates of the background are unlikely, as strong background
oversubtractions would occur in other regions. Consistency with a Poisson
background fluctuation (i.e.\ $P\gg 0.01$ for all four values) would require
local increases of the mean background by more than its estimated $1.6$--$3.2$\%
uncertainty (Appendix~\ref{sect:bkgsumappendix}). We conclude that Poisson
background fluctuations cannot entirely account for these measurements.

Only two sightlines in our sample probe $z>3.5$. Given the large observed variance
in $\tau_\mathrm{eff,HeII}$ at $z>3$, it is extremely difficult to draw firm
conclusions on the redshift evolution of the \ion{He}{2} absorption at the
highest redshifts. Moreover, the decreasing instrument sensitivity at the
corresponding wavelengths $\lambda>1350$\,\AA\ combined with the faintness of
the targets results in low sensitivity to high $\tau_\mathrm{eff,HeII}$ values,
some of which can be seen already at $z\simeq 3.4$. Statistically robust
constraints on the redshift evolution of the \ion{He}{2} effective optical depth
at $z>3.5$ will require a larger sample of \ion{He}{2} sightlines observed
at high S/N. Analysis of our recently obtained sample of three $z>3.6$ sightlines
is forthcoming (Program 13875).

We may compare the $\tau_\mathrm{eff,HeII}$ distributions with redshift
to test statistically for evolution in the \ion{He}{2} opacity.
The median value of $\tau_\mathrm{eff,HeII}$ is not well defined
at $z>3$ due to the frequent sensitivity limits and limited statistics.
In an attempt to better sample the underlying distribution of
$\tau_\mathrm{eff,HeII}$ at a given redshift, we assumed that contiguous
$\Delta z=0.04$ redshift bins of the same sightline are independent,
a strong approximation given the significant correlation between neighboring
redshift bins, especially at $z>3$. The median $\tau_\mathrm{eff,HeII}$
increases gradually from $1.94$ at $z=2.70$ (19 measurements at $2.66<z<2.74$)
to $5.17$ at $z\simeq 3.4$ (10 measurements at $3.34<z<3.50$),
although the latter is poorly constrained to the highest robustly measured
$\tau_\mathrm{eff,HeII}$ value (50\% of the data are sensitivity limits).
Nevertheless, this result highlights the trend described above: the effective
\ion{He}{2} Ly$\alpha$ opacity increases monotonically from $z=2.4$ to
$z=3.4$ by a factor of 2--3.

Armed with our statistical formalism to estimate the signal significance,
we combined the sightlines to estimate the overall significance of any residual
flux. While this dilutes the significance of individual detections, it also
averages out individual background errors.
For both high-redshift intervals $3.06<z<3.26$ and $3.34<z<3.50$, the probability
that all measured counts above the background are caused by Poisson background
fluctuations is very small ($P\simeq 10^{-7}$ and $P\simeq 2\times 10^{-6}$, respectively).
At $3.34<z<3.50$ the flux spike in SDSS~J1319$+$5202 dominates the signal.
Discarding this potentially rare transmission event, the signal becomes marginally
consistent with a Poisson background fluctuation ($P=0.015$).
We conclude that if the four sightlines sampling $z>3.3$ are representative of
the IGM, the average \ion{He}{2} effective optical depth is not much higher than
our typical sensitivity limit $\tau_\mathrm{eff,HeII}\simeq 5$.
Let us now consider implications for \ion{He}{2} reionization in the context of
several models.

\subsection{Comparison to Models}

\subsubsection{Semianalytic Modeling}
In \citet{worseck11b} we constructed a simple semianalyic model
for the post-reionization $\tau_\mathrm{eff,HeII}(z)$.
It relies on the fact that \ion{He}{2} is a hydrogenic ion, hence the Ly$\alpha$
optical depths 
\begin{equation}
\tau_i\left(z\right)=\frac{\pi e^2f_i\lambda_in_i\left(z\right)}{m_ecH\left(z\right)}
\end{equation}
\citep{gunn65} of species $i=$\ion{H}{1} or \ion{He}{2} are related to their number densities via
\begin{equation}\label{eq:he2h1ratio}
\frac{n_\mathrm{HeII}}{n_\mathrm{HI}}\simeq 4\frac{\tau_\mathrm{HeII}}{\tau_\mathrm{HI}}
\end{equation}
\citep{miralda-escude93}.
This relation is approximate, as it does not account for differences in thermal
broadening of \ion{H}{1} and \ion{He}{2} by the different masses of the two elements.
However, we showed in \citet{mcquinn14} that this approximation negligibly affects
the estimated $n_\mathrm{HeII}/n_\mathrm{HI}$ at the resolution of the COS G140L
grating, as the Hubble flow dominates the broadening in underdense regions.
In particular, this error should be small for our modeling of $\tau_\mathrm{eff,HeII}$ below.
For the resolved \ion{He}{2} Ly$\alpha$ forest in the post-reionization $z<2.7$
IGM the $n_\mathrm{HeII}/n_\mathrm{HI}$ number density ratio
has been traditionally approximated as the ratio of column
densities $\eta\equiv N_\mathrm{HeII}/N_\mathrm{HI}$
\citep[e.g.][]{kriss01,fechner06}, but the $z\ga 2.8$ \ion{He}{2} Gunn-Peterson
troughs cannot be decomposed into distinct lines even at high spectral
resolution \citep[e.g.][]{shull10,syphers14}, making the optical depth ratio
the optimal estimator for the $n_\mathrm{HeII}/n_\mathrm{HI}$ number density ratio.

If both H and He are highly photoionized and the IGM baryons follow
a temperature-density relation
$T\left(\Delta_\mathrm{b}=\rho_\mathrm{b}/\bar{\rho}_\mathrm{b}\right)=T_0\Delta_\mathrm{b}^{\gamma-1}$
\citep{hui97}, the \ion{H}{1} optical depth can be written as
\begin{eqnarray}\label{eq:fgpa}
\tau_\mathrm{HI}&\simeq&0.612\left(\frac{T_0}{20,000\mathrm{K}}\right)^{-0.724}
\left(\frac{\Gamma_\mathrm{HI}}{10^{-12}\mathrm{s}^{-1}}\right)^{-1}\\\nonumber
&\times&\Delta_\mathrm{b}^{2-0.724(\gamma-1)}\left(\frac{1+z}{4}\right)^{4.5}
\end{eqnarray}
\citep[e.g.][]{weinberg97}. In the same limit of high photoionization the
$n_\mathrm{HeII}/n_\mathrm{HI}$ number density ratio can be expressed with the primordial
helium mass fraction $Y=0.2477$ \citep{peimbert07} and the ratios of
photoionization rates $\Gamma_\mathrm{HI}/\Gamma_\mathrm{HeII}$ and Case A
recombination coefficients $\alpha_\mathrm{HII}/\alpha_\mathrm{HeIII}$ as
\begin{equation}\label{eq:he2h1ot}
\frac{n_\mathrm{HeII}}{n_\mathrm{HI}}=\frac{\alpha_\mathrm{HeIII}}{\alpha_\mathrm{HII}}\frac{Y}{4\left(1-Y\right)}\frac{\Gamma_\mathrm{HI}}{\Gamma_\mathrm{HeII}}\simeq 0.450\frac{\Gamma_\mathrm{HI}}{\Gamma_\mathrm{HeII}}
\end{equation}
at $T_0\sim 20,000$\,K \citep[e.g.][]{fardal98}. With the $n_\mathrm{HeII}/n_\mathrm{HI}$
number density ratio the \ion{He}{2} effective optical depth can be expressed
in terms of better constrained \ion{H}{1} quantities as
\begin{equation}\label{eq:h1he2taueff}
\tau_\mathrm{eff,HeII}=-\ln{\left[\int_0^\infty e^{-\frac{n_\mathrm{HeII}/n_\mathrm{HI}}{4}\tau_\mathrm{HI}}P\left(\tau_\mathrm{HI}\right)d\tau_\mathrm{HI}\right]}.
\end{equation}
Here $P\left(\tau_\mathrm{HI}\right)$ is the \ion{H}{1} optical depth
probability distribution function, which is related to the overdensity
probability distribution $P\left(\Delta_\mathrm{b}\right)$ as
$P\left(\tau_\mathrm{HI}\right)=P\left(\Delta_\mathrm{b}\right)\left| d\Delta_\mathrm{b}/d\tau_\mathrm{HI}\right|$.
Interpolating the fits of $P\left(\Delta_\mathrm{b}\right)$ by \citet{bolton09c}
in redshift and considering Equation~\ref{eq:fgpa}, $\tau_\mathrm{eff,HeII}$
depends on the temperature--density relation ($T_0,\gamma$) and the ionization
conditions ($\Gamma_\mathrm{HI},n_\mathrm{HeII}/n_\mathrm{HI}$).

Varying the parameters within their estimated accuracies, the predicted
$\tau_\mathrm{eff,HeII}(z)$ mostly depends on the ratio
$n_\mathrm{HeII}/n_\mathrm{HI}$, which depends on the ratio of
ionization rates (Equation~\ref{eq:he2h1ot}).
Adopting $T_0=15,000$\,K, a post-reionization asymptotic
value $\gamma=1.5$ \citep{hui97} and
$\Gamma_\mathrm{HI}=10^{-12}\,\mathrm{s}^{-1}$ \citep[e.g.][]{becker13b},
we obtain a set of curves of $\tau_\mathrm{eff,HeII}(z)$ for varying
$n_\mathrm{HeII}/n_\mathrm{HI}$, two of which are shown in Fig.~\ref{fig:he2tau}.
For the chosen set of parameters, $n_\mathrm{HeII}/n_\mathrm{HI}=60$ and
$n_\mathrm{HeII}/n_\mathrm{HI}=200$ represent the extremes of a hard
quasar-dominated UV background and a soft galaxy-dominated UV background
\citep{haardt12}, yielding an envelope of $\tau_\mathrm{eff,HeII}(z)$ in a
highly photoionized optically thin IGM. For a constant number density ratio
(and thus a fixed spectral shape of the UV background), the gradual
increase in $\tau_\mathrm{eff,HeII}(z)$ is due to density evolution
in the IGM. In fact, observations at $z<2.7$ and the lowest observed
$\tau_\mathrm{eff,HeII}$ values at $2.7<z<3$ are well fit by
$n_\mathrm{HeII}/n_\mathrm{HI}=60$--100, similar to the smaller sample of 5
\ion{He}{2} sightlines considered in \citet{worseck11b}. This suggests
that helium at $z<3$ is predominantly fully ionized. Strikingly, the
highest redshifts ($3.3<z<3.5$), covered by our sample statistically
for the first time (4 sightlines), show small and robust
\ion{He}{2} effective optical depths ($\tau_\mathrm{eff,HeII}\simeq 4$)
on 50\% of the pathlength.
The well-matching extrapolation from lower redshifts
suggests that at least parts of these \ion{He}{2} transmission patches
at $z>3.3$ are highly ionized. 

Density fluctuations in the IGM on the probed $\sim 10$\,Mpc scales, as well as
fluctuations in the UV background in the aftermath of \ion{He}{2} reionization
will cause scatter around our semianalytic model prediction for
$\tau_\mathrm{eff,HeII}(z)$. However, the frequent occurrence of
$\tau_\mathrm{eff,HeII}>3$ at $2.7<z<3$ and the upturn of
$\tau_\mathrm{eff,HeII}(z)$ at $z>3$ (frequent sensitivity limits) would require
spatial variations and redshift evolution in the model parameters,
e.g.\ $n_\mathrm{HeII}/n_\mathrm{HI}$.
A flattening in the temperature-density relation would lead to a decrease in
$\tau_\mathrm{eff,HeII}$, and a strong rise in $\tau_\mathrm{eff,HeII}$ would
require unreasonably low and redshift-dependent IGM temperatures.
Despite uncertainties in its normalization, the \ion{H}{1} photoionization rate
is found to be almost constant over the probed redshift range
\citep{bolton05,faucher08b,dallaglio08,becker13b} with only small spatial
fluctuations due to the large number of ionizing sources within a mean free path
of \ion{H}{1} Lyman continuum photons \citep{meiksin04,croft04,prochaska09,pontzen14,gontcho14}.
Thus, in a reionized IGM an increasing $n_\mathrm{HeII}/n_\mathrm{HI}$ ratio corresponds
to a decreasing \ion{He}{2} photoionization rate (Equation~\ref{eq:he2h1ot}). 

Making use of the primordial IGM abundances of hydrogen and helium, we may write
the \ion{He}{2} fraction as
\begin{equation}\label{eq:he2frac}
x_\mathrm{HeII}=\frac{n_\mathrm{HeII}}{n_\mathrm{He}}=\frac{4(1-Y)}{Y}\frac{n_\mathrm{HeII}}{n_\mathrm{HI}}x_\mathrm{HI}\simeq 12.15\frac{n_\mathrm{HeII}}{n_\mathrm{HI}}x_\mathrm{HI}.
\end{equation}
Similar to Equation~\ref{eq:fgpa}, the \ion{H}{1} fraction can be expressed as
\begin{eqnarray}\nonumber\label{eq:h1frac}
x_\mathrm{HI} &\simeq& 4.0\times 10^{-6}\left(\frac{T_0}{20,000\mathrm{K}}\right)^{-0.724}\Delta_\mathrm{b}^{1-0.724(\gamma-1)}\left(\frac{1+z}{4}\right)^3\\
& &\times\left(0.062(2-x_\mathrm{HeII})+0.752\right)\left(\frac{\Gamma_\mathrm{HI}}{10^{-12}\mathrm{s}^{-1}}\right)^{-1},
\end{eqnarray}
with a weak dependence on the temperature-density relation, the \ion{H}{1}
photoionization rate $\Gamma_\mathrm{HI}\simeq\mathrm{const.}$ \citep{becker13b},
and the ionization state of helium (the above equation assumes $x_\mathrm{HeI}\simeq 0$).
Integration of Equation~\ref{eq:h1frac}
over all densities with the overdensity probability distribution
$P\left(\Delta_\mathrm{b}\right)$ \citep{bolton09c} gives the mean
\ion{H}{1} fraction as a function of redshift. The weak implicit dependence on
$x_\mathrm{HeII}$ can be neglected given the assumptions of Equation~\ref{eq:h1frac}
with fixed parameters ($\Gamma_\mathrm{HI}=10^{-12}$\,s$^{-1}$, $T_0=15,000$\,K, $\gamma=1.5$).

At $z\simeq 2.7$ our measured $\tau_\mathrm{eff,HeII}\simeq 2$ implies
$n_\mathrm{HeII}/n_\mathrm{HI}\simeq 100$ in 8/10 sightlines (Fig.~\ref{fig:he2tau})
at $\bar{x}_\mathrm{HI}\simeq 2.6\times 10^{-6}$, corresponding to
$\bar{x}_\mathrm{HeII}\simeq 0.003$ (Equation~\ref{eq:he2frac}),
confirming that helium is highly ionized. At higher redshifts, however,
the scatter in $\tau_\mathrm{eff,HeII}$ implies fluctuations in
$\bar{x}_\mathrm{HeII}$ by a factor $\sim 4$ if we assume that our measurements
on scales of $\simeq 10$\,Mpc statistically sample all relevant densities.
At $z\simeq 2.8$ the ionization level is still quite homogeneous, but the
Gunn-Peterson troughs in HE~2347$-$4342 and SDSS~J0936$+$2927 imply
$\bar{x}_\mathrm{HeII}\simeq 0.015$ on $\simeq 10$\% of the probed pathlength,
suggesting that we sample the tail end of the \ion{He}{2} reionization process.
The lack of strong $x_\mathrm{HeII}$ fluctuations at $z\la 2.7$ suggests
that \ion{He}{2} reionization ended at $z\simeq 2.7$, in agreement with previous
results based on less than half of the data analyzed here
\citep{shull10,furlanetto10,worseck11b}.
At $z\simeq 3.16$ the low values $\tau_\mathrm{eff,HeII}\simeq 3.3$
still imply \ion{He}{2} fractions of $\bar{x}_\mathrm{HeII}\simeq 0.006$
over 30\% of the pathlength, while 65\% of the covered pathlength is highly
saturated at $\tau_\mathrm{eff,HeII}\ga 5.5$, implying $\bar{x}_\mathrm{HeII}\ga 0.013$.
The spread in the inferred \ion{He}{2} fractions persists out to the highest redshifts
statistically probed by our sample.
The 5 statistically significant detections of \ion{He}{2} transmission at $z\simeq 3.4$
imply $n_\mathrm{HeII}/n_\mathrm{HI}<200$ or equivalently $\bar{x}_\mathrm{HeII}<0.01$
over 50\% of the covered pathlength.
However, the \ion{He}{2} effective optical depths for $n_\mathrm{HeII}/n_\mathrm{HI}\ga 200$ 
reach our sensitivity limits (Fig.~\ref{fig:he2tau}), leaving the \ion{He}{2}
fractions poorly constrained to $\bar{x}_\mathrm{HeII}\ga 0.01$ for the other
half of the sample at $z\simeq 3.4$.

The strong variance in $\tau_\mathrm{eff,HeII}$ with implied factor $\ga 4$ 
variations in the \ion{He}{2} fraction suggests that \ion{He}{2} reionization was
inhomogeneous and extended. The small scatter of $\tau_\mathrm{eff,HeII}$
at low redshifts confines the end of the process to $z\simeq 2.7$, although we
note that lower redshifts are currently probed just by two quasar sightlines.
On the other hand, the high fraction of low $\tau_\mathrm{eff,HeII}$ values at
$z\simeq 3.4$ suggests that \ion{He}{2} reionization was well underway at
these redshifts and must have begun at $z>4$.
The low $\tau_\mathrm{eff,HeII}$ values at $z>3$ indicate substantially ionized
regions, expected for gradual \ion{He}{2} reionization \citep{furlanetto09},
and our measurements yield the first statistical constraints on their frequency.
If the four sightlines at $z\simeq 3.4$ yield a representative sample of the
density field (which is likely given our averaging over $\approx 10$\,Mpc scales),
a low $\bar{x}_\mathrm{HeII}<1$\% over half of the pathlength translates
to a \ion{He}{2} photoionization rate $\Gamma_\mathrm{HeII}\sim 2\times 10^{-15}$\,s$^{-1}$,
while for the other half $\Gamma_\mathrm{HeII}$ may be much lower.
At $z<3$, fluctuations in $\Gamma_\mathrm{HeII}$ are better constrained
due to lower IGM densities and better statistical sampling.
The locus of low $\tau_\mathrm{eff,HeII}\simeq 2$ values at $2.66<z<2.80$,
i.e.\ the bulk of the IGM at these redshifts, is consistent with
$\Gamma_\mathrm{HeII}\sim 5\times 10^{-15}$\,s$^{-1}$,
whereas $\sim 10$\% of the pathlength is exposed to a factor $\sim 4$
lower $\Gamma_\mathrm{HeII}$.

Fluctuations in the \ion{He}{2} photoionization rate are expected during and
after \ion{He}{2} reionization \citep{furlanetto09,furlanetto10,davies14,mcquinn14}.
However, downward fluctuations by a factor of $\sim 4$ on $\sim 10$\,Mpc
scales are unlikely to occur in a post-reionization IGM, indicating that
\ion{He}{2} reionization is ongoing at $z\simeq 2.8$ \citep{furlanetto10}. 
Recent work tried to reproduce the evolution in the the mean \ion{He}{2}
absorption at $z<3$ with a steeply evolving $\Gamma_\mathrm{HeII}\left(z\right)$
in a post-reionization IGM \citep{khaire13,davies14,puchwein15}, corresponding
to a steeply evolving $n_\mathrm{HeII}/n_\mathrm{HI}$ ratio in Fig.~\ref{fig:he2tau}.
However, the steeply evolving mean free path to \ion{He}{2}-ionizing photons
implied by these models becomes comparable to the typical quasar separation at
$z>3$, leading to large fluctuations in $\Gamma_\mathrm{HeII}$ that likely make
\ion{He}{2} reionization unavoidable \citep{davies14}. Moreover, these models
cannot predict the variance in the \ion{He}{2} effective optical depth,
such that they cannot be straightforwardly compared to our measurements.
Essentially, homogeneous UV background models \citep[e.g.][]{haardt12} cannot
predict the redshift evolution of the \ion{He}{2} effective optical depth in
the presence of UV background fluctuations and during \ion{He}{2} reionization
\citep[see also][]{puchwein15}.

Finally, we can estimate the mean free path to \ion{He}{2}-ionizing photons from
the well-constrained \ion{H}{1} column density distribution and
$n_\mathrm{HeII}/n_\mathrm{HI}$ \citep[e.g.][]{mcquinn14}.
The inferred fluctuations in $n_\mathrm{HeII}/n_\mathrm{HI}$ translate to
fluctuations in the mean free path.
At $z\sim 3.4$ we infer a \ion{He}{2} mean free path of $\sim 50$ comoving Mpc
for 50\% of the pathlength, while it may be substantially shorter for the rest of
the pathlength, which may signal ongoing \ion{He}{2} reionization. On the other hand,
as the mean free path becomes comparable to the mean separation between luminous
($\nu_BL_B>10^{11}L_\sun$) quasars at $z>3$ \citep{furlanetto08,davies14},
quasar clustering will result in spatial variations of the mean free path if such
sources dominate \ion{He}{2} reionization. Further insights require
numerical simulations of \ion{He}{2} reionization.

\subsubsection{Comparison to Numerical Simulations of \ion{He}{2} Reionization}

We also compared our measurements to a numerical simulation of \ion{He}{2}
reionization. As in \citet{worseck11b} we calculated $\tau_\mathrm{eff,HeII}(z)$
from 1000 skewers per snapshot from the L3 run of \citet{mcquinn09a} in which
\ion{He}{2} reionization completes at $z\simeq 2.7$. The thick curves in
Fig.~\ref{fig:he2tau} show the median and $1\sigma$ deviation of
$\tau_\mathrm{eff,HeII}(z)$ obtained on the same scale as our measurements
$\Delta z=0.04$. With much improved statistics at $z\simeq 2.7$, we confirm the
turnover to the optically thin post-reionization IGM at $z\simeq 2.7$ found in
\citet{worseck11b}. Model L3 is the only model considered in \citet{mcquinn09a}
that reproduces the observed $\tau_\mathrm{eff,HeII}$ at $z\simeq 2.7$ with
similar scatter. At higher redshifts the increase in the median modeled
$\tau_\mathrm{eff,HeII}$ and its scatter are tracers of ongoing \ion{He}{2}
reionization. However, the model is inconsistent with the data in several ways:
(1) It does not reproduce the large scatter seen in the data, most notably
the dark trough ($\tau_\mathrm{eff,HeII}=5.05^{+0.09}_{-0.08}$) at $z=2.76$ towards
HE~2347$-$4342 that is inconsistent with the L3 model at $\sim 98$\%
confidence \citep{worseck11b};
(2) It somewhat overpredicts the median $\tau_\mathrm{eff,HeII}$ at $2.8\la z\la 2.9$
that is robustly measured in the current data; 
(3) It fails to reproduce the frequent low effective optical depths at $z>3$.
The \ion{He}{2} reionization model predicts that $\simeq 17$\% of the data at
$z=3.14$ should have $\tau_\mathrm{eff,HeII}<4$, but we observe almost twice
as many of these low effective optical depths (5/17 values at $3.06<z<3.26$).
Similarly, at $z\sim 3.4$ the model predicts a low fraction of
$\tau_\mathrm{eff,HeII}<5$ patches (11\%), but we observe a much larger fraction (50\%).
To ease the discrepancy in the median effective optical depth,
all $\tau_\mathrm{eff,HeII}$ sensitivity limits would need to correspond to
highly opaque regions, yielding a bimodal \ion{He}{2} fraction consistent with
the $\bar{x}_\mathrm{HeII}\simeq 0.4$ predicted by the model at $z\simeq 3.4$.
While such a reionization scenario may be plausible, the mismatch in the tail to
low effective optical depths remains. However, we note that the models presented
in \citet{mcquinn09a} were not tested for convergence of $\tau_\mathrm{eff,HeII}$
with simulation resolution, and higher resolution will likely produce lower
opacities in ionized regions.

Likewise, our observations are inconsistent with predictions from
numerical radiative transfer simulations of \ion{He}{2} reionization by
\citet{compostella13}, who predict a similarly steep
$\tau_\mathrm{eff,HeII}\left(z\right)$ evolution as \citet{mcquinn09a},
with large fluctuations at $z>3$ due to ongoing \ion{He}{2} reionization.
None of their three models of a rapid \ion{He}{2} reionization between $z=4$ and
$z=2.7$ matches our measured $\tau_\mathrm{eff,HeII}$ distribution.
At $z\simeq 3.2$ and on the same scale $\Delta z=0.04$, the observed frequency
of low values $\tau_\mathrm{eff,HeII}\simeq 3.3$ is much higher (30\%) than
predicted by their simulations ($\la 2$\%).
The large discrepancy in the occurrence of low effective optical depths
$\tau_\mathrm{eff,HeII}\simeq 4$ at $z\simeq 3.4$
($\sim 50$\% observed vs.\ $\la 15$\% predicted) is inconsistent with the bulk
of \ion{He}{2} in the IGM being ionized by $z<4$ quasars
unless they had significantly different properties
(e.g.\ characteristic halo mass, lifetime, opening angle) than assumed by
\citet{compostella13}.
Recently, \citet{compostella14} showed that this tension is alleviated for models
in which \ion{He}{2} reionization around quasars starts at $z=5$--6.
Indeed, our refined measurements at $z\simeq 3.4$ are in good agreement with
the simulations by \citet{compostella14} in the tail to low \ion{He}{2} effective
optical depths. However, the \citet{compostella14} simulations fail to reproduce
the tail to $\tau_\mathrm{eff,HeII}>4$ at $z\simeq 2.8$, likely because
\ion{He}{2} reionization completes at $z\simeq 3$ in their models.
Numerical resolution effects leading to overestimates of $\tau_\mathrm{eff,HeII}$
in low-density regions exacerbate the tension between these models and the data at
$z\simeq 2.8$.
We reiterate that our $\tau_\mathrm{eff,HeII}$
measurements are not centered on specific features in the sightlines,
enabling a one-to-one comparison to the models.

Our observations call for refined models of a more extended \ion{He}{2}
reionization that achieve small \ion{He}{2} fractions $x_\mathrm{HeII}<0.01$
throughout the IGM at $z\simeq 3.4$, while still leaving a few $\sim 10$\,Mpc
patches of incompletely ionized \ion{He}{2} ($x_\mathrm{HeII}\simeq 0.015$) at
$z\simeq 2.8$. In particular, if \ion{He}{2} reionization was driven by quasars,
it must have started at $z>4$ to yield substantial ionization levels at
$z\simeq 3.4$ \citep{mcquinn09a,compostella14}.
Furthermore, the observed coherence in \ion{He}{2} absorption
along the sightlines constrain scenarios in which only small regions within
each probed $\sim 10$\,Mpc patch are highly ionized.
However, the coherence may arise due to correlations in the density field,
which we will explore next with coeval \ion{H}{1} Ly$\alpha$ forest spectra.

\begin{figure*}[!t]
\includegraphics[width=\textwidth]{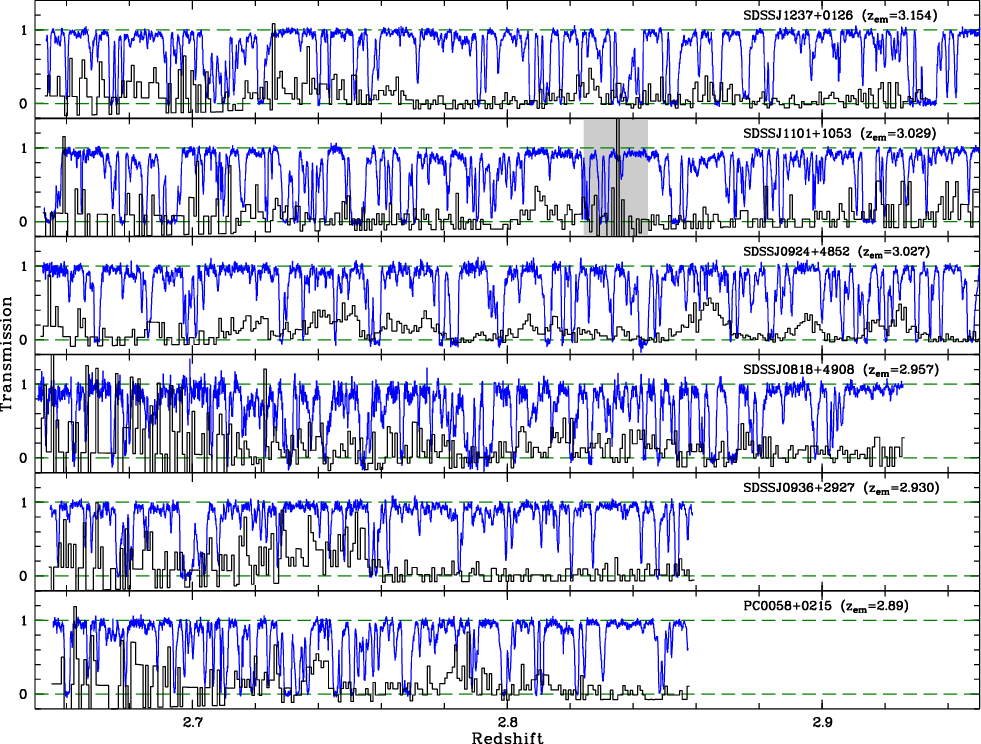}
\caption[]{\label{fig:h1he2spc}
Normalized coeval optical echelle \ion{H}{1} (blue) and \textit{HST}/COS UV
\ion{He}{2} (black) Ly$\alpha$ absorption spectra of the six \ion{He}{2}
sightlines from our Cycle~17 survey as a function of redshift. For display
purposes the \ion{H}{1} spectra have been rebinned to $7.8$\,km\,s$^{-1}$\,pixel$^{-1}$.
Proximity zones of the background quasars are not shown. The dashed lines
mark the zero and the continuum level, respectively. The shaded region in the
spectrum of SDSS~J1101$+$1053 is impacted by foreground absorption
\citep{worseck11b}.
} 
\end{figure*}

\section{Comparing \ion{He}{2} Absorption with the Coeval \ion{H}{1}
  Ly$\alpha$ Forest}\label{sect:he2h1}

In the previous section we presented results based on $\tau_\mathrm{eff,HeII}$
statistics which suggest that \ion{He}{2} reionization was incomplete at $z\sim 2.8$
but that it was fully in progress by $z=3.5$. These conclusions rely, in part,
on our expectation that the distribution of $\tau_\mathrm{eff,HeII}$ values
correspond to typical regions of the $z\sim 3$ IGM.
This assumption may be assessed through an evaluation of and comparison with the
coeval \ion{H}{1} Ly$\alpha$ forest. The coeval \ion{H}{1} Ly$\alpha$ forest
traces the underlying density field, constraining to which degree the fluctuating
\ion{He}{2} absorption can be explained by variations in density and ionization
level, respectively. Here we analyze the subset of 10 \ion{He}{2} sightlines with
available optical echelle spectroscopy of the \ion{H}{1} Ly$\alpha$
forest. We then compare the results with simple models of completely reionized
\ion{He}{2}.

\subsection{Qualitative Comparison of the Coeval \ion{He}{2} and
  \ion{H}{1} Ly$\alpha$ Spectra}

\begin{figure*}[t]
\includegraphics[width=\textwidth]{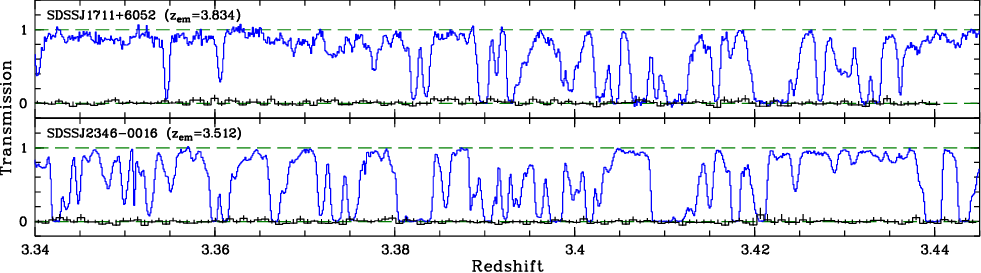}
\caption[]{\label{fig:h1he2spc2}
Normalized coeval optical echelle \ion{H}{1} (blue) and \textit{HST}/COS UV \ion{He}{2} (black)
Ly$\alpha$ absorption spectra of the two high-redshift \ion{He}{2} sightlines
toward SDSS~J1711$+$6052 and SDSS~J2346$-$0016. For display purposes the
\ion{H}{1} spectra have been rebinned to $7.8$\,km\,s$^{-1}$\,pixel$^{-1}$.
The \ion{He}{2} spectra have been binned to two pixels per resolution element
($\simeq 0.24$\,\AA\,pixel$^{-1}$) and individual Poisson errors are overplotted.
 The dashed lines mark the zero and the continuum level, respectively.
Toward SDSS~J1711$+$6052 there is low-level \ion{He}{2} transmission in some regions of
high \ion{H}{1} transmission, whereas the SDSS~J2346$-$0016 sightline is fully saturated in \ion{He}{2}.
} 
\end{figure*}

Figure~\ref{fig:h1he2spc} presents the coeval \ion{H}{1} and \ion{He}{2}
Ly$\alpha$ absorption spectra of the 6 \ion{He}{2}-transparent sightlines from
our Cycle~17 survey, highlighting the variance in \ion{He}{2} absorption at
similar redshifts (e.g.\ the $2.76<z<2.86$ \ion{He}{2} Gunn-Peterson
trough in SDSS~J0936$+$2927 vs.\ the very patchy localized absorption
in SDSS~J0924$+$4852 over the full covered redshift range $2.66<z<2.94$).
Some \ion{He}{2} transmission regions show substructure with corresponding
\ion{H}{1} lines, for example toward SDSS~J0924$+$4852 at $z\simeq 2.86$
and $z\simeq 2.92$. A few regions have high \ion{H}{1} and \ion{He}{2}
transmission, e.g.\ toward SDSS~J0924$+$4852 at $z\simeq 2.71$ and $z\simeq 2.89$.
On average, however, \ion{H}{1} and \ion{He}{2} absorption do not appear 
well correlated. While several instances of strong \ion{H}{1} Ly$\alpha$
absorption with non-zero \ion{He}{2} transmission might be explained by the
low resolution of the \ion{He}{2} spectra, there are numerous regions of
weak \ion{H}{1} absorption but strong \ion{He}{2} absorption, e.g.\ at
$2.76<z<2.86$ toward SDSS~J0936$+$2927.
Together, these examples indicate a significant dispersion in the
number density ratio $n_\mathrm{HeII}/n_\mathrm{HI}$.
They further suggest that the different \ion{He}{2} absorption patterns are
primarily due to variations in the ionization level as a sign of ongoing \ion{He}{2}
reionization \citep{shull10,worseck11b}. While we do not expect a one-to-one
relation between the \ion{H}{1} and \ion{He}{2} fluxes even for a constant
$n_\mathrm{HeII}/n_\mathrm{HI}$ due to thermal broadening, low COS resolution
and Poisson noise, the indication for an anti-correlation of \ion{H}{1} and
\ion{He}{2} absorption deserves further exploration
(see Section~\ref{sect:h1he2tau} below).

The Keck/HIRES spectra of SDSS~J1711$+$6052 and SDSS~J2346$-$0016 extend this
comparison to $z\simeq 3.4$ (Fig.~\ref{fig:h1he2spc2}).
The sightline to SDSS~J1711$+$6052 shows some low-level \ion{He}{2} transmission
in \ion{H}{1} transmission regions, although the quality of the \ion{He}{2}
spectrum is poor. At $3.34<z<3.38$ and $3.38<z<3.42$ we measure \ion{He}{2}
effective optical depths $\tau_\mathrm{eff,HeII}=5.17^{+0.90+0.30}_{-0.49-0.04}$
and $\tau_\mathrm{eff,HeII}=4.55^{+0.44+0.07}_{-0.31-0.03}$, respectively.
Both measurements are rather unlikely to result from a Poisson background fluctuation
($P=0.043$ and $0.001$, respectively). This is in sharp contrast with the
SDSS~J2346$-$0016 sightline that remains highly saturated at all levels of
\ion{H}{1} absorption (formally negative flux over the same redshift ranges,
$1\sigma$ sensitivity limit $\tau_\mathrm{eff,HeII}>5.6$).
The somewhat stronger \ion{H}{1} forest absorption toward SDSS~J2346$-$0016 can
only partly account for the difference. Qualitatively, our limited sample of two
$z>3.3$ sightlines with high-quality \ion{He}{2} and \ion{H}{1} spectra
indicates that the patchiness in \ion{He}{2} absorption persists up to $z\simeq 3.4$,
albeit at a lower level of transmission due to the higher density of the IGM.  

\subsection{\ion{H}{1} Ly$\alpha$ Statistics along the \ion{He}{2} Sightlines}

With coeval \ion{H}{1} Ly$\alpha$ spectroscopy for the majority of our \ion{He}{2}
sightlines in hand, we may assess whether the IGM probed by these quasars
has properties consistent with those derived in previous studies.
Specifically, we measure the \ion{H}{1} Ly$\alpha$ effective optical depth
$\tau_\mathrm{eff,HI}=-\ln{\left<F_\mathrm{HI}\right>}$, with
$\left<F_\mathrm{HI}\right>$ the average \ion{H}{1} transmission in a given
redshift window, and compare against measurements from much larger, statistical
studies of the IGM. Before proceeding, we should note that all \ion{He}{2}
sightlines are biased in at least one manner: the quasars were chosen to have
significant flux at rest frame wavelength 304\,\AA, demanding the absence of
\ion{H}{1} Lyman limit systems with column densities
$N_\mathrm{HI}\ga 10^{19}\,\mathrm{cm}^{-2}$ at
$z_\mathrm{LLS}>0.33\left(1+z_\mathrm{em}\right)-1$.
Quasars with lower column density Lyman limit systems may sufficiently recover
in the FUV \citep[e.g.][]{worseck11}. As most statistical studies of the
\ion{H}{1} Ly$\alpha$ opacity exclude or correct for optically thick absorbers
\citep{kirkman05,faucher08,becker13} that are naturally missing in
our sightlines, we do not expect a significant bias in the overall \ion{H}{1}
Ly$\alpha$ absorption. However, we stress that models constructed to compare
against the \ion{He}{2} sample should exclude/avoid gas with large \ion{H}{1}
Lyman limit opacity along the sightline.

\begin{figure}[t]
\includegraphics[width=\linewidth]{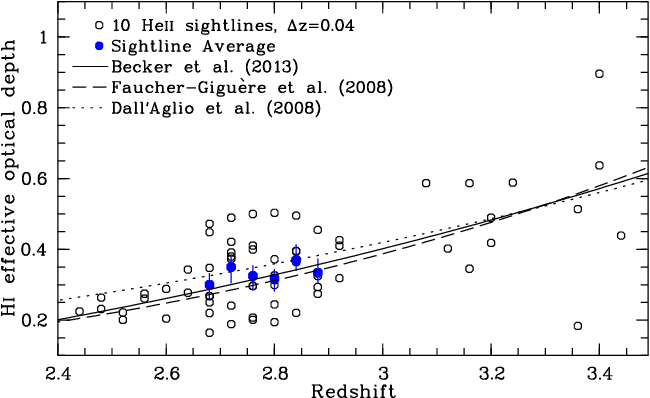}
\caption[]{\label{fig:taueffh1}
\ion{H}{1} effective optical depths $\tau_\mathrm{eff,HI}$ of the 10
\ion{He}{2}-transparent sightlines with coeval echelle \ion{H}{1} spectra,
measured in the same regular $\Delta z=0.04$ bins as the \ion{He}{2} effective
optical depths. All measured values have been continuum-corrected, but not
metal-corrected. Filled circles show the combined $\tau_\mathrm{eff,HI}$ with
$1\sigma$ errors estimated from bootstrap analysis of the contributing sightlines
($\ge 4$). The curves show $\tau_\mathrm{eff,HI}\left(z\right)$ from recent
studies based on larger samples \citep{faucher08,dallaglio08,becker13}.
} 
\end{figure}

Figure~\ref{fig:taueffh1} presents $\tau_\mathrm{eff,HI}$ measurements in
$\Delta z=0.04$ bins for our 10 \ion{He}{2} sightlines with echelle
\ion{H}{1} spectra. These are restricted to the redshifts where \ion{He}{2}
Ly$\alpha$ analysis is performed (e.g.\ avoiding geocoronal gaps and proximity
zones, see Section~\ref{sect:taueff}). The uncertainty in $\tau_\mathrm{eff,HI}$
is dominated by systematic error from continuum placement, which we estimate to
be 3\%, in agreement with published work on high-resolution samples of similar quality
\citep{kirkman05,kim07b,faucher08}. Applying this continuum correction increases
$\tau_\mathrm{eff,HI}$ by $\simeq 0.03$. Metal absorption might bias
$\tau_\mathrm{eff,HI}$ high by a few percent, but statistical corrections
\citep[e.g.][]{faucher08} are likely too high due to the avoidance of low-redshift
Lyman limit systems along our sightlines. For redshift bins with more than three
contributing sightlines we also show $\tau_\mathrm{eff,HI}$ evaluations based on
their average \ion{H}{1} transmission. Overplotted are fits to
$\tau_\mathrm{eff,HI}$ as measured in a sample of 6065 SDSS spectra
\citep{becker13} and two samples of high-resolution
spectra \citep{faucher08,dallaglio08}.
We find that the \ion{He}{2} sightline-averaged values of $\tau_\mathrm{eff,HI}$
are in very good agreement with previous IGM measurements. We may confidently
conclude that these \ion{He}{2} sightlines provide a small, but representative
sampling of the $z\sim 3$ Universe. On the other hand, significant
$\tau_\mathrm{eff,HI}$ variations occur on the $\sim 10$\,Mpc scales probed by
our measurements, as expected. The symmetric scatter of $\tau_\mathrm{eff,HI}$
around the fits suggests that metal contamination does not contribute
significantly to the sightline-to-sightline variance.

\subsection{The Coeval Effective Optical Depths of \ion{He}{2} and \ion{H}{1}}
\label{sect:h1he2tau}

\subsubsection{Measurement Technique}

There have been many attempts to compare the coeval \ion{He}{2} and \ion{H}{1}
absorption to estimate the number density ratio $n_\mathrm{HeII}/n_\mathrm{HI}$
via the ratio of the respective optical depths \citep{shull04,shull10,syphers14,mcquinn14}
or column densities \citep[e.g.][]{reimers97,heap00,kriss01,zheng04,fechner06}.
While the latter analysis should be restricted to the post-reionization IGM at
$z<2.7$ with well-resolved Ly$\alpha$ forests in both species such
that column densities can be measured, the former is
affected by the generally low \ion{He}{2} data quality and remaining
percent-level continuum uncertainties in the \ion{H}{1} data even at
exceptionally high S/N$\sim 100$ \citep{mcquinn14,syphers14}.
Physically, the affected spectral regions at $\tau_\mathrm{HI}\simeq 0.01$
correspond to the most underdense regions of the IGM
($\Delta_\mathrm{b}\simeq 0.1$; Equation~\ref{eq:fgpa}) that give rise to much
of the \ion{He}{2} opacity \citep[e.g.][]{croft97,mcquinn09b}. As one needs to model
$\tau_\mathrm{HeII}\left(n_\mathrm{HeII}/n_\mathrm{HI},\tau_\mathrm{HI}\right)$
with $\tau_\mathrm{HI}$ resolved and well-defined at every position along the
sightline, a continuum error of $\simeq 2$\% causes a factor $\sim 2$ uncertainty
in the $n_\mathrm{HeII}/n_\mathrm{HI}$ estimates from G140L data \citep{mcquinn14},
and ill-defined $n_\mathrm{HeII}/n_\mathrm{HI}$ values in G130M data wherever
$\tau_\mathrm{HI}$ is comparable to the continuum uncertainty \citep{syphers14}.

With the spectra presented in Figures~\ref{fig:h1he2spc} and \ref{fig:h1he2spc2},
we may examine the number density ratio of \ion{He}{2} to \ion{H}{1} at redshifts
$z\approx 2.5$--$3.5$. However, limited S/N for the majority of the COS data and
\ion{H}{1} continuum uncertainty precludes forward-modeling of the resolved
\ion{H}{1} data to directly estimate the underlying optical depth ratio in each
COS pixel, as was done in \citet{mcquinn14}.
Instead, we consider effective optical depths of \ion{H}{1} and \ion{He}{2}
evaluated in fixed regular redshift bins $\Delta z$. Selecting a value for
$\Delta z$ represents a compromise between maximizing the sensitivity to high
\ion{H}{1} optical depths (i.e.\ high-density regions), minimizing the impact of
\ion{H}{1} continuum error, minimizing the impact of photon counting noise, and the
(related) desire to minimize the frequency of limits in the \ion{He}{2}
effective optical depths. In these respects, the choice of $\Delta z$ is data
driven, not science driven.  After exploring a range of values, we identified
$\Delta z=0.01$ as a good compromise, corresponding to $\approx 2.5$\,Mpc at $z=3$.
We kept the redshift bins fixed, and excluded contaminated/biased regions
(e.g.\ proximity zones, geocoronal emission, identified low-$z$ \ion{H}{1} absorption).

The evaluation of the \ion{H}{1} effective optical depth $\tau_\mathrm{eff,HI}$
is straightforward. Because we evaluate $\tau_\mathrm{eff,HI}$ in redshift
windows that are much larger than the COS G140L line-spread function, we perform
the measurements without smoothing the high-resolution spectra. On the chosen
scale $\Delta z=0.01$ ($\approx 250$ pixels in the \ion{H}{1} spectra)
there are no examples of complete \ion{H}{1} absorption in
our spectra and the highest $\tau_\mathrm{eff,HI}$ value recorded is $1.40$.
Similarly, we measure no lower limits to $\tau_\mathrm{eff,HI}$ in these
windows. The uncertainty in $\tau_\mathrm{eff,HI}$ is dominated by systematic
error due to \ion{H}{1} continuum uncertainties around the applied mean 3\%
correction and is only of importance in the few bins where
$\tau_\mathrm{eff,HI}\ll 0.1$. The uncertainty in $\tau_\mathrm{eff,HeII}$
is dominated by statistical Poisson error.

\subsubsection{Observational Results}

\begin{figure*}[t]
\includegraphics[width=\textwidth]{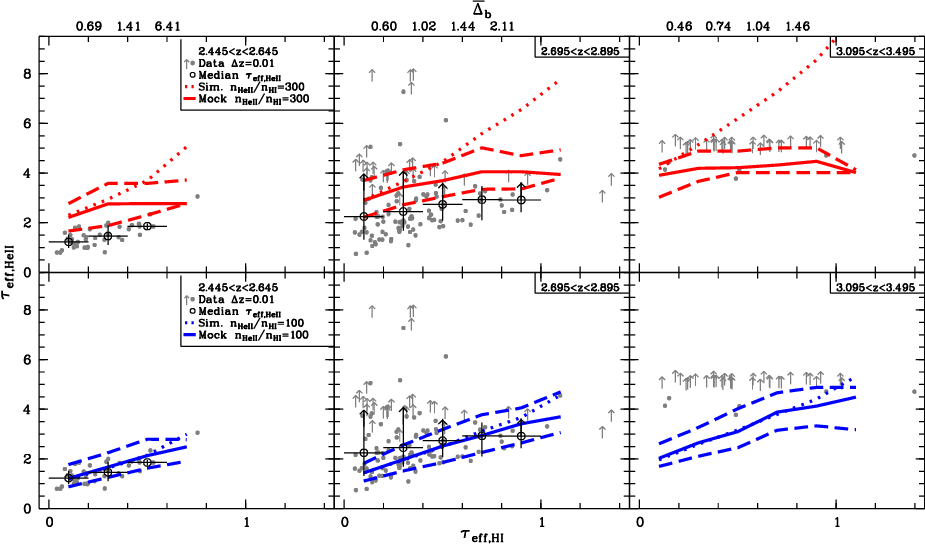}
\caption[]{\label{fig:tauh1he2}
\ion{He}{2} effective optical depth as a function of the \ion{H}{1} effective
optical depth, measured in regular $\Delta z=0.01$ bins
($\approx 2.5$\,Mpc at $z=3$) in three broad redshift intervals
(gray; \emph{left:}$2.445<z<2.645$, \emph{middle:}$2.695<z<2.895$,
\emph{right:}$3.095<z<3.495$). For clarity of presentation, error bars have been
omitted. Filled circles indicate measurements with a finite upper $1\sigma$
confidence limit on $\tau_\mathrm{eff,HeII}$ ($P<0.1587$), whereas arrows show
sensitivity limits ($P=0.1587$) determined if the upper confidence limit includes
infinite $\tau_\mathrm{eff,HeII}$ or if the flux is formally negative.
Errors in $\tau_\mathrm{eff,HI}$ are negligible for $\tau_\mathrm{eff,HI}\ga 0.1$.
We also show the median $\tau_\mathrm{eff,HeII}$ in $\Delta\tau_\mathrm{eff,HI}=0.2$
intervals (open circles with horizontal bars), and the $1\sigma$ deviation in the covered
subsample (vertical bars, shown as arrows if the 84th percentile in the
distribution contains sensitivity limits). Overplotted are results from
mock data generated from our numerical simulations (see text) for two
representative constant number density ratios $n_\mathrm{HeII}/n_\mathrm{HI}=300$
(upper panels) and $n_\mathrm{HeII}/n_\mathrm{HI}=100$ (lower panels). The dotted lines
show the median model $\tau_\mathrm{eff,HeII}$ as a function of $\tau_\mathrm{eff,HI}$
in noise-free simulations degraded to our instrumental resolution.
The solid lines show the median $\tau_\mathrm{eff,HeII}$ for noisy mock data in regular
$\Delta\tau_\mathrm{eff,HI}=0.2$ intervals, whereas the dashed lines indicate
the $1\sigma$ deviation in the distribution (16th and 84th percentile).
The scale on the top indicates the mean density contrast $\bar{\Delta}_\mathrm{b}$
in a $\Delta z=0.01$ bin that yields the corresponding $\tau_\mathrm{eff,HI}$
value on the lower axis, as determined from our cosmological simulation.
}
\end{figure*}

Figure~\ref{fig:tauh1he2} displays the measured effective optical depths, grouped
into three redshift intervals to account for IGM density evolution between
$z=2.5$ and $3.5$. Our sample of coeval
\ion{H}{1} and \ion{He}{2} absorption mostly covers $z\simeq 2.8$ (7 sightlines),
whereas only two sightlines sample $z\simeq 2.55$ and $z\sim 3.3$, respectively.
At low redshifts $z\simeq 2.55$ the effective optical depths of both species are
correlated, but their ratio does not yield a direct estimate of
$n_\mathrm{HeII}/n_\mathrm{HI}$ since $\tau_\mathrm{eff}\ne \tau$
(Equation~\ref{eq:he2h1ratio}). To better describe this correlation we computed
the median $\tau_\mathrm{eff,HeII}$ and the $1\sigma$ deviation (16th and 84th percentile)
of the distribution about the median for $\Delta\tau_\mathrm{eff,HI}=0.2$ intervals
with sufficient data. Thus, the $1\sigma$ deviation plotted in Fig.~\ref{fig:tauh1he2}
is not the error on the median, but the variance of the process about the median
estimated from the data. As all low-$z$ $\tau_\mathrm{eff,HeII}$ values are well
defined (no limits), the accuracy of these estimates is limited by sample size,
especially at high $\tau_\mathrm{eff,HI}$. We find that the median
$\tau_\mathrm{eff,HeII}$ steadily increases. The scatter among the values might
decrease with $\tau_\mathrm{eff,HI}$, but the limited sample size at high
$\tau_\mathrm{eff,HI}$ does not allow for definite conclusions.

As shown in the middle panels of Fig.~\ref{fig:tauh1he2}, this correlation starts
to disappear at higher redshifts. At $z\simeq 2.8$ there is a locus of well
determined $\tau_\mathrm{eff,HeII}$ values that increase with increasing
$\tau_\mathrm{eff,HI}$, but $>11$\% of the data clearly separate from this
locus (either measured or sensitivity limit $\tau_\mathrm{eff,HeII}>4$).
This fraction might be as high as 29\% if all sensitivity limits intrinsically have
$\tau_\mathrm{eff,HeII}>4$.
The median $\tau_\mathrm{eff,HeII}$ is still well defined, as more than 50\% of
the values in each $\Delta\tau_\mathrm{eff,HI}=0.2$ bin are finite.
However, the $1\sigma$ deviation is not, as the 84th percentiles of the
distributions include lower limits in four out of five $\tau_\mathrm{eff,HI}$ bins.
We include the lower limits in the estimation of the variance about the median,
such that the variance in these four $\tau_\mathrm{eff,HI}$ bins is in fact a
lower limit (denoted as an arrow in Fig.~\ref{fig:tauh1he2}).
The scatter might decrease with increasing $\tau_\mathrm{eff,HI}$, but
\ion{He}{2} saturation and the small sample size at $\tau_\mathrm{eff,HI}>0.4$
preclude firm conclusions.
However, the dispersion in the data is definitely larger than at lower
redshifts, indicating a change in the IGM between $z=2.5$ and $z=2.8$.
In Section~\ref{sect:taueff} we argued from the decreasing scatter in the
\ion{He}{2} effective optical depth that \ion{He}{2} reionization ended at
$z\simeq 2.7$ \citep[see also][]{shull10,furlanetto10,worseck11b}.
Here we show that the scatter in $\tau_\mathrm{eff,HeII}$ at $z\simeq 2.8$ is
not driven by IGM density variations, as the relative frequency of high
$\tau_\mathrm{eff,HeII}$ values increases toward low $\tau_\mathrm{eff,HI}$.
Moreover, the scatter in the $\tau_\mathrm{eff,HeII}$ distributions at
$\tau_\mathrm{eff,HI}<0.4$ increases significantly from $z\simeq 2.55$ to
$z\simeq 2.8$. At $z\simeq 2.55$ the scatter is small without any limit,
but at $z\simeq 2.8$ it is much larger and in fact a lower limit due to the
frequent high $\tau_\mathrm{eff,HeII}$ values. This increasing scatter probably
indicates that \ion{He}{2} reionization is ongoing.

At $z>3$, limited sample size and depth of the \ion{He}{2} spectra do not allow
for a density-dependent analysis on small scales. The locus of regions with low
\ion{He}{2} opacity seems to have disappeared between $z\simeq 2.8$ and
$z\simeq 3.3$, and in the following we shall quantify how much of that is driven
by IGM density evolution.

\subsubsection{Realistic Mock Spectra from Cosmological Simulations}

To further interpret these observations we created realistic mock spectra from
cosmological simulations of a fully reionized IGM and a uniform number density
ratio $n_\mathrm{HeII}/n_\mathrm{HI}$ (equivalent to a uniform radiation field
with a constant ratio of photoionization rates; Equation~\ref{eq:he2h1ot}).
We derived models for hydrogen in the IGM using a suite of cosmological
simulations. We used a $2\times 512^3$ particle, 25 Mpc$/h$
smooth particle hydrodynamics simulation using the Gadget-3 code
\citep{springel05}, run with the \citet{faucher09} UV background model and the
full atomic chemistry for gas of primordial composition.
To speed up the calculation, gas particles with $1000$ times the cosmic mean
density are efficiently turned into collisionless stellar particles.
This simulation was initialized at $z=100$ with second order Lagrangian
perturbation theory initial conditions for a flat $\Lambda$CDM cosmology with
$h=0.7$, $\Omega_\mathrm{m}=0.27$, $\Omega_\mathrm{b}=0.046$, $\sigma_8=0.8$,
and $n_s=0.96$. Mock \ion{H}{1} Ly$\alpha$ spectra were generated by randomly
tracing skewers across the simulation volume. These skewers were renormalized
so that the average transmission in 2000, $25/h$~comoving Mpc skewers matches
the mean flux measurement of \citet{faucher08}. Spectra were generated from four
snapshots sampling the redshift windows used in the above analysis
($z=2.55$, $2.75$, $2.85$, $3.30$).

We then generated simulated \ion{He}{2} Ly$\alpha$ absorption spectra by assuming
a constant number density ratio $n_\mathrm{HeII}/n_\mathrm{HI}=100$.
The spectra include the effects of thermal broadening, inducing a scatter of our
estimator $4\tau_\mathrm{HeII}/\tau_\mathrm{HI}$ (Equation~\ref{eq:he2h1ratio})
around the prescribed value of $n_\mathrm{HeII}/n_\mathrm{HI}=100$.
With the simulations we confirm that thermal broadening marginally affects the estimated
$n_\mathrm{HeII}/n_\mathrm{HI}$ since the distribution is very peaked.
The 16th, 50th and 84th percentile of the distribution are 75, 97, and 106, respectively.
Other values of $n_\mathrm{HeII}/n_\mathrm{HI}$ were modeled by rescaling
$\tau_\mathrm{HeII}$ for simplicity (Equation~\ref{eq:he2h1ratio}), i.e.\
not accounting for the different thermal broadening at different $\Gamma_\mathrm{HeII}$.
This is a reasonable approximation, since we are interested in the \ion{He}{2}
effective optical depth on a scale $\Delta z=0.01$ rather than the resolved
$\tau_\mathrm{HeII}$.
The \ion{He}{2} spectra were convolved with the COS line spread function and
rebinned to the pixel size of our data. With the parameters from the observations
(exposure time, continuum, background) we then simulated Poisson-distributed
counts, yielding realistic mock COS \ion{He}{2} spectra for each observed
sightline. The corresponding mock \ion{H}{1} Ly$\alpha$ spectra have
$R=40,000$ and S/N$=20$. The specific value of the \ion{H}{1} S/N is not very
important considering that our redshift bins contain $\approx 250$ pixels each.
Lastly, we evaluated the coeval
\ion{H}{1} and \ion{He}{2} effective optical depths in redshift bins
$\Delta z=0.01$ as for the data, using the same analysis software.

The adopted scale $\Delta z=0.01$ implies a minimum length scale over which we
probe fluctuations in the UV radiation field ($\approx 2.5$ proper Mpc at $z=3$).
Even after the completion of \ion{He}{2} reionization large-amplitude fluctuations
in $n_\mathrm{HeII}/n_\mathrm{HI}$ may occur on smaller scales due to the small
space density of quasars \citep{fardal98,bolton06,furlanetto09,furlanetto10,davies14}
with varying quasar spectral energy distributions \citep{telfer02,stevans14,tilton16},
in addition to radiative transfer in the IGM
\citep{kriss01,shull04,shull10,zheng04,bolton06,fechner07,syphers13,mcquinn14}.
Given the low spectral resolution and quality of our \ion{He}{2} spectra, we do not
model the UV radiation field from discrete quasars, but rather assume a homogeneous
UV background over $\Delta z=0.01$, which is somewhat larger than the length scale
of UV background fluctuations in the post-reionization IGM \citep{mcquinn14}.

\subsubsection{Comparison to Models}

The upper and lower panels of Fig.~\ref{fig:tauh1he2} show simulation results
for two representative constant values of the $n_\mathrm{HeII}/n_\mathrm{HI}$
number density ratio of 100 and 300, corresponding to a hard and a soft UV
background in a highly ionized IGM, respectively (Equation~\ref{eq:he2h1ot}).
For each of these, we created 100 mock realizations of our dataset with redshift
coverage of a given simulation snapshot. To account for the small effect of
redshift evolution between $z=2.7$ and $z=2.9$ where our observed sample is
largest, we merged the mock spectra from the snapshots at $z=2.75$ and $z=2.85$.

As for the observed data, we computed the median $\tau_\mathrm{eff,HeII}$ and
its scatter (16th and 84th percentile) in $\Delta\tau_\mathrm{eff,HI}=0.2$
intervals. Due to the inclusion of sensitivity limits the mock median \ion{He}{2}
effective optical depths (solid lines in Fig.~\ref{fig:tauh1he2})
are lower than the ones computed from noise-free spectra (dotted lines),
mostly at high $\tau_\mathrm{eff,HI}$ and high $n_\mathrm{HeII}/n_\mathrm{HI}$.
The mock $1\sigma$ scatter (dashed lines) flattens similarly, and turns into a
lower limit at high $\tau_\mathrm{eff,HI}$ and high $n_\mathrm{HeII}/n_\mathrm{HI}$.
At constant $n_\mathrm{HeII}/n_\mathrm{HI}$, $\tau_\mathrm{eff,HeII}$
increases with $\tau_\mathrm{eff,HI}$ as expected, until saturation in the
\ion{He}{2} spectra causes the relation to flatten, preventing constraints on
the underlying $n_\mathrm{HeII}/n_\mathrm{HI}$ at large $\tau_\mathrm{eff,HI}$.
For $n_\mathrm{HeII}/n_\mathrm{HI}=100$ the agreement between the median values of
$\tau_\mathrm{eff,HeII}$ in the simulations with and without noise shows that
our data are sensitive to $n_\mathrm{HeII}/n_\mathrm{HI}\sim 100$ at all redshifts
at $\tau_\mathrm{eff,HI}<0.8$.
The top abscissa of Fig.~\ref{fig:tauh1he2} shows the mean density contrast
encountered along skewers with a specific $\tau_\mathrm{eff,HI}$ evaluated in
$\Delta z=0.01$ redshift bins (bottom abscissa of Fig.~\ref{fig:tauh1he2}).
Thus, our numerical simulations indicate that our $\Delta z=0.01$ redshift bins
($\approx 2.5$\,Mpc at $z=3$) retain some sensitivity to the typical gas
density, although a range of densities will be encountered.

At $z\simeq 2.55$ the data are consistent with a slightly lower value than
our chosen $n_\mathrm{HeII}/n_\mathrm{HI}=100$, as indicated by
the lower observed median $\tau_\mathrm{eff,HeII}$.
Values of $n_\mathrm{HeII}/n_\mathrm{HI}\gg 100$ are ruled out.
Our results are in good agreement with a refined analysis of the same data
(HE~2347$-$4342 and HS~1700$+$6416) to estimate $n_\mathrm{HeII}/n_\mathrm{HI}$ for
individual COS pixels via a different technique (forward-modeling of the \ion{He}{2} spectrum),
but with necessarily larger systematic uncertainties \citep{mcquinn14}.

At higher redshifts $z\simeq 2.8$, the data are highly inconsistent with a
constant number density ratio $n_\mathrm{HeII}/n_\mathrm{HI}$. Mock spectra with
$n_\mathrm{HeII}/n_\mathrm{HI}=100$ result in a tight relation between the
effective optical depths, characterized by a small scatter and sufficient
sensitivity in high-density regions. At a given $\tau_\mathrm{eff,HI}$ the
locus of low $\tau_\mathrm{eff,HeII}$ values is well described by
$n_\mathrm{HeII}/n_\mathrm{HI}\approx 100$, but $\ga 10$\% of the
$\tau_\mathrm{eff,HeII}$ values are much higher than predicted by the model,
indicating a higher $n_\mathrm{HeII}/n_\mathrm{HI}$ in a subset of the data.
On the other hand, $n_\mathrm{HeII}/n_\mathrm{HI}=300$
does not fit the locus of low $\tau_\mathrm{eff,HeII}$ values, visible as a
strong mismatch between the observed and the simulated median
$\tau_\mathrm{eff,HeII}$. The scatter in $\tau_\mathrm{eff,HeII}$ at a given
$\tau_\mathrm{eff,HI}$ estimated from the data is much larger than predicted
by any constant $n_\mathrm{HeII}/n_\mathrm{HI}$ model, in particular for the well
sampled low-density regions in the IGM ($\tau_\mathrm{eff,HI}<0.4$ corresponding
to $\bar{\Delta}_\mathrm{b}<1$). Our simple modeling indicates that in underdense
regions the number density ratio $n_\mathrm{HeII}/n_\mathrm{HI}$ varies from
$\sim 100$ to $>300$, giving rise to the patchwork of weak and strong \ion{He}{2}
absorption at similar levels of \ion{H}{1} absorption (Fig.~\ref{fig:h1he2spc}).

At $z>3$ saturation in the \ion{He}{2} spectra generally sets a lower limit
$n_\mathrm{HeII}/n_\mathrm{HI}\ga 300$ at all densities.
Still, the two bins at $z=3.36$ and $z=3.39$ toward SDSS~J1711$+$6052 are
consistent with $100<n_\mathrm{HeII}/n_\mathrm{HI}<300$ (Fig.~\ref{fig:h1he2spc2}).
This shows that there are isolated ionized patches at $z\simeq 3.4$,
but constraints on their abundance will require high-quality echelle data of the
remaining high-$z$ \ion{He}{2} sightlines.

It is evident from Fig.~\ref{fig:tauh1he2} that even in low-density regions at
$z\simeq 2.8$ the variance in $\tau_\mathrm{eff,HeII}$ is difficult to estimate
due to limited sensitivity to high $\tau_\mathrm{eff,HeII}$ values.
Our technique of adopting the limits as values allows to calculate a well-defined
quantity, but this nevertheless results in lower limits on the true variance of
the process. However, the median $\tau_\mathrm{eff,HeII}$ steadily increases with
$\tau_\mathrm{eff,HI}$ even for the small subsamples covering high-density
regions (6--10 individual measurements). Therefore, we also estimated the
$1\sigma$ statistical error on the median $\tau_\mathrm{eff,HeII}$ in the
defined $\Delta\tau_\mathrm{eff,HI}=0.2$ intervals by bootstrapping the
subsamples, noting that convergence is rather poor due to the limited sample
size, except for $\tau_\mathrm{eff,HI}<0.4$ at $z\simeq 2.8$
(40 and 50 measurements in the two $\Delta\tau_\mathrm{eff,HI}=0.2$ intervals).
We also computed -- by virtue of the large number of simulated skewers
-- converged bootstrap errors on the median $\tau_\mathrm{eff,HeII}$ in our mock
spectra, considering the different subsample sizes in the
$\Delta\tau_\mathrm{eff,HI}=0.2$ intervals. A significantly larger bootstrap
error on the mock median $\tau_\mathrm{eff,HeII}$ than on the observed one
indicates that the bootstrap error in the observations has been underestimated
due to insufficient sample size.

\begin{figure}[t]
\includegraphics[width=\linewidth]{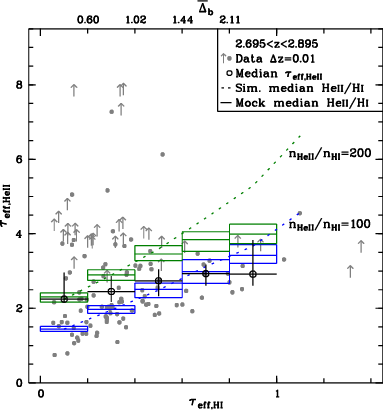}
\caption[]{\label{fig:tauh1he2med}
Similar to Fig.~\ref{fig:tauh1he2}, but showing only the $z\simeq 2.8$ data and
showing the $1\sigma$ error on the median as estimated from a bootstrap instead
of the $1\sigma$ deviation in the subsample per $\Delta\tau_\mathrm{HI}=0.2$
interval. We consider two models with constant number density ratios
$n_\mathrm{HeII}/n_\mathrm{HI}=100$ (blue) and 200 (green).
The boxes indicate the $1\sigma$ error of the median $\tau_\mathrm{eff,HeII}$
(vertical line in the box) in the mock spectra, accounting for \ion{He}{2}
data quality and size of the observed subsample per $\Delta\tau_\mathrm{HI}=0.2$
interval.  The dotted lines show the median model $\tau_\mathrm{eff,HeII}$ as a
function of $\tau_\mathrm{eff,HI}$ in noise-free simulations degraded to our
instrumental resolution.
}
\end{figure}

In Figure~\ref{fig:tauh1he2med} we compare our $z\simeq 2.8$ sample
to two models with constant $n_\mathrm{HeII}/n_\mathrm{HI}= 100$ and 200, respectively.
The model with $n_\mathrm{HeII}/n_\mathrm{HI}= 300$ has been omitted, as it poorly
describes the data (Fig.~\ref{fig:tauh1he2}). We see clear evidence
that the observed median $\tau_\mathrm{eff,HeII}$ has a shallower dependence on
$\tau_\mathrm{eff,HI}$ than predicted by the models. At the lowest densities
$\bar{\Delta}_\mathrm{b}\la 0.6$ $n_\mathrm{HeII}/n_\mathrm{HI}\approx 200$ is required,
but with large $n_\mathrm{HeII}/n_\mathrm{HI}$ variations to reproduce the
variance in the observations.
In the next interval $0.2\le \tau_\mathrm{eff,HI}<0.4$, corresponding to mildly
underdense regions $0.6\la\bar{\Delta}_\mathrm{b}\la 1.0$, the median
$\tau_\mathrm{eff,HeII}$ is between the two considered models,
neither of which can explain the large variance in the observed sample (Fig.~\ref{fig:tauh1he2}).
We conclude that the number density ratio $n_\mathrm{HeII}/n_\mathrm{HI}$
needs to vary to explain the data, but around a lower value than in the most
underdense regions.

The trend of a lower $n_\mathrm{HeII}/n_\mathrm{HI}$ required to match the observed
median $\tau_\mathrm{eff,HeII}$ continues to higher $\tau_\mathrm{eff,HI}$,
although the two bins at $\tau_\mathrm{eff,HI}>0.6$ are affected by \ion{He}{2}
saturation and small sample size. Overall, the flatter relation between the
observed and simulated effective optical depths toward 7 sightlines argues for
an anticorrelation of $n_\mathrm{HeII}/n_\mathrm{HI}$ and overdensity at $z\simeq 2.8$,
confirming earlier indications from a single \ion{He}{2} sightline \citep{shull04},
the \textit{FUSE} data of which were likely affected by systematic errors \citep{fechner07,mcquinn14}.
Since for an almost fully reionized IGM a lower $n_\mathrm{HeII}/n_\mathrm{HI}$
corresponds to a higher \ion{He}{2} photoionization rate (Equation~\ref{eq:he2h1ot}),
an anticorrelation of $n_\mathrm{HeII}/n_\mathrm{HI}$ and overdensity
implies a higher \ion{He}{2} photoionization rate in overdense regions which may be
more strongly correlated with locations of quasars than the underdense IGM
\citep{shull04,bolton06}. Alternatively, the high $\tau_\mathrm{eff,HeII}$ values
at low $\tau_\mathrm{eff,HI}$ might indicate regions that are not yet in
photoionization equilibrium, since it takes longer to reach the equilibrium
\ion{He}{2} fraction in underdense regions.

Statistics on the individual measurements provide additional evidence for this
anticorrelation between $n_\mathrm{HeII}/n_\mathrm{HI}$ and overdensity.
First, the fraction of lower limits decreases from $27.5$\% at $\tau_\mathrm{eff,HI}\simeq 0.1$
to $17.4$\% at the still well-sampled $\tau_\mathrm{eff,HI}\simeq 0.5$,
contrary to the expected higher fraction of lower limits at high
$\tau_\mathrm{eff,HI}$ in any model with a constant $n_\mathrm{HeII}/n_\mathrm{HI}$.
Second, the fraction of very high \ion{He}{2} effective optical depths
$\tau_\mathrm{eff,HeII}>4$ (measurements or limits) drops from $\simeq 14$\% at
$\tau_\mathrm{eff,HI}<0.4$ to $\simeq 5$\% at higher $\tau_\mathrm{eff,HI}$,
contrary to the expected steep evolution in $\tau_\mathrm{eff,HeII}$ with
$\tau_\mathrm{eff,HI}$ if $n_\mathrm{HeII}/n_\mathrm{HI}$ were constant.
Moreover, we verified that the anticorrelation is not due to a few peculiar
sightlines in our sample.

In summary, our data indicate that the two main assumptions of the above
numerical modeling -- a fully reionized IGM in photoionization equilibrium with
a uniform UV radiation field -- do not hold at $z\simeq 2.8$.
The mismatch between the data and the models indicates that while helium is
mostly reionized by $z\simeq 2.8$
(i.e.\ the locus of IGM patches consistent with $n_\mathrm{HeII}/n_\mathrm{HI}\simeq 100$
corresponding to $\bar{x}_\mathrm{HeII}\simeq 0.003$ at mean density),
there may be significant fluctuations in the \ion{He}{2}
photoionization rate (Figs.~\ref{fig:tauh1he2} and \ref{fig:tauh1he2med},
Equation~\ref{eq:he2h1ot} with $\Gamma_\mathrm{HI}\simeq\mathrm{const.}$),
at least in underdense regions where we can probe such large variations.
The large variations in $n_\mathrm{HeII}/n_\mathrm{HI}$ would directly
correspond to variations in the \ion{He}{2} fraction by a factor $\sim 4$
(Equation~\ref{eq:he2frac}). Thus, we may observe the end phase of \ion{He}{2}
reionization in the underdense regions of the IGM
(large $\tau_\mathrm{eff,HeII}$ variations), whereas the higher-density regions
were already ionized earlier (small $\tau_\mathrm{eff,HeII}$ variations).
After the end of \ion{He}{2} reionization, the \ion{He}{2} photoionization rate
may vary due to the rarity of quasars
\citep{fardal98,bolton06,furlanetto09,furlanetto10,davies14}, but intersected
quasar proximity zones will rarely produce upward fluctuations by a factor $>2$
\citep{furlanetto10,mcquinn14}.
While our current observations at $z\simeq 2.55$ favor the presence of such a
quasi-homogeneous UV radiation field, large \emph{downward} fluctuations of the
UV radiation field may occur in 10--20\% of all $\simeq 2.8$\,Mpc patches of the
underdense IGM at $z\simeq 2.8$, in order to explain the large number of lower
limits that we observe at $\tau_\mathrm{eff,HI}<0.4$ ($\bar{\Delta}_\mathrm{b}<1$).
At face value, this might suggest an `inside-out' evolution to \ion{He}{2} reionization,
in which underdense regions far from the sources (quasars) are reionized last.

Similarly, if these low-density regions are not yet in photoionization equilibrium
at $z\simeq 2.8$, one concludes that \ion{He}{2} reionization is ongoing.
\citet{furlanetto10} used Monte Carlo simulations to argue for the presence of
incompletely ionized IGM patches far from ionizing sources to explain the large
variance in $\tau_\mathrm{eff,HeII}$ at $z\sim 2.8$ seen in the handful of previously
analyzed sightlines, but their calculations did not include the density field or
a realistic source distribution. While current numerical simulations naturally
produce a strong evolution in $n_\mathrm{HeII}/n_\mathrm{HI}$ during
\ion{He}{2} reionization \citep{mcquinn09a,tittley12,compostella13,compostella14},
the emergence of saturated \ion{He}{2} absorption in underdense regions
-- the clearest observational signature for incomplete \ion{He}{2} reionization --
has received little attention thus far.

%\vspace*{3ex}

\section{Summary and Conclusions}
\label{sect:concl}

We have presented the first results from a systematic survey to probe the timing
and morphology of \ion{He}{2} reionization in the IGM, dubbed the Helium
Reionization Survey (HERS). The core of its first public data
release\footref{note:hers} is a sample of 17 homogeneously reduced
%\footnotemark[\ref{note:hers}] is a sample of 17 homogeneously reduced
high-quality \textit{HST} far-UV quasar spectra probing intergalactic \ion{He}{2}
Ly$\alpha$ absorption at $2.3<z<3.5$, collected in various programs before \textit{HST}
Cycle~20. For a subset of 10 \ion{He}{2} sightlines we have analyzed
complementary Keck/HIRES and VLT/UVES spectra of the coeval \ion{H}{1}
Ly$\alpha$ forest. Our results and their implications can be summarized as follows:
\begin{enumerate}
\item The \ion{He}{2} effective optical depth increases with redshift between
$z=2.3$ ($\tau_\mathrm{eff,HeII}\simeq 1$) and $z=3.4$
($\tau_\mathrm{eff,HeII}\ga 4.5$), but with significant sightline-to-sightline
variance at $z>2.7$ (Fig.~\ref{fig:he2tau}). On the adopted scale
$\Delta z=0.04$ ($\simeq 10$ proper Mpc at $z=3$) many sightlines reveal
surprisingly low \ion{He}{2} absorption ($\tau_\mathrm{eff,HeII}\simeq 3$--4) out
to $z=3.5$. We have carefully recalibrated the COS background including
a correction for scattered light, and a battery of tests has been
conducted to ensure its robustness that is critical to our analysis
(Appendix~\ref{sect:backgroundappendix}).

\item Over the full covered redshift range the locus of low
$\tau_\mathrm{eff,HeII}$ values is in very good agreement with a semianalytic
model of a fully reionized IGM (Fig.~\ref{fig:he2tau}). Much of the very gradual
redshift evolution is consistent with density evolution in a predominantly fully
ionized IGM with a mean \ion{He}{2} fraction $\bar{x}_\mathrm{HeII}\simeq 0.003$
in highly ionized regions.
The diminishing variance in $\tau_\mathrm{eff,HeII}$ at $z\simeq 2.7$ supports
the end of \ion{He}{2} reionization at that epoch.
However, the variance in $\tau_\mathrm{eff,HeII}$ on $\simeq 10$\,Mpc scales at
$2.8\la z<3.5$ implies that the \ion{He}{2} fraction
(\ion{He}{2} photoionization rate) is a factor $\sim 4$ higher (lower)
for 10--20\% of the probed pathlength, indicating that our observations
probe the tail end of \ion{He}{2} reionization.
Likewise, current numerical radiative transfer simulations of rapid
quasar-driven \ion{He}{2} reionization fail to reproduce the observed
$\tau_\mathrm{eff,HeII}$ distribution either at $z\simeq 3.4$
\citep{mcquinn09a,compostella13} or at $z\simeq 2.8$ \citep{compostella14},
pointing to a very extended epoch of \ion{He}{2} reionization.
Our observations probe the last 600\,Myr of \ion{He}{2} reionization
($2.7<z<3.5$) that must have begun at $z>4$ to result in the observed
$\tau_\mathrm{eff,HeII}$ distribution at $z\simeq 3.4$ \citep{compostella14}.

\item By measuring the coeval \ion{H}{1} and \ion{He}{2} effective optical depths
on small scales ($\Delta z=0.01$ corresponding to $\simeq 2.5$\,Mpc at $z=3$)
and by comparing them to realistic mock spectra from a hydrodynamical simulation
of the optically thin post-reionization IGM, we disentangle fluctuations in the
\ion{He}{2} fraction from those in the density field
(Figs.~\ref{fig:tauh1he2} \& \ref{fig:tauh1he2med}).
At $z\simeq 2.5$ the effective optical depths of both species are highly
correlated, as expected for an optically thin IGM with a quasi-homogeneous
UV radiation field \citep{mcquinn14}.
At $z\simeq 2.8$ $\tau_\mathrm{eff,HeII}$ strongly varies at
$\tau_\mathrm{eff,HI}<0.4$ corresponding to underdense regions in the IGM.
This may be explained by a \ion{He}{2} photoionization rate that is
anticorrelated with density, perhaps indicating that \ion{He}{2} reionization
proceeded in an inside-out fashion around sources embedded in overdense regions.
Alternatively, these underdensities might be not yet be in photoionization
equilibrium due to recent reionization.

\end{enumerate}

At face value, the mild redshift evolution of the \ion{He}{2} absorption demands
that \ion{He}{2} reionization proceeded very gradually between $z\simeq 3.4$ and
$z\simeq 2.7$, and that the bulk of the intergalactic helium was ionized at $z>4$.
It remains to be tested what sources could have powered such an early reionization.

In current models of quasar-driven \ion{He}{2} reionization
\citep{mcquinn09a,compostella13,compostella14} a rapidly evolving quasar
luminosity density translates into a steeply evolving \ion{He}{2} absorption
that fails to reproduce our data at $z=2.8$ and/or $z=3.4$. Our observations
call for refined numerical models that achieve a slowly evolving
\ion{He}{2} fraction, with $\bar{x}_\mathrm{HeII}<0.01$ in $\sim 50$\%
($\sim 90$\%) of the $z\simeq 3.4$ ($z\simeq 2.8$) IGM.
Similarly, the highest \ion{He}{2} effective optical depths at $z\simeq 2.8$ are
inconsistent with early \ion{He}{2} reionization by numerous faint quasars at
$z>4$ \citep{madau15}, unless the post-reionization UV background is strongly
fluctuating on large scales \citep{furlanetto10,mcquinn14}.
We note that much of the difference in current model predictions for the timing
of \ion{He}{2} reionization is due to the uncertain faint end of the $z\ga 4$
quasar luminosity function \citep{glikman11,masters12,mcgreer13,giallongo15}
that leads to a factor $\sim 6$ spread in the quasar ionizing emissivity for
a fixed spectral energy distribution \citep{madau15}.
Since faint ($M_{1450}>-24$) quasars can be verified with current instruments
even at $z\sim 6$ \citep{kashikawa15,kim15}, a systematic survey for faint $z>4$
quasars may test early helium reionization scenarios.

An explanation of the slowly evolving \ion{He}{2} effective optical depth may
require additional, more exotic sources of hard photons at high redshift,
such as Bremsstrahlung from gas shock heated by cosmic structure formation
\citep{miniati04} or X-ray emission from stellar binaries \citep{power09}
or black holes in high-redshift galaxies \citep{ricotti04}.
Likewise, current theoretical models do not readily produce a density-dependent
\ion{He}{2} photoionization rate in a predominantly ionized IGM.

In the present study we have challenged current reionization scenarios with two
sensitive tests: (1) the frequency of regions with low \ion{He}{2} opacity,
and (2) the fraction of underdense regions with significant \ion{He}{2} opacity.
With refined models and a legacy sample of $z>3$ \ion{He}{2} sightlines
collected with \textit{HST}/COS in its remaining lifetime, we may be able to resolve the
physics of \ion{He}{2} reionization.

\acknowledgments
We thank Cora Fechner for providing us with the \textit{FUSE} spectra of
HS~1700$+$6416 and HE~2347$-$4342 and the combined Keck/HIRES spectrum of HS~1700$+$6416.
The UVES spectra were kindly reduced by Aldo Dall'Aglio.
We also thank John O'Meara for standing in for us for the Keck/HIRES observations
of SDSS~J1711$+$6052 and SDSS~J2346$-$0016.\\
We acknowledge support by an NSF CAREER grant (AST-0548180) and by NSF grants
AST-0908910, AST-1109447 and AST-1412981.
Support for Program GO\,11742 was provided by NASA through a grant from
the Space Telescope Science Institute, which is operated by the Association of
Universities for Research in Astronomy, Inc., under NASA contract NAS5-26555.
G.W. has been partially supported by the Deutsches Zentrum f\"ur Luft- und
Raumfahrt (DLR) under contrsct 50\,OR\,1317.
J.F.H. acknowledges generous support from the Alexander von Humboldt foundation
in the context of the Sofja Kovalevskaja Award. The Humboldt foundation is
funded by the German Federal Ministry for Education and Research.
M.M. acknowledges support from NASA through a grant from the Space Telescope
Science Institute (HST-AR-13903). 

{{\it Facilities:} \facility{\textit{GALEX}}, \facility{\textit{HST} (STIS, COS)}, \facility{Keck:I (HIRES)}, \facility{VLT:Kueyen (UVES)}}

\bibliographystyle{apj}
\bibliography{he2cos2}
\appendix

%\begin{appendix}

\section{Notes on individual objects}
\label{sect:noteappendix}

\subsection{Extreme UV Continua}

\begin{figure*}[t]
\includegraphics[width=\textwidth]{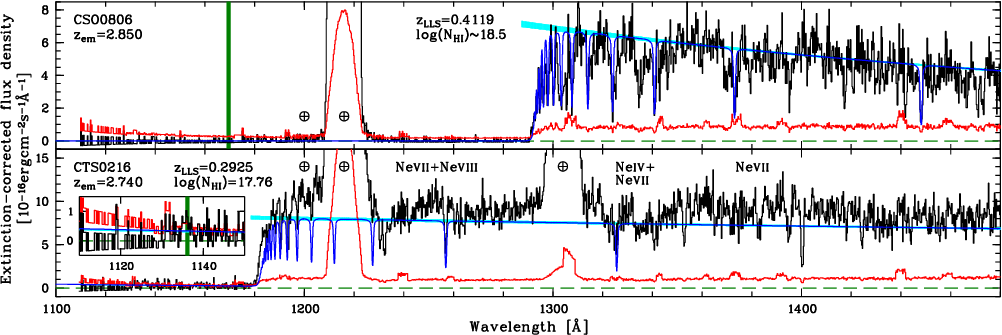}
\caption[]{\label{fig:cycle17lls}Extinction-corrected Nyquist-sampled
\textit{HST}/COS G140L spectra (black) and corresponding $1\sigma$ error arrays
(red) of the two quasars from our Cycle~17 survey unsuitable for \ion{He}{2}
absorption studies due to strong intervening \ion{H}{1} Lyman limit systems.
The green dashed lines mark the zero level, while the vertical bars indicate
\ion{He}{2} Ly$\alpha$ in the quasar rest frame. Geocoronal emission (Earth symbols)
has been reduced or eliminated with shadow data where possible (CSO~0806).
The blue lines show power-law continuum fits including the Lyman limit systems,
whose Lyman series lines have been convolved with the COS line spread function.
The cyan shaded regions show the $1\sigma$ error of the continuum fits
(power law $+$ Lyman limit break). The flux of CTS~0216 recovers blueward of a
strong partial Lyman limit system. The excess flux near the break is likely due to
geocoronal \ion{N}{1}$\,\lambda$1200\AA\ emission. We see indications for a
proximity zone and intergalactic \ion{He}{2} absorption (inset), but the low S/N
precludes a detailed analysis. In comparison to the majority of the analyzed quasars,
CTS~0216 shows strong neon emission in several transitions and ionization levels.}
\end{figure*}

In the covered wavelength range the quasar continua are well
described by power-laws. However, the slopes given in Table~\ref{tab:he2qsolist}
do not represent the intrinsic quasar continuum slopes due to unknown intervening
\ion{H}{1} Lyman continuum absorption in the near UV \citep{moller90,worseck11}.
\ion{H}{1} Lyman continuum absorption occurring in the covered spectral range
has been accounted for by searching for strong Lyman series transitions
redward of \ion{He}{2} Ly$\alpha$ in the quasar rest frame.
In particular, we searched for Lyman series transitions of absorbers whose Lyman
limit breaks might fall into the \ion{He}{2} absorption region, thereby leading
to a potential overestimate of the continuum and the \ion{He}{2} effective optical depth.
Our improved reduction revealed a previously unnoticed partial Lyman limit system
toward SDSS~J1101$+$1053 ($z=0.3177$, $N_\mathrm{HI}\sim 10^{16.5}$\,cm$^{-2}$;
see Table~\ref{tab:he2qsolist}) that results in a modest decrease in the inferred
\ion{He}{2} effective optical depths compared to the initial analysis presented
in \citet{worseck11b}. Other spectra show similar partial Lyman limit systems at
lower redshifts, such that their Lyman limit breaks occur outside the covered
spectral range. From our analysis we estimate that we can robustly detect
$N_\mathrm{HI}\ga 10^{16.5}$\,cm$^{-2}$ absorbers in S/N$=4$ COS spectra
if their Lyman limit occurs redward of \ion{He}{2} Ly$\alpha$.
For lower-redshift absorbers the sensitivity depends on the particular Lyman
series coverage, line blending, and the presence of \ion{He}{2} transmission
regions that provide upper limits to the column density.
For an unambiguous detection of strong \ion{He}{2} absorption
it is essential to `clear' the available spectral range for low-redshift
\ion{H}{1} Lyman continuum absorption, which restricts a detailed scientific
analysis to S/N$\ga 4$ COS spectra.

This search is particularly important for the high-redshift sightlines with
complete coverage of the \ion{He}{2} Ly$\alpha$ absorption.
Toward SDSS~J1711$+$6052 we find a strong system at $z=0.4370$ detected in
Ly$\alpha$ and Ly$\beta$, whose Lyman continuum potentially blacks out
residual \ion{He}{2} flux at $z<3.31$. Unfortunately, its column density is
not well constrained due to the lack of higher-order Lyman series transitions
covered in the quasar continuum, but the strength of Ly$\alpha$ and Ly$\beta$
suggest that the absorber is optically thick.
Therefore, we have not considered the impacted redshift range for our
measurements of the \ion{He}{2} absorption. Toward HS~0911$+$4809 we detect
an absorber at $z=0.3028$ in Ly$\alpha$ and Ly$\beta$, the column density
of which is constrained to $N_\mathrm{HI}<10^{16.8}$\,cm$^{-2}$ by a
\ion{He}{2} flux spike at 1160\,\AA. Due to the uncertainty
in the actual column density we have not included its Lyman continuum absorption
in the continuum fit (Fig.~\ref{fig:he2spc_archive}), hence \ion{He}{2} effective
optical depths at $z<2.91$ may have been overestimated by $\delta\tau_\mathrm{eff}<0.4$.

Four of the quasars shown in Figs.~\ref{fig:he2spc_cycle17} and
\ref{fig:he2spc_archive} show at least tentative evidence for extreme UV
emission lines, visible as local continuum departures that are not due to
\ion{H}{1} Lyman continuum breaks. These features agree well with extreme UV
transitions of neon and nitrogen in various ionization states that have been
detected at longer rest frame wavelengths \citep{shull12,stevans14,tilton16}.
SDSS~J0936$+$2927 shows prominent \ion{He}{2} Ly$\alpha$ emission, occurring
just outside the spectral region contaminated by geocoronal \ion{H}{1} Ly$\alpha$
and \ion{N}{1}\,$\lambda 1200$\,\AA\ emission in the COS spectrum.
The observed weakness of extreme UV emission lines may be a challenge for quasar
accretion disk models \citep[e.g.][]{syphers12}. However, we note that the line
fluxes are inevitably reduced due to intervening \ion{H}{1} Lyman continuum absorption,
making low-equivalent width features less discernable from a smooth continuum in
low-S/N spectra. Detailed constraints on quasar accretion disk models will likely
require a full reconstruction of the intrinsic quasar spectral energy distribution
with high-quality near UV and optical spectra.

\subsection{Quasars with Strong Lyman Limit Systems Redward of \ion{He}{2} Ly$\alpha$}

Two UV-bright targets from our Cycle~17 \ion{He}{2} absorption survey reveal
strong intervening low-$z$ \ion{H}{1} Lyman limit systems, rendering these
targets unusable for \ion{He}{2} absorption studies. Figure~\ref{fig:cycle17lls}
presents their COS spectra. The spectrum of CSO~0806 is truncated by a strong
Lyman limit system at $z=0.4119$, for which we infer a lower limit on the
column density of $N_\mathrm{HI}>10^{18.14}$\,cm$^{-2}$ from parametric
bootstraps including the background uncertainty. However, the high Lyman series
transitions of the system indicate that its column density cannot be much larger
than this limit
($N_\mathrm{HI}\sim 10^{18.5}$\,cm$^{-2}$ for Doppler parameters $b\simeq 30$\,km\,s$^{-1}$).
Its Lyman continuum was used for our assessment of scattered light in the COS
instrument (Appendix~\ref{sect:scatterappendix}).

The sightline to CTS~0216 has a strong Lyman limit system at $z=0.2925$ with
a column density $N_\mathrm{HI}=10^{17.76}$\,cm$^{-2}$ that is well
determined from the residual flux below the Lyman limit break.
The inset of Fig.~\ref{fig:cycle17lls} shows that there are fewer counts detected
blueward of \ion{He}{2} Ly$\alpha$ than redward of it, indicating the presence
of intergalactic \ion{He}{2} absorption. However, with this low remaining
continuum flux, CTS~0216 is too faint for \ion{He}{2} Ly$\alpha$ absorption
studies with \textit{HST}. The continuum redward of the Lyman limit break is not
well described by a simple power law, but might instead show broad emission
features of neon, similar to 5 of the 17 analyzed \ion{He}{2}-transmitting
quasars (Figs.~\ref{fig:he2spc_cycle17} and \ref{fig:he2spc_archive}).

\section{COS background components}
\label{sect:backgroundappendix}

\subsection{Detector Dark Current}
\label{sect:darkappendix}

The COS FUV detector is a two-segment windowless photon-counting device,
each composed of an opaque CsI photocathode on a stack of 3 microchannel plates
(MCPs) and a cross-delay line anode \citep{mcphate00}. The MCP stack acts as a
bundle of photomultipliers yielding spatial and/or spectral information at a
total gain of $\sim 10^7$ secondary electrons per photoelectron.
The anode measures the location and the total charge (pulse height) of the
electron shower, and onboard electronics digitize the signal in position
($16,384\times 1,024$ pixels) and pulse height amplitude (PHA; $0\le\mathrm{PHA}\le 31$).
The PHA value non-linearly depends on the number of secondary electrons and does
not indicate the photon energy.
The pulse height distribution depends on the local count rate, such that photon events
occurring in the COS aperture tend to have a different pulse height distribution than
dark current events\footnote{The 3-layer Z-stack MCPs
developed for high-gain applications typically have distinct PHA distributions
for the signal and the dark current when operated in the limit of space charge
saturation, in which the electrostatic repulsion of secondary electrons limit
the total gain \citep[e.g.][]{siegmund88}. Space charge saturation results in
a quasi-Gaussian peak in the signal PHA distribution, adjustable via the applied
MCP voltage. On the ground, radioactive decay within the MCP glass and cosmic
rays typically result in a negative exponential dark PHA distribution,
as expected for background events uniformly created throughout the MCP stack
\citep{siegmund88,siegmund89}.}.
However, as these distributions still overlap with each other and depend on time
in several ways, it is necessary to subtract the appropriate COS dark current.

Due to its windowless design the detector is subject to the ambient conditions
at \textit{HST}'s altitude (e.g.\ thermosphere, South Atlantic Anomaly).
Thermospheric charged particles not repelled by the grid wires above the
detector contribute to the detector dark current, in addition to cosmic rays
and the radioactive inventory of the instrument \citep[e.g.][]{green12}.
Furthermore, secondary electrons may ionize neutral thermospheric gas atoms within
the MCP channels, which then hit the MCP channel walls, releasing unwanted
secondary electrons that may be recorded as a secondary low-gain pulse
(so-called ion feedback).

At a given MCP voltage, the detector dark current varies on several timescales,
most of which originate in the solar cycle. First, the cosmic ray flux is
anticorrelated with the solar cycle due to interaction with the Sun's magnetic
field \citep[e.g.][]{parker65}. Second, on much shorter timescales the cosmic
ray flux is modulated by the Earth's magnetic field traversed by \textit{HST}
on its orbit. Third, the Earth's thermospheric density correlates with solar
activity \citep[e.g.][]{qian12}, leading to a higher dark current due to ion
feedback in the windowless COS FUV MCPs. Fourth, at higher average thermospheric
temperature, ions might pass the ion repeller grid (voltage $+15$\,V) and
interact with the photocathode. In combination, these effects naturally result
in a complex time-dependent behavior of the COS detector in orbit.

Analysis of the dark current is exacerbated by the degrading detector sensitivity
\citep[the so-called gain sag,][]{sahnow11}. The secondary electron emission
coefficient of an MCP decreases with time due to elemental migration in the MCP
glass caused by electron bombardment, i.e.\ cumulative exposure at a given
detector location. Over time the electron shower generated by an incident photon
decreases in amplitude, and the recorded PHA distribution shifts to lower values,
resulting in sensitivity loss once the gain drops below the rejection threshold
of the COS electronics. COS science exposures always illuminate the same
MCP region (the so-called lifetime position). Initially, gain sag is most
prominent at detector locations where geocoronal Ly$\alpha$ emission is recorded
in the various grating and detector offset positions \citep{sahnow11}.
Continuous exposure, particularly on bright targets, eventually results in gain sag
across the entire spectral range, such that the spectra need to be moved
to a pristine location \citep{sahnow12}. 

At COS lifetime position 1, CALCOS systematically overestimates the actual dark current
in the COS aperture by up to 20\%, as the offset dark current estimation
windows had not experienced gain sag. This subtle time-dependent effect was first
noted in dark current monitoring exposures by \citet{syphers12}.
The trend may reverse at subsequent lifetime positions where initial science exposures
are taken at a pristine detector location, but the dark current estimation windows
may include the gain-sagged previous lifetime positions.
Evidently, an accurate dark subtraction of COS data requires an estimate of the
dark current in the gain-sagged COS science aperture via ancillary data from the
COS dark monitoring programs. However, the overall small dark current
($2.5$--$6\times 10^{-6}$\,counts\,s$^{-1}$\,pixel$^{-1}$ from 2009 to 2011
varying with solar activity) and infrequent dark monitoring preclude a straightforward
estimate with unbinned dark frames. \citet{syphers13} created a coadded and smoothed
master dark frame and scaled it to the dark current measured in their science data.
Still, to fully match the science observations, the dark calibration exposures must not
only match the gain sag, but also the ambient conditions.

\begin{figure}[t]
\includegraphics[width=\linewidth]{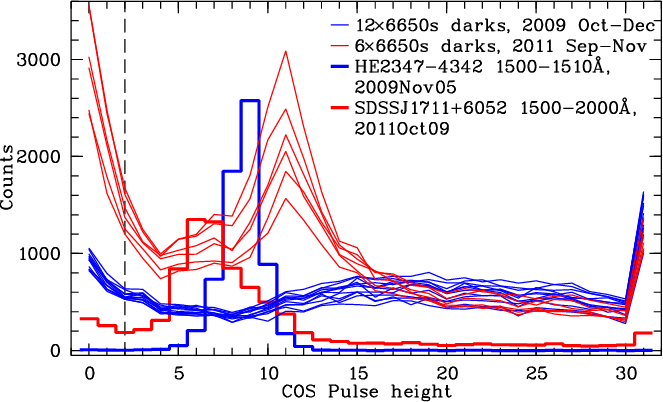}
\caption[]{\label{fig:phadist}
COS detector segment A pulse height distributions from dark and science exposures.
The thin unbinned lines show the PHA distributions of dark monitoring visits
(exposure time 6650\,s each), obtained in two 3-month intervals in 2009 (blue)
and 2011 (red). These distributions have been measured in unilluminated regions outside the COS aperture,
i.e.\ they are not affected by gain sag. The thick binned lines show representative pulse height
distributions from two science targets observed contemporarily to the dark exposures.
Wavelength ranges have been chosen to show the peaked PHA profile, i.e.\ the
number of counts has not been scaled to the darks. The offset in the science PHA
distributions indicates gain sag in the science aperture. The dashed line marks
the CALCOS default lower pulse height rejection threshold PHA$<2$ at lifetime position 1.
}
\end{figure}

To better characterize these effects we obtained all observations taken in the
COS dark current monitoring programs (\textit{HST} Programs 11895, 12423, 12716)
between October 2009 and November 2011, spanning the COS observations discussed
in this paper. During that time, COS spectra were collected at the first lifetime
position and the detector segment A voltage was kept constant, enabling a
monitoring of gain sag and the ambient conditions.
Figure~\ref{fig:phadist} presents the PHA distributions of COS darks obtained in
two 3-month intervals in 2009 and 2011. Each dark `visit' consists of five 1330\,s
exposures usually taken within a time span of several hours during Earth
occultation periods, with more frequent monitoring immediately after installation
of COS in 2009. The PHA distributions are for two spatial windows outside the COS
aperture across the length of detector segment A, and excluding grid wire shadows
and other obvious detector blemishes.
Despite reflecting the initial state of the detector, the PHA distributions
differ significantly between 2009 and 2011, and also within the considered
3-month intervals. In general, variations between visits are due to the varying
cosmic ray flux in \textit{HST}'s orbit, whereas the overall increase in the COS
dark current between 2009 and 2011 is due to increasing solar activity \citep[e.g.][]{sahnow12}.
Specifically, the relative increase of low-gain pulses is likely due to ion
feedback in the MCPs operating at higher thermospheric density, while the peak at
PHA$=11$ is likely caused by space charge saturation of the dark current.

For illustration, Fig.~\ref{fig:phadist} also shows the PHA distributions of
HE~2347$-$4342 and SDSS~J1711+6052 that were observed in the chosen time periods.
Their object flux dominates over the dark current, resulting in a distinctive
peak due to space charge saturation in the MCP channels.
With time this peak shifted to lower PHA values due to gain sag at COS lifetime
position 1.
Space charge saturation in MCPs is the basis for any pulse height screening to
lower the dark current in post-processing, i.e.\ by excluding the tails of the
pulse height distribution that are thought to contain negligible photon signal.
CALCOS employs rigid PHA thresholds across the detector that have been adjusted
multiple times to account for gain sag and the changing ambient conditions.
While these cuts reduce the dark current by 20--25\%, they cannot account for
the strong local gain sag caused by geocoronal emission lines that is
unnecessarily spread on the detector due to the four detector offset positions.
To prevent a loss of object flux near geocoronal Ly$\alpha$
(1155\,\AA$\la\lambda\la$1246\,\AA\ for coadded G140L spectra in the 1105\AA\ setup)
we chose to refrain from CALCOS pulse height screening.
Any local pulse height optimization requires a careful analysis of the detector
state and the ambient conditions, which is barely possible with the sparse
COS calibration data.

We used 3-month stacks of dark monitoring data to investigate the spatial
distribution of the dark current at different pulse heights.
Detector Segment A is much less affected by cosmetics than Segment B,
but up to now only the most obvious features at very low and very high pulse
heights have been documented \citep{mcphate10}. We find that intermediate pulse
heights $5<$PHA$<17$ show inhomogeneities across detector Segment A that
slowly increased in strength between 2009 and 2011, likely due to increasing
solar activity. The trend continued in subsequent
years\footnote{See the dark monitoring animations at \anchor{http://www.stsci.edu/hst/cos/cos_monitors/fuv_dark}{http://www.stsci.edu/hst/cos/cos\_monitors/fuv\_dark}.},
but apart from \citet{syphers12} there have been surprisingly little efforts to
characterize and calibrate these spatial variations that are crucial for science
near the sensitivity limit of COS
($f_\lambda\la 10^{-17}$\,erg\,cm$^{-2}$\,s$^{-1}$\,\AA$^{-1}$).
Our investigations suggest that the pulse height distribution is the most
sensitive indicator of the variable spatial structure of the dark current and
the general ambient conditions (Fig.~\ref{fig:phadist}).
Thus, we obtained an improved estimate of the dark current in post-processing by
matching the pulse height distributions of dark and science exposures in the
unilluminated detector area, restricting that comparison to contemporary dark
exposures to account for gain sag.

Specifically, for our science observations at COS lifetime position 1 we chose
two unilluminated regions below and above the trace (geometrically corrected
spatial coordinates $399.5<y<459.5$ and $527.5<y<587.5$ for Segment A,
$458.5<y<518.5$ and $590.5<y<650.5$ for Segment B) avoiding the zones impacted
by scattered geocoronal Ly$\alpha$ emission and detector blemishes, and compared
the cumulative pulse height distributions of the science and dark frames.
In order to minimize differential gain sag we considered only darks obtained
within a 3-month time window around the date of the science observation.
This ensured a sufficient sampling of the ambient conditions in periods of
infrequent dark monitoring, while still yielding a sufficiently accurate
estimate of the dark current in the COS science aperture
(gain sag increases on a timescale of several months).
We stacked only those dark exposures whose cumulative pulse height distributions
in the calibration windows were sufficiently similar to the one in the science
exposure (maximum absolute difference $D<0.05$). The threshold $D<0.05$ was
chosen to get the best possible match of the ambient conditions with sparsely
sampled dark calibrations. To limit Poisson noise the threshold was increased
until at least three matching dark exposures had been identified. Then the dark
current in the COS science aperture was extracted and smoothed with a 500 pixel
running average, again avoiding grid wire shadows. The 500 pixel smoothing scale
was chosen to limit Poisson noise while still tracking real dark current
variations across the detector. Finally, this modeled `master dark' was scaled
to the science exposure using the ratio between the total counts in the dark
calibration windows. The model dark current was estimated for each science
exposure and coadded while coadding exposures. Focal plane offsets were accounted
for via cross-correlation with the grid wires. Statistical errors of the scaling
factors and Poisson count errors in the smoothed darks were propagated,
yielding an estimate of the uncertainty in our final dark current model for
every science spectrum.

As the above dark current modeling procedure relies on post-processing of
independently obtained data, we performed extensive validation tests to check
its accuracy. Specifically, we considered all dark exposures taken in a 3-month
interval, and treated random subsets of these as science data while the
remaining darks served as the calibration datasets. The test exposures were
coadded like the science data using the same parameters and including focal
plane offsets to match their fixed-pattern noise properties and pixel exposure
times. Two 3-month intervals, each with a realistic science exposure time,
sampled the slowly increasing solar activity during \textit{HST} Cycle~17.
Unfortunately, the infrequent dark monitoring during phases of high solar
activity in Cycles~18 and 19 precluded validation tests tailored to the
highest-redshift sightlines that were observed for more than 20\,ks
(Table~\ref{tab:he2qsolist}). For the two chosen time intervals and exposure
times we created 100 test spectra, and measured the relative deviation between
the measured dark counts and the model dark counts in the $\Delta z=0.04$
windows considered for our $\tau_\mathrm{eff,HeII}$ measurements. 

\begin{figure*}
\includegraphics[width=\textwidth]{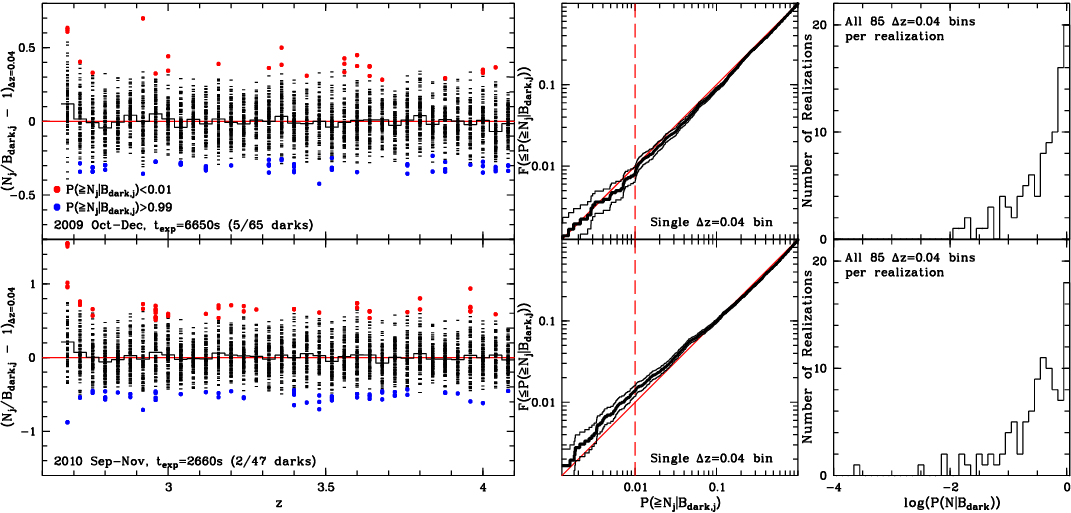}
\caption[]{\label{fig:darksubstat}
Validation tests of our dark current subtraction procedure treating subsets of
darks as data. \emph{Left:} Relative deviation between the measured dark counts
$N_j$ and the model dark counts $B_{\mathrm{dark},j}$ in $\Delta z=0.04$ bins
for 100 realizations (dashes), considering two representative exposure times of
our dataset and matching our setup ($\Delta z=0.04$ bins corresponding to 150
native G140L pixels, detector Segment A, two focal plane offset positions).
The black binned line shows the average of the 100 realizations, while the red
line marks zero deviation. Red (blue) filled circles show the most significant
positive (negative) deviations for which the Poisson probability
$P\left(\ge N_j|B_{\mathrm{dark},j}\right)$ to detect at least $N_j$ dark counts
given the mean dark current $B_{\mathrm{dark},j}$ is $<0.01$ ($>0.99$),
indicating a spurious signal (unphysical negative signal).
\emph{Middle: } Cumulative fraction of the computed Poisson probabilities per
$\Delta z=0.04$ bin (thick black lines) and its $1\sigma$ statistical error
(thin black lines) arising from the limited number of realizations.
The red solid and dashed lines mark identity and
$P\left(\ge N_j|B_{\mathrm{dark},j}\right)=0.01$, respectively.
\emph{Right: } Histograms of the probability that the model dark current globally
matches the measured dark counts, estimated for every realization by parametric bootstraps.
}
\end{figure*}

The results of our validation tests are summarized in Fig.~\ref{fig:darksubstat}.
Apart from a $\sim 15$\% underestimation in the first bin at $z=2.68$ due to
the nearby detector edge (subsequently corrected in the science observations),
the average model dark current is consistent with the observed dark current.
The absolute mean relative difference in the remaining redshift bins is $<1$\%,
indicating that the global systematic error of our dark estimation procedure is
negligible. The scatter in the mean relative difference in our $\Delta z=0.04$
bins provides an upper limit to small-scale systematics (likely caused by
fixed-pattern noise and flatfield effects) of $\la 3$\%. The scatter between
individual measurements per redshift bin is dominated by Poisson fluctuations of
the measured dark counts around the mean implied by the model.
We tested this by computing the Poisson probability
\begin{equation}
P\left(\ge N_j|B_{\mathrm{dark},j}\right) = 1-\sum_{k=0}^{N_j-1}\frac{B_{\mathrm{dark},j}^k e^{-B_{\mathrm{dark},j}}}{k!}
\end{equation}
of detecting at least $N_j$ dark counts in redshift bin $j$ given the model with
mean $B_{\mathrm{dark},j}$, and comparing that number to the actual
fraction of measurements fulfilling this condition. The result of this exercise
is shown in the middle panels of Fig.~\ref{fig:darksubstat}. The mild deviations
from the identity line indicate mild deviations from pure Poisson statistics
due to detector effects (fixed-pattern noise and uncorrected flatfield) or
imperfections in the dark current model (smoothing and scaling required,
possibly poor match to ambient conditions). For our adopted reduction parameters
the number of strong background undersubtractions indicating a spurious
statistically significant signal
($P\left(\ge N_j|B_{\mathrm{dark},j}\right)<0.01$ highlighted in red in the left panels)
is very similar to the one expected from Poisson fluctuations around the modeled
mean dark. This also holds for the high-probability tail for significant
oversubtractions of the dark current. We conclude that the COS detector counts
are reasonably described by a Poisson distribution, enabling statistical
significance tests for our \ion{He}{2} effective optical depth measurements. 

Finally, we also performed parametric bootstraps to check the overall accuracy
of each dark current model as a function of redshift. For each of the 100 
validation datasets with 85 $\Delta z=0.04$ bins and a slowly varying mean
$B_{\mathrm{dark},j}$ per bin $j$ we drew $10^5$ Poisson samples
$\left\{M_1,\cdots,M_{85}\right\}$. For these $10^5$ mock datasets
$\left\{M_1,\cdots,M_{85}\right\}$ we computed the respective
likelihood of the model
\begin{equation}
L\left(B_\mathrm{dark}|M\right) = \prod_{j=1}^{85}\frac{B_{\mathrm{dark},j}^{M_j} e^{-B_{\mathrm{dark},j}}}{M_j!}
\end{equation}
to sample the expected likelihood distribution. We then computed the actual
likelihood of the validation dataset $L\left(B_\mathrm{dark}|N\right)$
and evaluated its consistency with the likelihood distribution by computing its
two-sided probability to $L\left(B_\mathrm{dark}|M\right)$. 
The right panels of Fig.~\ref{fig:darksubstat} shows the resulting histogram of
probabilities for the considered 100 validation datasets. While the high probabilities
indicate a reasonable agreement between the model and the data, the tail toward low
probabilities is due to the realizations with the strongest individual outliers
highlighted in the left panels. Shifting the dark current model by a few percent
results in a much stronger low-probability tail, indicating that the test is
sensitive and that the modeled dark current is accurate to a few percent,
in agreement with the propagated statistical error.

\subsection{The UV (extra-)Galactic Sky Background}
\label{sect:galuvbappendix}
A commonly overlooked source of open-shutter background in \textit{HST}/COS
observations is the diffuse UV `sky' background that is due to dust-scattered
Galactic UV starlight, \ion{H}{2} two-photon emission, and a small extragalactic component
\citep[e.g.][]{seon11,hamden13,murthy14}.
In addition, emission lines from warm-hot Galactic halo gas
\citep[e.g.][]{martin90a,korpela06,welsh07} and H$_2$ Lyman-Werner fluorescence lines
\citep[e.g.][]{martin90b,korpela06} have been detected. The rich H$_2$ fluorescence spectrum
\citep[e.g.][]{sternberg89} will appear as quasi-continuous emission in COS G140L spectra
($R\simeq 160$ for extended sources), while the G130M grating may preserve some spectral structure.

\citet{murthy14} tabulated the sky background in the \textit{GALEX} FUV and NUV bands at
a resolution of $2\arcmin$. We extracted all individual visit FUV tiles centered around
our targets within a radius of $2\arcmin$, and estimated the FUV sky background for each target by
calculating the exposure-time-weighted mean FUV flux\footnote{
It is customary in this field to express fluxes in photons\,cm$^{-2}$\,s$^{-1}$\,\AA$^{-1}$\,sr$^{-1}$.
At the effective wavelength of the \textit{GALEX} FUV filter ($\lambda_\mathrm{eff}=1539$\,\AA)
1\,photon\,cm$^{-2}$\,s$^{-1}$\,\AA$^{-1}$\,sr$^{-1}$$=3.03\times 10^{-22}$\,erg\,cm$^{-2}$\,s$^{-1}$\,\AA$^{-1}$\,arcsec$^{-2}$.}
to minimize the impact of photon noise in shallow \textit{GALEX} observations.
We then converted these fluxes to sky background counts in the COS aperture
($B_\mathrm{sky}$) via the target exposure times and COS sensitivity curves,
assuming $f_\lambda=\mathrm{const.}$ in agreement with spectroscopic
observations at high Galactic latitude \citep{seon11}.
Assuming this background spectrum we account for the diffuse UV emission and (approximately)
the quasi-continuous H$_2$ fluorescence, as the \textit{GALEX} FUV filter includes the H$_2$
Lyman band fluorescence. A more accurate sky subtraction would require either extensive COS
blank-sky calibrations near every target or STIS slit spectra (almost infeasible for our targets
due to low instrument sensitivity).
The fluxes that we subtracted from the COS spectra are small (4--$11\times 10^{-19}$\,erg\,cm$^{-2}$\,s$^{-1}$\,\AA$^{-1}$)
but non-negligible, as they would set a floor to the \ion{He}{2} effective
optical depth of $\tau_\mathrm{eff,HeII}\sim 5$ for our faint targets with
continuum fluxes $f_\lambda\simeq 10^{-16}$\,erg\,cm$^{-2}$\,s$^{-1}$\,\AA$^{-1}$.

We accounted for a random flux error of 10\% in our total error budget, somewhat
larger than the percent-level variations between individual \textit{GALEX} visits
that are due to Poisson noise and possible structure below the 2\arcmin\ tiling
scale \citep{murthy14}. There may be systematic errors due to imperfectly blocked
geocoronal emission in the \textit{GALEX} FUV filter. \citet{murthy14b} separated
this contamination into a time-varying component with minimum at local midnight
and a sun-angle-dependent local-midnight background with an offset of
$\sim 1.6\times 10^{-19}$\,erg\,cm$^{-2}$\,s$^{-1}$\,\AA$^{-1}$\,arcsec$^{-2}$
that could still include some residual geocoronal emission.
Indeed, the minimum background of $\sim 0.9\times 10^{-19}$\,erg\,cm$^{-2}$\,s$^{-1}$\,\AA$^{-1}$\,arcsec$^{-2}$
measured by \citet{seon11} in spectra that excluded geocoronal emission suggests
some residual geocoronal contamination in the \citet{murthy14} sky background values.
However, our \textit{HST}/COS measurements of scattered geocoronal emission
(Section~\ref{sect:scatterappendix}) indicate that the \citet{murthy14} values
are not strongly overestimated. 

\subsection{COS Aperture Sky Acceptance and Scattered Geocoronal Ly$\alpha$ Emission}
\label{sect:scatterappendix}

In the COS FUV channel the grating performs the diffraction, aberration correction
(\textit{HST}'s spherical aberration and aberration internal to COS) and focusing
of the incoming light, making it the most sensitive UV spectrograph ever flown on
\textit{HST} \citep{green12}. As a consequence, the COS entrance aperture is out
of focus, such that the sky acceptance region does not have a sharp
edge\footnote{The commonly quoted nominal diameter of $2.5\arcsec$ approximately accounts for the increasing vignetting with off-axis angle \citep{green12}.}.
This has important ramifications for the determination of low-level flux in COS
spectra, as intense geocoronal emission causes significant contamination even
far away from the line center due to the small but non-zero COS aperture
transmission. The shape and width of the geocoronal line profile is governed by
the COS aperture transmission, with minor modifications due to the COS
line-spread function. 

\citet{green12} used dispersed-light acquisitions to determine the COS aperture
transmission in the dispersion direction out to radii $\pm 3\arcsec$ from the
aperture center, corresponding to $\pm 10.5$\,\AA\ for the COS G140L grating.
While their transmission profile approximately matches the shape of geocoronal
Ly$\alpha$ emission in our G140L spectra, the observed Ly$\alpha$ profiles are
more extended (at least to $\pm 15$\,\AA, maybe further if Ly$\alpha$ is
particularly strong). We obtained $>700$ FUV ACQ/PEAKD peak-up sequences taken
until October 2014 from the \textit{HST} archive, and constructed the
transmission profile of the COS Primary Science Aperture by normalizing
individual sequences at zero offset, followed by averaging individual
measurements at each offset. To obtain meaningful results at offsets $>2\arcsec$
from the aperture center only the 300 sequences with more than 5000 counts at
zero offset were considered. Similarly to \citet{green12} we measure an aperture
transmission $<2$\% at offsets $>2\arcsec$, but we see tentative evidence for
non-zero transmission at the maximum covered offset of $\pm 4\arcsec$
($-4\arcsec$: mean $0.0012\pm 0.0003$, median $0.0009$;
$+4\arcsec$: mean $0.0004\pm 0.0003$, median zero),
corresponding to $\pm 14$\,\AA\ around geocoronal Ly$\alpha$ in a G140L spectrum.
The large contrast (1--4$\times 10^5$) between the geocoronal Ly$\alpha$ peak
flux and the residual flux in highly saturated \ion{He}{2} absorption regions
requires further characterization of the COS aperture sky acceptance at even
larger off-axis angles. For our observations we replaced the spectral range
around geocoronal Ly$\alpha$ with shadow data to limit the impact of the profile
wings caused by the out-of-focus aperture, and flagged the region with remaining
geocoronal contamination.

The extended wings of the COS line-spread function also contribute to the wings
of the observed geocoronal Ly$\alpha$ profile. \citet{kriss11} modeled the COS
G130M and G160M line-spread functions including mid-frequency wavefront errors
and scattering due to micro-roughness on the \textit{HST} primary mirror.
The published COS G140L line-spread function only includes the former effect
\citep{ghavamian09}. Scattered light from the gratings is below the maximum
specified level, but only upper limits were obtained in ground tests
\citep{osterman02}, and grating scatter was never quantified on orbit.

\begin{deluxetable}{llll}
\tabletypesize{\footnotesize}
\tablewidth{0pt}
\tablecaption{\label{tab:lightscatterflux}Systematic Flux Variations near Geocoronal Ly$\alpha$}
\tablehead{
\colhead{Object} &\colhead{$f_{1165}$\tablenotemark{a}} &\colhead{$f_{1269}$\tablenotemark{b}} &\colhead{Note}}
\startdata
SDSS~J1319$+$5202 &$1.11_{-0.27}^{+0.27}$ &$0.73_{-0.13}^{+0.14}$ &visit 1, sun alt. $>60\degr$\\
SDSS~J1319$+$5202 &$0.13_{-0.09}^{+0.10}$ &$0.11_{-0.06}^{+0.06}$ &visit 1, sun alt. $<0\degr$\\
SDSS~J1319$+$5202 &$<0.36$                &$0.47_{-0.23}^{+0.23}$ &visit 2, sun alt. $>30\degr$\\
SDSS~J1319$+$5202 &$<0.13$                &$<0.08$                &visit 2, sun alt. $<-40\degr$\\
SDSS~J1711$+$6052 &$0.77_{-0.15}^{+0.14}$ &$0.68_{-0.08}^{+0.09}$ &visit 1, sun alt. $>80\degr$\\
SDSS~J1711$+$6052 &$<0.30$                &$<0.17$                &visit 1, sun alt. $<0\degr$\\
SDSS~J1711$+$6052 &$1.56_{-0.30}^{+0.33}$ &$0.89_{-0.15}^{+0.15}$ &visit 2, sun alt. $>80\degr$\\
SDSS~J1711$+$6052 &$<0.14$                &$0.14_{-0.08}^{+0.09}$ &visit 2, sun alt. $<0\degr$\\
CSO~0806          &$<0.16$                &$0.35_{-0.11}^{+0.12}$ &sun alt. $>0\degr$\\
CSO~0806          &$<0.12$                &$<0.06$                &sun alt. $<0\degr$
\enddata
%\tablecomments{}
\tablenotetext{a}{Mean 1150--1180\,\AA\ flux density in $10^{-17}$\,erg\,s$^{-1}$\,cm$^{-2}$\,\AA$^{-1}$ with $1\sigma$ statistical error or $1\sigma$ upper limit.}
\tablenotetext{b}{Mean 1249--1289\,\AA\ flux density in $10^{-17}$\,erg\,s$^{-1}$\,cm$^{-2}$\,\AA$^{-1}$ with $1\sigma$ statistical error or $1\sigma$ upper limit.}
\end{deluxetable}

Our long exposures reaching fluxes
$f_\lambda\sim 10^{-18}$\,erg\,cm$^{-2}$\,s$^{-1}$\,\AA$^{-1}$ combined with our
accurate dark subtraction enabled an on-orbit determination of the G140L grating
scatter by testing for residual geocoronal Ly$\alpha$ emission in regions of
likely negligible COS aperture transmission.
Specifically, we considered three targets with intrinsically low flux near
geocoronal Ly$\alpha$ after dark and sky subtraction, either due to
strong \ion{He}{2} Ly$\alpha$ or \ion{H}{1} Lyman continuum absorption.
We measured the mean flux in two wavelength windows unaffected by geocoronal
\ion{O}{1} and \ion{N}{1} (1150--1180\,\AA\ and 1249--1289\,\AA, corresponding
to offsets of $\sim 15\arcsec$ from the COS aperture center for geocoronal Ly$\alpha$),
restricting each dataset in sun altitude (i.e.\ geocoronal Ly$\alpha$ flux).
Our measurements listed in Table~\ref{tab:lightscatterflux} revealed
systematically higher fluxes during orbital day, and also statistically
significant variations between separate visits of the same target.
This strongly suggested the presence of geocoronal Ly$\alpha$ emission in the
low-flux regions, as Ly$\alpha$ contributes $\sim 90$\% of the incoming photons
in a typical observation of a faint target. These photons are unlikely to have
been transmitted through the out-of-focus COS aperture due to their large
offset angle, but are more likely to originate from the G140L grating.

\begin{deluxetable}{llllll}
\tabletypesize{\footnotesize}
\tablewidth{0pt}
\tablecaption{\label{tab:lightscatterobs}Archival Datasets for COS G140L Scattered Light Calibration}
\tablehead{
\colhead{Dataset} &\colhead{Prog.} &\colhead{RA (J2000)} &\colhead{DEC (J2000)} &\colhead{[$f_\lambda^\mathrm{Gal}$]\tablenotemark{a}} &\colhead{Note\tablenotemark{c}}}
\startdata
\dataset[lb8710060]{\texttt{lb8710060}} &11860 &$21^\mathrm{h}51^\mathrm{m}27\fs14$ &$+28\degr45\arcmin16\farcs7$ &$25.36$\tablenotemark{b} &A\\
\dataset[lbn6e4060]{\texttt{lbn6e4060}} &12414 &$21^\mathrm{h}51^\mathrm{m}43\fs31$ &$+28\degr49\arcmin30\farcs6$ &$26.65$\tablenotemark{b} &A\\
\dataset[lbw3e4060]{\texttt{lbw3e4060}} &12775 &$21^\mathrm{h}51^\mathrm{m}45\fs10$ &$+28\degr51\arcmin23\farcs7$ &$26.89$\tablenotemark{b} &A\\
\dataset[lc1va8010]{\texttt{lc1va8010}} &12870 &$15^\mathrm{h}44^\mathrm{m}53\fs61$ &$+25\degr53\arcmin48\farcs8$ &$10.70$\tablenotemark{b} &A\\
\dataset[lc6201010]{\texttt{lc6201010}} &13108 &$07^\mathrm{h}48^\mathrm{m}33\fs73$ &$-67\degr45\arcmin07\farcs9$ &$37.63$ &B\\
\dataset[lc6202010]{\texttt{lc6202010}} &13108 &$07^\mathrm{h}48^\mathrm{m}33\fs73$ &$-67\degr45\arcmin07\farcs9$ &$37.63$ &B\\
\dataset[lb1s02010]{\texttt{lb1s02010}} &11742 &$13^\mathrm{h}04^\mathrm{m}11\fs99$ &$+29\degr53\arcmin48\farcs8$ &$4.40$ &C\\
\dataset[lbj8b1010]{\texttt{lbj8b1010}} &12249 &$17^\mathrm{h}11^\mathrm{m}34\fs41$ &$+60\degr52\arcmin40\farcs3$ &$9.51$ &D\\
\dataset[lbj8b2010]{\texttt{lbj8b2010}} &12249 &$17^\mathrm{h}11^\mathrm{m}34\fs41$ &$+60\degr52\arcmin40\farcs3$ &$9.51$ &D\\
\dataset[lbj8d1010]{\texttt{lbj8d1010}} &12249 &$13^\mathrm{h}19^\mathrm{m}14\fs20$ &$+52\degr02\arcmin00\farcs1$ &$6.00$ &D\\
\dataset[lbj8d2010]{\texttt{lbj8d2010}} &12249 &$13^\mathrm{h}19^\mathrm{m}14\fs20$ &$+52\degr02\arcmin00\farcs1$ &$6.00$ &D
\enddata
%\tablecomments{}
\tablenotetext{a}{\textit{GALEX} diffuse Galactic FUV emission in the COS aperture in $10^{-19}$\,erg\,s$^{-1}$\,cm$^{-2}$\,\AA$^{-1}$ \citep{murthy14}.}
\tablenotetext{b}{Pointing not covered by \textit{GALEX}. Average \citet{murthy14} measurements within $r<20$--30\arcmin\ of the pointing instead of $r<2\arcmin$.}
\tablenotetext{c}{A: Deliberate airglow observation. B: Blank sky observed due to acquisition failure. C: Optically thick Lyman limit system. D: Strong \ion{He}{2} absorption contaminated by sky emission and scattered light (i.e.\ assume $\tau_\mathrm{eff,HeII}=\infty$).}
\end{deluxetable}

\begin{figure*}
\includegraphics[width=\textwidth]{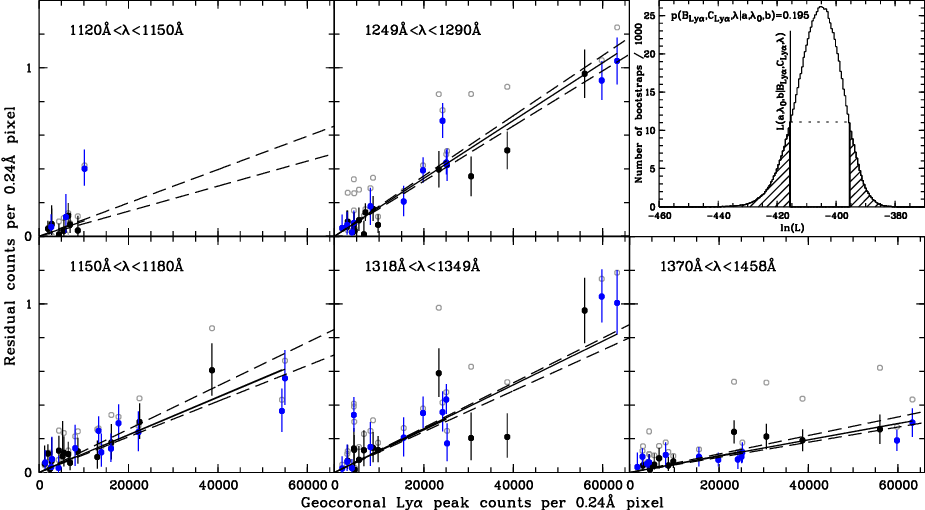}
\caption[]{\label{fig:lightscatter}
Residual counts above the background as a function of geocoronal Ly$\alpha$ peak
counts in 5 indicated wavelength ranges without clearly noticable geocoronal
contamination for the time-split G140L dataset in Table~\ref{tab:lightscatterobs}.
Open symbols mark dark-subtracted data (not used), whereas filled symbols show
the data after dark and sky subtraction (Appendix~\ref{sect:galuvbappendix})
with Poisson error bars ($1\sigma$, $68.26$\% confidence).
Black (blue) filled circles mark data from sky and Lyman limit system observations
(\ion{He}{2} trough observations). Solid lines show our best-fit model
(Equation~\ref{eq:lightscatter} fitted to the filled symbols),
whereas the dashed lines show the projected $1\sigma$ error estimated from
1000 parametric bootstraps. The upper right panel shows the likelihood
distribution estimated from $10^6$ parametric bootstraps.
The two-sided $p$-value of the maximum likelihood of the actual data is
$0.195$, indicating that our empirical model reasonably describes the data.
}
\end{figure*}

Because this so far unquantified systematic effect would clearly affect our
measurements, we embarked on a closer investigation using archival
\textit{HST}/COS G140L spectra of faint targets covering geocoronal Ly$\alpha$
in the 1105\,\AA\ setup. We used the few G140L blank sky observations taken as
part of the COS calibrations or if a science target had been too bright to
observe\footnote{\anchor{http://www.stsci.edu/hst/cos/calibration/airglow.html}{http://www.stsci.edu/hst/cos/calibration/airglow.html}},
and added two sufficiently deep observations that had blank acquisition exposures
(target too faint to acquire). The quasars from Table~\ref{tab:lightscatterflux}
were added to this sparse dataset assuming $\tau_\mathrm{eff,HeII}=\infty$
for the two \ion{He}{2} sightlines. Multi-orbit observations of our final
calibration dataset (Table~\ref{tab:lightscatterobs}) were split into two or
three exposures according to sun altitude to sample a sufficient range in the
accumulated geocoronal Ly$\alpha$ flux while still reaching sufficient depth.
The calibration data were reduced with the same routines as our science data.
Our coaddition routine tracked the total geocoronal Ly$\alpha$ peak counts
present during the uneven exposure at every wavelength. We then selected five
wavelength ranges with vanishing source flux and without clearly noticable
geocoronal emission, and measured the residual counts above the background
for the maximum number of homogeneously exposed pixels. The average residual
counts per considered pixel gave an estimate of the scattered light.

Figure~\ref{fig:lightscatter} shows the measured residual counts per pixel as a
function of the total geocoronal Ly$\alpha$ peak counts for the five considered
wavelength ranges. Despite significant scatter due to limited depth and small
sample size there is a clear relation between the residual counts and the
Ly$\alpha$ counts that varies with proximity to Ly$\alpha$.
This points to the presence of residual geocoronal Ly$\alpha$ photons that
dominate the total geocoronal flux at all times. The two high-$z$ \ion{He}{2}
sightlines (SDSS~J1319$+$5202 and SDSS~J1711$+$6052, blue symbols in
Fig.~\ref{fig:lightscatter}) show a similar behavior as the blank sky observations,
indicating that their \ion{He}{2} absorption regions are affected by scattered light.
Fitting the data with linear functions in every wavelength range, we obtain
positive intercepts if we only subtract dark current (open circles in
Fig.~\ref{fig:lightscatter}), whereas the intercepts are consistent with zero
once we subtract the sky background
(Appendix~\ref{sect:galuvbappendix}, filled circles in Fig.~\ref{fig:lightscatter}).
Thus, our measurements confirm the existence of a sky background at an
amplitude consistent with the \textit{GALEX} measurements by \citet{murthy14}.

We fitted the dark- and sky-subtracted residuals with an empirical model that is
linear in Ly$\alpha$ counts $C_{\mathrm{Ly}\alpha}$ and Gaussian in wavelength $\lambda$,
\begin{equation}\label{eq:lightscatter}
B_{\mathrm{Ly}\alpha}\left(C_{\mathrm{Ly}\alpha},\lambda\right) = a C_{\mathrm{Ly}\alpha} e^{-\frac{\left(\lambda-\lambda_0\right)^2}{2b^2}},
\end{equation}
with the scattered-light amplitude $a$, the central wavelength of the Gaussian
$\lambda_0$ and its standard deviation $b$ as free parameters.
The Poisson likelihood for the counts (Equation~\ref{eq:likelihood}) was
maximized for $a=1.7348\times 10^{-5}$, $\lambda_0=1254.6$\,\AA\ and $b=100.9$\,\AA.
Statistical errors on the fit parameters were estimated by a
parametric bootstrap, refitting Poisson deviates of the total model counts
(i.e.\ all three background components $B_\mathrm{\mathrm{Ly}\alpha}+B_\mathrm{dark}+B_\mathrm{sky}$
including a typical 2\% statistical error in $B_\mathrm{dark}+B_\mathrm{sky}$)
and the Ly$\alpha$ counts.
The 1000 bootstrap parameter samples $\left\{a,\lambda_0,b\right\}$ provided an
estimate of the uncertainty of the scattered-light correction applied to the
\ion{He}{2} dataset, yielding a relative error of $\simeq 6$\% at 1200\,\AA$<\lambda<$1380\,\AA\
that increases to $\simeq 20$\% at 1100\,\AA.
A larger set of $10^6$ parametric bootstrap samples was used to evaluate the
goodness of fit via the likelihood distribution, shown in the top-right panel of
Fig.~\ref{fig:lightscatter}. The maximum likelihood of our fit
$L\left(a,\lambda_0,b|B_{\mathrm{Ly}\alpha},C_{\mathrm{Ly}\alpha},\lambda\right)$
lies within the likelihood distribution (two-sided $p$-value $0.195$),
indicating negligible variance in the data in addition to the assumed Poisson
count errors and the typical background subtraction error. There are likely
remaining systematic uncertainties due to the assumed simple model for the total
residual Ly$\alpha$ flux that is actually a combination of flux transmitted
through the COS aperture and scattered light from the grating.
Future more sophisticated modeling will require better knowledge of the COS
aperture transmission at large offsets from the center and a large set of deep
blank sky calibrations.

\subsection{The Total COS Background in Science Spectra}
\label{sect:bkgsumappendix}

The total background $B=B_\mathrm{dark}+B_\mathrm{sky}+B_{\mathrm{Ly}\alpha}$
accumulated in a science exposure was estimated and coadded while coadding the spectra.
In this way we accounted for the changing background conditions,
i.e.\ cosmic ray intensity, scattered airglow, solar activity and gain sag
(for the two separate visits of SDSS~J1711$+$6052).
To model the scattered light (Equation~\ref{eq:lightscatter}) we measured the
geocoronal Ly$\alpha$ peak counts $C_{\mathrm{Ly}\alpha}$ in every exposure
taken with the G140L grating in the 1105\,\AA\ setup. For the targets observed in
the 1230\,\AA\ and 1280\,\AA\ setups we had to proceed differently, as geocoronal
Ly$\alpha$ falls in the gap between the two detector segments.
For these exposures we either estimated the geocoronal Ly$\alpha$ counts with
portions of 1105\,\AA\ exposures of the same visit matching \textit{HST}'s orbit
(SDSS~J1253$+$6817 and SDSS~J2346$-$0016), used these exposures only in the quasar
continuum where scattered light is insignificant (HS~0911$+$4809),
or carefully compared the day and night portions of the spectra to assess the
contamination (HE~2347$-$4342 and HS~1700$+$6416). For HE~2347$-$4342 scattered
light was insignificant, whereas for HS~1700$+$6416 we considered only nighttime
data in the \ion{He}{2} absorption region due to significant earthshine at low
limb angles during orbital day. Background uncertainties were estimated and
propagated during coadding. In the final coadded spectra the relative
statistical error of the dark current ranges between 1 and $2.5$\%\ depending
on the number of dark monitoring exposures matching the ambient conditions
during the science observations (Appendix~\ref{sect:darkappendix}).
For the small sky background we assumed a relative error of 10\%
(Appendix~\ref{sect:galuvbappendix}), whereas the relative error of our
scattered-light correction is 6--20\% depending on wavelength
(Appendix~\ref{sect:scatterappendix}).

\begin{figure}
\includegraphics[width=\linewidth]{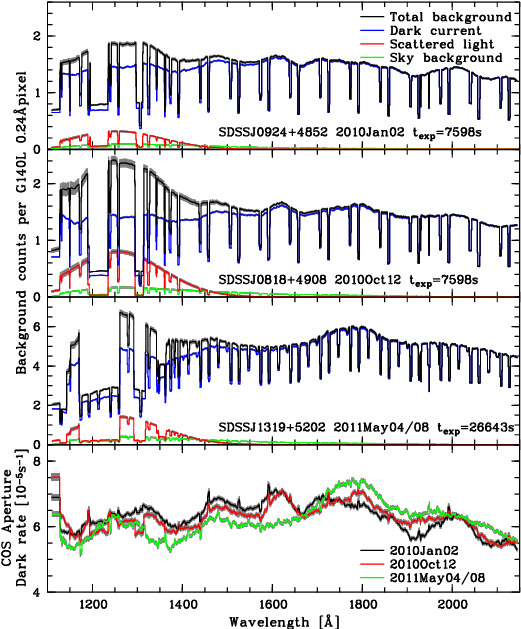}
\caption[]{\label{fig:bkgcomp}
\emph{Upper three panels:} Background components of three representative COS
G140L \ion{He}{2} spectra as a function of wavelength, labeled with observation
date and exposure time. We show the total background accumulated in the 25 pixel
extraction aperture in the respective exposure time (black), decomposed into
dark current (blue), scattered geocoronal Ly$\alpha$ emission (red),
and sky background (green). The respective $1\sigma$ errors are shown
in gray. Local decreases in the background components are due to different pixel
exposure times (differences in spectral coverage and grid wires at the four
detector offset positions, restriction to nighttime data around geocoronal lines).
\emph{Bottom panel:} Dark current rate in the COS aperture for these observations
(see labels) and the respective $1\sigma$ error (gray), illustrating the
effects of gain sag and changing environmental conditions.
}
\end{figure}

Figure~\ref{fig:bkgcomp} shows three representative examples of our final
post-processed background estimates and their components as a function of
wavelength. The total background is dominated by the dark current that
mildly varies with wavelength in the 25 pixel science aperture, and also
as a function of time (during each visit and between visits).
The sharp drops in the background are caused by different pixel exposure times
that are either due to grid wires, our restriction to nighttime data around
geocoronal lines, or limited spectral coverage at the shortest wavelengths.
The bottom panel of Fig.~\ref{fig:bkgcomp} shows the dark current rates of the
respective spectra. The dark rates of SDSS~J0924$+$4852 (January 2010) and
SDSS~J0818$+$4908 (October 2010) are directly comparable, as both were
observed in the same focal plane offset positions for the same amount of time.
The differences in the dark rates are manifestations of gain sag and changes
in solar activity. Gain sag is most prominent at
1300\,\AA$\la\lambda\la$1600\,\AA\ where the dark rate steadily
decreased with time. Longer wavelengths were less affected by gain sag due to
the lower G140L sensitivity, but eventually, exposures on bright standard
stars and partial overlap with the G130M and G160M traces caused gain sag
along the entire trace. Nevertheless, the dark rate increased with time at
$\lambda\ga 1700$\,\AA, likely due to increasing solar activity.
Our dark smoothing scale of 500 native pixels (40\,\AA\ for the G140L) was
chosen to capture these variations.

The modeled scattered light directly depends on the intensity of geocoronal
Ly$\alpha$ emission that mostly depends on the particular orbit configuration,
but also on solar activity. For instance, although SDSS~J0818$+$4908 and
SDSS~J0924$+$4852 were observed for the same amount of time,
their modeled scattered light background varies by a factor $\simeq 2.5$,
mostly due to the different daytime fractions of their orbits
($71.5$\% for SDSS~J0818$+$4908, 51\% for SDSS~J0924$+$4852) and the maximum
solar altitude. To preserve the sensitivity to high \ion{He}{2} effective
optical depths it was crucial to subtract the scattered light instead of just
minimizing it with nighttime data. In turn, we recommend faint \textit{HST}/COS
targets be observed in orbits with maximum nighttime fraction.

The uncertainty on the total background was estimated by propagating the errors
of the background components. In the wavelength range covering the \ion{He}{2}
Ly$\alpha$ absorption the relative background error varies between 2 and 8\%
depending on the particular observation. It is dominated either by the
uncertainty in the modeled dark current, or the uncertainty in the modeled
scattered light if the target was observed mainly during orbital day.
As detailed in Section~\ref{sect:tauml} we incorporated the background
uncertainty as a systematic error to our \ion{He}{2} effective optical depth
measurements.

\clearpage

\section{Measured \ion{He}{2} effective optical depths}
\LongTables
\begin{deluxetable}{lcrrr}
\tabletypesize{\footnotesize}
\tablewidth{0pt}
\tablecaption{\label{tab:he2tau}Measured \ion{He}{2} effective optical depths}
\tablehead{
\colhead{Quasar}&\colhead{$z$}&\colhead{$\tau_\mathrm{eff,HeII}$}&\colhead{stat. 1$\sigma$ error}&\colhead{sys. error}
}
\startdata
HS~1700$+$6416   &$2.32$ &$1.52$ &$^{+0.60}_{-0.40}$   &$^{+0.00}_{-0.00}$\\
                 &$2.44$ &$1.43$ &$^{+0.22}_{-0.21}$   &$^{+0.01}_{-0.00}$\\
                 &$2.48$ &$1.20$ &$^{+0.21}_{-0.18}$   &$^{+0.00}_{-0.00}$\\
                 &$2.52$ &$1.23$ &$^{+0.21}_{-0.18}$   &$^{+0.00}_{-0.00}$\\
                 &$2.56$ &$1.79$ &$^{+0.16}_{-0.17}$   &$^{+0.02}_{-0.00}$\\
                 &$2.60$ &$1.78$ &$^{+0.08}_{-0.09}$   &$^{+0.00}_{-0.00}$\\
                 &$2.64$ &$1.33$ &$^{+0.04}_{-0.04}$   &$^{+0.00}_{-0.00}$\\
HS~1024$+$1849   &$2.68$ &$1.83$ &$^{+0.26}_{-0.23}$   &$^{+0.00}_{-0.00}$\\
                 &$2.72$ &$2.49$ &$^{+0.26}_{-0.20}$   &$^{+0.00}_{-0.00}$\\
                 &$2.76$ &$2.78$ &$^{+0.23}_{-0.19}$   &$^{+0.00}_{-0.00}$\\
Q~1602$+$576     &$2.68$ &$1.67$ &$^{+0.14}_{-0.13}$   &$^{+0.00}_{-0.00}$\\
                 &$2.72$ &$1.95$ &$^{+0.10}_{-0.10}$   &$^{+0.00}_{-0.00}$\\
                 &$2.76$ &$1.52$ &$^{+0.06}_{-0.06}$   &$^{+0.00}_{-0.00}$\\
                 &$2.80$ &$2.53$ &$^{+0.11}_{-0.11}$   &$^{+0.00}_{-0.00}$\\
HE~2347$-$4342   &$2.32$ &$1.08$ &$^{+0.14}_{-0.13}$   &$^{+0.00}_{-0.00}$\\
                 &$2.44$ &$1.51$ &$^{+0.10}_{-0.10}$   &$^{+0.00}_{-0.00}$\\
                 &$2.48$ &$1.21$ &$^{+0.09}_{-0.09}$   &$^{+0.00}_{-0.00}$\\
                 &$2.52$ &$1.32$ &$^{+0.09}_{-0.10}$   &$^{+0.00}_{-0.00}$\\
                 &$2.56$ &$1.60$ &$^{+0.07}_{-0.07}$   &$^{+0.00}_{-0.00}$\\
                 &$2.60$ &$1.25$ &$^{+0.03}_{-0.04}$   &$^{+0.00}_{-0.00}$\\
                 &$2.64$ &$1.48$ &$^{+0.03}_{-0.02}$   &$^{+0.00}_{-0.00}$\\
                 &$2.68$ &$1.88$ &$^{+0.02}_{-0.03}$   &$^{+0.00}_{-0.00}$\\
                 &$2.72$ &$1.80$ &$^{+0.02}_{-0.02}$   &$^{+0.00}_{-0.00}$\\
                 &$2.76$ &$5.05$ &$^{+0.07}_{-0.09}$   &$^{+0.02}_{-0.00}$\\
                 &$2.80$ &$1.93$ &$^{+0.01}_{-0.01}$   &$^{+0.00}_{-0.00}$\\
                 &$2.84$ &$3.42$ &$^{+0.01}_{-0.03}$   &$^{+0.01}_{-0.00}$\\
PC~0058$+$0215   &$2.68$ &$1.95$ &$^{+0.37}_{-0.31}$   &$^{+0.01}_{-0.00}$\\
                 &$2.72$ &$1.87$ &$^{+0.22}_{-0.15}$   &$^{+0.00}_{-0.01}$\\
                 &$2.76$ &$2.53$ &$^{+0.22}_{-0.21}$   &$^{+0.03}_{-0.00}$\\
                 &$2.80$ &$1.96$ &$^{+0.13}_{-0.13}$   &$^{+0.01}_{-0.00}$\\
SDSS~J0936$+$2927&$2.68$ &$2.12$ &$^{+0.37}_{-0.32}$   &$^{+0.04}_{-0.00}$\\
                 &$2.72$ &$1.37$ &$^{+0.16}_{-0.11}$   &$^{+0.00}_{-0.01}$\\
                 &$2.76$ &$1.80$ &$^{+0.12}_{-0.15}$   &$^{+0.03}_{-0.00}$\\
                 &$2.80$ &$4.48$ &$^{+\infty}_{-0.00}$ &$^{+0.00}_{-0.00}$\\
SDSS~J0818$+$4908&$2.68$ &$1.48$ &$^{+0.29}_{-0.25}$   &$^{+0.02}_{-0.00}$\\
                 &$2.72$ &$1.96$ &$^{+0.23}_{-0.20}$   &$^{+0.02}_{-0.00}$\\
                 &$2.76$ &$2.31$ &$^{+0.24}_{-0.19}$   &$^{+0.01}_{-0.01}$\\
                 &$2.80$ &$2.13$ &$^{+0.19}_{-0.15}$   &$^{+0.00}_{-0.01}$\\
                 &$2.84$ &$1.94$ &$^{+0.14}_{-0.13}$   &$^{+0.01}_{-0.00}$\\
                 &$2.88$ &$2.51$ &$^{+0.21}_{-0.18}$   &$^{+0.02}_{-0.00}$\\
HS~1157$+$3143   &$2.80$ &$2.54$ &$^{+0.30}_{-0.20}$   &$^{+0.00}_{-0.00}$\\
                 &$2.84$ &$2.34$ &$^{+0.14}_{-0.13}$   &$^{+0.00}_{-0.00}$\\
                 &$2.88$ &$3.42$ &$^{+0.47}_{-0.26}$   &$^{+0.00}_{-0.00}$\\
                 &$2.92$ &$2.10$ &$^{+0.13}_{-0.10}$   &$^{+0.00}_{-0.00}$\\
SDSS~J0924$+$4852&$2.68$ &$3.03$ &$^{+0.42}_{-0.32}$   &$^{+0.00}_{-0.00}$\\
                 &$2.72$ &$1.89$ &$^{+0.10}_{-0.08}$   &$^{+0.00}_{-0.00}$\\
                 &$2.76$ &$1.58$ &$^{+0.06}_{-0.06}$   &$^{+0.00}_{-0.00}$\\
                 &$2.80$ &$2.37$ &$^{+0.09}_{-0.08}$   &$^{+0.00}_{-0.00}$\\
                 &$2.84$ &$2.63$ &$^{+0.10}_{-0.09}$   &$^{+0.00}_{-0.00}$\\
                 &$2.88$ &$1.92$ &$^{+0.06}_{-0.06}$   &$^{+0.00}_{-0.00}$\\
                 &$2.92$ &$2.50$ &$^{+0.09}_{-0.09}$   &$^{+0.00}_{-0.00}$\\
SDSS~J1101$+$1053&$2.68$ &$3.39$ &$^{+\infty}_{-0.00}$   &$^{+0.00}_{-0.00}$\\  
                 &$2.72$ &$2.54$ &$^{+0.32}_{-0.27}$   &$^{+0.03}_{-0.00}$\\  
                 &$2.76$ &$3.94$ &$^{+1.06}_{-0.58}$   &$^{+0.37}_{-0.00}$\\  
                 &$2.80$ &$2.98$ &$^{+0.35}_{-0.26}$   &$^{+0.00}_{-0.00}$\\  
                 &$2.88$ &$3.89$ &$^{+1.05}_{-0.57}$   &$^{+0.26}_{-0.00}$\\  
                 &$2.92$ &$2.27$ &$^{+0.24}_{-0.21}$   &$^{+0.00}_{-0.00}$\\  
SDSS~J1237$+$0126&$2.68$ &$1.82$ &$^{+0.24}_{-0.21}$   &$^{+0.01}_{-0.00}$\\  
                 &$2.72$ &$1.94$ &$^{+0.18}_{-0.12}$   &$^{+0.00}_{-0.02}$\\  
                 &$2.76$ &$2.25$ &$^{+0.14}_{-0.15}$   &$^{+0.01}_{-0.00}$\\  
                 &$2.80$ &$3.78$ &$^{+0.61}_{-0.37}$   &$^{+0.00}_{-0.00}$\\  
                 &$2.84$ &$2.33$ &$^{+0.13}_{-0.13}$   &$^{+0.00}_{-0.00}$\\  
                 &$2.88$ &$3.09$ &$^{+0.25}_{-0.21}$   &$^{+0.00}_{-0.00}$\\    
                 &$3.08$ &$3.04$ &$^{+0.21}_{-0.19}$   &$^{+0.00}_{-0.00}$\\  
Q~0302$-$003     &$2.80$ &$2.07$ &$^{+0.18}_{-0.15}$   &$^{+0.00}_{-0.00}$\\  
                 &$2.84$ &$2.06$ &$^{+0.12}_{-0.11}$   &$^{+0.00}_{-0.00}$\\  
                 &$2.88$ &$4.69$ &$^{+1.25}_{-0.74}$   &$^{+0.00}_{-0.00}$\\  
                 &$2.92$ &$4.17$ &$^{+0.72}_{-0.44}$   &$^{+0.00}_{-0.00}$\\  
                 &$3.08$ &$5.11$ &$^{+1.22}_{-0.63}$   &$^{+0.00}_{-0.00}$\\  
                 &$3.12$ &$5.38$ &$^{+1.32}_{-0.73}$   &$^{+0.00}_{-0.00}$\\  
                 &$3.16$ &$5.51$ &$^{+\infty}_{-0.00}$ &$^{+0.00}_{-0.00}$\\
HS~0911$+$4809   &$2.72$ &$2.14$ &$^{+0.24}_{-0.15}$   &$^{+0.00}_{-0.02}$\\
                 &$2.76$ &$2.89$ &$^{+0.22}_{-0.23}$   &$^{+0.04}_{-0.00}$\\
                 &$2.80$ &$2.03$ &$^{+0.13}_{-0.10}$   &$^{+0.00}_{-0.00}$\\
                 &$2.84$ &$2.78$ &$^{+0.17}_{-0.17}$   &$^{+0.02}_{-0.00}$\\
                 &$2.88$ &$4.46$ &$^{+1.13}_{-0.42}$   &$^{+0.00}_{-0.05}$\\
                 &$3.08$ &$3.42$ &$^{+0.22}_{-0.23}$   &$^{+0.03}_{-0.00}$\\
                 &$3.12$ &$5.38$ &$^{+\infty}_{-0.00}$ &$^{+0.00}_{-0.00}$\\
                 &$3.16$ &$3.11$ &$^{+0.21}_{-0.15}$   &$^{+0.00}_{-0.00}$\\
                 &$3.20$ &$4.28$ &$^{+0.49}_{-0.36}$   &$^{+0.03}_{-0.00}$\\
                 &$3.24$ &$5.43$ &$^{+\infty}_{-0.00}$ &$^{+0.00}_{-0.00}$\\
SDSS~J1253$+$6817&$2.84$ &$2.89$ &$^{+0.15}_{-0.13}$   &$^{+0.01}_{-0.00}$\\
                 &$2.88$ &$2.76$ &$^{+0.13}_{-0.12}$   &$^{+0.01}_{-0.00}$\\
                 &$3.12$ &$3.41$ &$^{+0.20}_{-0.18}$   &$^{+0.02}_{-0.00}$\\
                 &$3.16$ &$3.11$ &$^{+0.17}_{-0.15}$   &$^{+0.00}_{-0.00}$\\
                 &$3.20$ &$5.27$ &$^{+\infty}_{-0.00}$ &$^{+0.00}_{-0.00}$\\
                 &$3.24$ &$5.39$ &$^{+\infty}_{-0.00}$ &$^{+0.00}_{-0.00}$\\
                 &$3.36$ &$5.39$ &$^{+\infty}_{-0.00}$ &$^{+0.00}_{-0.00}$\\
SDSS~J2346$-$0016&$2.84$ &$3.13$ &$^{+0.16}_{-0.15}$   &$^{+0.00}_{-0.00}$\\
                 &$2.88$ &$2.52$ &$^{+0.10}_{-0.09}$   &$^{+0.00}_{-0.00}$\\
                 &$3.12$ &$5.56$ &$^{+\infty}_{-0.00}$ &$^{+0.00}_{-0.00}$\\
                 &$3.16$ &$5.56$ &$^{+\infty}_{-0.00}$ &$^{+0.00}_{-0.00}$\\
                 &$3.20$ &$5.54$ &$^{+\infty}_{-0.00}$ &$^{+0.00}_{-0.00}$\\
                 &$3.36$ &$5.59$ &$^{+\infty}_{-0.00}$ &$^{+0.00}_{-0.00}$\\
                 &$3.40$ &$5.68$ &$^{+\infty}_{-0.00}$ &$^{+0.00}_{-0.00}$\\
                 &$3.44$ &$5.57$ &$^{+\infty}_{-0.00}$ &$^{+0.00}_{-0.00}$\\
SDSS~J1711$+$6052&$3.36$ &$5.17$ &$^{+0.90}_{-0.49}$   &$^{+0.30}_{-0.04}$\\
                 &$3.40$ &$4.55$ &$^{+0.44}_{-0.31}$   &$^{+0.07}_{-0.03}$\\
SDSS~J1319$+$5202&$3.20$ &$5.26$ &$^{+\infty}_{-0.00}$ &$^{+0.00}_{-0.00}$\\
                 &$3.36$ &$4.88$ &$^{+\infty}_{-0.00}$ &$^{+0.00}_{-0.00}$\\
                 &$3.40$ &$4.31$ &$^{+0.86}_{-0.51}$   &$^{+0.17}_{-0.00}$\\
                 &$3.44$ &$2.41$ &$^{+0.15}_{-0.14}$   &$^{+0.00}_{-0.00}$\\
                 &$3.48$ &$3.59$ &$^{+0.58}_{-0.35}$   &$^{+0.00}_{-0.00}$\\

\tablecomments{Sensitivity lower limits on $\tau_\mathrm{eff,HeII}$ are marked with infinite upper error.}
\end{deluxetable}

%\end{appendix}

\end{document}